\documentclass[twocolumn, resetfootnote]{aastex701}

\usepackage{xspace}
\usepackage[version=4]{mhchem} 
\graphicspath{{./}{figures/}}

\usepackage[T1]{fontenc}
\usepackage[inline]{enumitem}
\usepackage{amssymb}

\newcommand{\poseidon}{\texttt{POSEIDON}\xspace}

\newcommand\teff{T$_{\rm{eff}}$}

\newcommand\solmass{$M_{\odot}$}

\hypersetup{linkcolor=violet,citecolor=teal}

\newcommand{\Goddard}{NASA Goddard Space Flight Center, 8800 Greenbelt Road, Greenbelt, MD 20771, USA}
\newcommand{\Carnegie}{Earth and Planets Laboratory, Carnegie Science, 5241 Broad Branch Road, NW, Washington, DC 20015, USA}
\newcommand{\Zurich}{Department of Astrophysics, University of Z\"urich, Winterthurerstrasse 190, 8057 Z\"urich, Switzerland}
\newcommand{\PSUAA}{Department of Astronomy \& Astrophysics, The Pennsylvania State University, 525 Davey Laboratory, University Park, PA 16802, USA}
\newcommand{\PSUCEHW}{Center for Exoplanets and Habitable Worlds, The Pennsylvania State University, 525 Davey Laboratory, University Park, PA 16802, USA}
\newcommand{\APL}{Johns Hopkins APL, 11100 Johns Hopkins Rd, Laurel, MD 20723, USA}

\received{February 10, 2025}
\revised{January 16, 2026}
\accepted{January 27, 2026}

\submitjournal{AJ}

\newcommand{\figsetcapnum}{}
\newcommand{\figsetcaptitle}{}
\renewcommand{\figsetstart}{{\bf Fig. Set~}}
\renewcommand{\figsetend}{}
\renewcommand{\figsetnum}[1]{{\bf #1.}}
\renewcommand{\figsettitle}[1]{ {\bf #1}}
\renewcommand{\figsetplot}[1]{\epsscale{1.17}\plotone{#1}}
\renewcommand{\figsetgrpnote}[1]{\caption{#1}}
\renewcommand{\figsetgrpnum}[1]{ \renewcommand{\figsetcapnum}{#1}}
\renewcommand{\figsetgrptitle}[1]{ \renewcommand{\figsetcaptitle}{#1} }
\renewcommand{\figsetgrpstart}{\begin{figure*}[!ht]\renewcommand{\figurename}{Fig.}\renewcommand{\thefigure}{Set \figsetcapnum~$-$ \figsetcaptitle}
}
\renewcommand{\figsetgrpend}{\end{figure*}}
\begin{document}

\title{GEMS JWST: Transmission spectroscopy of TOI-5205b reveals significant stellar contamination and a metal-poor atmosphere}

\correspondingauthor{Caleb Ca\~nas}
\email{c.canas@nasa.gov}

\author[0000-0003-4835-0619]{Caleb I. Ca\~nas}
\altaffiliation{NASA Postdoctoral Program Fellow}
\affiliation{\Goddard}
\email{c.canas@nasa.gov}

\author[0000-0002-0746-1980]{Jacob Lustig-Yaeger}
\affiliation{\APL}
\email{jacob.lustig-yaeger@jhuapl.edu}

\author[0000-0002-8163-4608]{Shang-Min Tsai}
\affiliation{Department of Earth and Planetary Sciences, University of California, Riverside, CA, USA}
\email{shami.bern@gmail.com}

\author[0000-0002-8278-8377]{Simon M\"{u}ller}
\affiliation{\Zurich}
\email{simonandres.mueller@uzh.ch}

\author[0000-0001-5555-2652]{Ravit Helled}
\affiliation{\Zurich}
\email{rhelled@physik.uzh.ch}

\author[0000-0001-8401-4300]{Shubham Kanodia}
\affiliation{\Carnegie}
\email{skanodia@carnegiescience.edu}

\author[0000-0002-2457-272X]{Dana R. Louie}
\affiliation{Catholic University of America, Department of Physics, Washington, DC, 20064, USA}
\affiliation{\Goddard}
\affiliation{Center for Research and Exploration in Space Science and Technology II, NASA/GSFC, Greenbelt, MD 20771, USA}
\email{dana.r.louie@nasa.gov}

\author[0000-0001-6340-8220]{Giannina Guzm\'an Caloca}
\affiliation{Department of Astronomy, University of Maryland, College Park, MD 20742, USA}
\affiliation{\Goddard}
\email{gguzmanc@umd.edu}

\author[0000-0002-8518-9601]{Peter Gao}
\affiliation{\Carnegie}
\email{pgao@carnegiescience.edu}

\author[0000-0002-2990-7613]{Jessica Libby-Roberts}
\affiliation{\PSUAA}
\affiliation{\PSUCEHW}
\email{jer5346@psu.edu}

\author[0000-0003-3702-0382]{Kevin K.\ Hardegree-Ullman}
\affiliation{Caltech/IPAC-NASA Exoplanet Science Institute, 1200 E. California Blvd., MC 100-22, Pasadena, CA 91125, USA}
\email{khullman@ipac.caltech.edu}

\author[0000-0001-8020-7121]{Knicole D. Col\'on}
\affil{\Goddard}
\email{knicole.colon@nasa.gov}

\author[0000-0002-1483-8811]{Ian Czekala}
\affiliation{School of Physics \& Astronomy, University of St. Andrews, North Haugh, St. Andrews KY16 9SS, UK}
\email{iancze@gmail.com}

\author[0000-0003-1439-2781]{Megan Delamer}
\affiliation{\PSUAA}
\affiliation{\PSUCEHW}
\email{mmd6393@psu.edu}

\author[0000-0002-7127-7643]{Te Han}
\affil{Department of Physics \& Astronomy, The University of California, Irvine, Irvine, CA 92697, USA}
\email{teh2@uci.edu}

\author[0000-0002-9082-6337]{Andrea S.J. Lin}
\affiliation{Department of Astronomy, California Institute of Technology, 1200 E California Blvd, Pasadena, CA 91125, USA}
\email{asjlin@caltech.edu}

\author[0000-0001-9596-7983]{Suvrath Mahadevan}
\affiliation{\PSUAA}
\affiliation{\PSUCEHW}
\email{suvrath@astro.psu.edu}

\author[0000-0002-2739-1465]{Erin M. May}
\affiliation{\APL}
\email{Erin.May@jhuapl.edu}

\author[0000-0001-8720-5612]{Joe P. Ninan}
\affiliation{Department of Astronomy and Astrophysics, Tata Institute of Fundamental Research, Homi Bhabha Road, Colaba, Mumbai 400005,
India}
\email{indiajoe@gmail.com}

\author[0000-0002-4487-5533]{Anjali A. A. Piette}
\affiliation{School of Physics and Astronomy, University of Birmingham, Edgbaston, Birmingham B15 2TT, UK}
\email{a.a.a.piette@bham.ac.uk}

\author[0000-0001-7409-5688]{Gu\dh mundur Stef\'ansson}
\affil{Anton Pannekoek Institute for Astronomy, University of Amsterdam, Science Park 904, 1098 XH Amsterdam, The Netherlands}
\email{g.k.stefansson@uva.nl}

\author[0000-0002-7352-7941]{Kevin B. Stevenson}
\affiliation{\APL}
\email{kevin.stevenson@jhuapl.edu}

\author[0009-0008-2801-5040]{Johanna Teske}
\affil{\Carnegie}
\email{jteske@carnegiescience.edu}

\author[0000-0003-0354-0187]{Nicole L. Wallack}
\affil{\Carnegie}
\email{nwallack@carnegiescience.edu}

\begin{abstract}
Recent discoveries of transiting giant exoplanets ($R_p\gtrsim8\mathrm{~R_\oplus}$) around M dwarfs (GEMS) present an opportunity to investigate their atmospheric compositions and explore how such massive planets form around low-mass stars contrary to the prediction from formation models. We present the first transmission spectra of TOI-5205b, a short-period ($P=1.63~\mathrm{days}$) Jupiter-like planet ($M_p=1.08~\mathrm{M_J}$ and $R_p=0.94~\mathrm{R_J}$) orbiting an M4 dwarf ($M_\star=0.392~\mathrm{M_\odot}$, $R_\star=0.394~\mathrm{R_\odot}$). We obtained three transits using the PRISM mode of the JWST Near Infrared Spectrograph (NIRSpec) spanning $0.6-5.3$~\textmu{}m. The data reveal significant stellar contamination that is evident in the light curves as spot-crossing events and in the transmission spectra as a larger transit depth at bluer wavelengths. Atmospheric retrievals demonstrate that stellar contamination from unocculted starspots and faculae is the dominant component of the transmission spectrum at wavelengths $\lambda\lesssim3.0$~\textmu{}m, reducing the sensitivity to the presence of clouds or hazes in our models and preventing detection of $\mathrm{H_2O}$. The wavelength coverage enabled a robust detection of $\mathrm{CH_4}$ and $\mathrm{H_2S}$, which have detectable molecular features between $3.0-5.0$~\textmu{}m. For both clear or cloudy atmospheres, Bayesian retrievals consistently favored an atmosphere with sub-solar metallicity ($3\sigma$ upper limit of $\log\mathrm{[M/H]}\lesssim-1.24$) and super-solar C/O ratio ($3\sigma$ lower limit of $\log\mathrm{[C/O]}\gtrsim0.09$), although this may partly be driven by the non-detection of water due to stellar contamination. Planetary interior models predict a bulk metallicity of 10--20\%, which is larger than the atmospheric metallicity and suggests that the interior of TOI-5205b is decoupled from its atmosphere.
\end{abstract}

\section{Introduction} \label{sec:intro}
Short-period ($P<10$ days), Jupiter-mass planets were the first type of exoplanet discovered around main-sequence Sun-like stars \citep{mayor_jupiter-mass_1995}, but their formation process remains uncertain. The growing number of short-period \uline{G}iant \uline{E}xoplanets around \uline{M} dwarf \uline{S}tars (GEMS) presents additional complications to the theories of gas giant formation \citep[e.g.,][]{kanodia_toi-5205b_2023, delamer_toi-4201_2024}. GEMS are difficult to form through core accretion because the low disk masses and long orbital timescales for M dwarfs impede the efficient formation of massive planetary cores ($\sim10~\mathrm{M_\oplus}$) capable of initiating runaway gas accretion \citep[e.g.,][]{laughlin_core_2004, burn_new_2021}. These planets represent an extreme regime of planet formation for the mid- to late-M dwarfs because the high planet-to-star mass ratios require core masses that exceed the estimated dust mass in the protoplanetary disk \citep[e.g.,][]{morales_giant_2019,Quirrenbach2022}. 

Understanding giant planet formation around all types of stars is crucial for explaining the architectures of exoplanetary systems. Studies both inside \citep{raymond_building_2009, brasser_analysis_2016} and outside \citep{mulders_why_2021-1} our Solar System show that the presence of gas giant planets affects the formation and evolution of smaller terrestrial planets. Close-in giant exoplanets have inspired multiple theories of planet formation and evolution \citep[e.g.,][]{dawson_origins_2018,fortney_hot_2021}, with the dominant origin channel suggesting that these planets form at large separations from their host star and subsequently migrate to their present-day observed location through various mechanisms \citep[e.g.,][]{rice_origins_2022, jackson_statistical_2022,Wu2023}.  Alternatively, they could have formed at their present-day location, \citep[i.e., \textit{in-situ};][]{bodenheimer_models_2000}. The degeneracies among both migration and \textit{in-situ} formation limit direct connections between the present-day observations of a giant planet and its primordial location in the disk \citep{molliere_interpreting_2022}.

Studying the atmospheres of GEMS offers an opportunity to both characterize their chemical compositions and compare them with giant planets orbiting FGK stars. Any trends in the present-day atmospheric and bulk metallicities may provide insight into possible formation and evolution pathways, and thus a better understanding of giant planet formation processes for M dwarfs. This is especially useful for GEMS orbiting the lowest mass stars, such as TOI-5205b, a Jupiter-like planet ($M_p\sim1.08~\mathrm{M_J}$ and $R_p\sim1.03~\mathrm{R_J}$) orbiting an M4 dwarf. \cite{kanodia_toi-5205b_2023} used disk mass ratio scaling laws, based on results by \cite{2013ApJ...771..129A} for the Taurus region, to determine that there should be insufficient dust in a median Class II disk around such a star (estimated at $\sim4-5~\mathrm{M_\oplus}$) to initiate runaway gas accretion and form TOI-5205b. 

In this work, we characterize the atmosphere of TOI-5205b using three transits obtained with JWST as part of our large Cycle 2 program (GO 3171) --- \textit{Red Dwarfs and the Seven Giants: First Insights Into the Atmospheres of Giant Exoplanets around M dwarf Stars} \citep{2023jwst.prop.3171K}. In \S\ref{sec:surveydesign}, we describe the design of our survey, followed by descriptions of the observations and data reduction in \S\ref{sec:observation}. We recovered the atmospheric properties from the spectra using two independent approaches: forward modeling (\S\ref{sec:forwardmodelling}) and Bayesian retrievals (\S\ref{sec:retrievals}). In \S\ref{sec:stellarmetallicity}, we provide a new empirical measurement for the stellar metallicity of TOI-5205 and estimate the bulk metallicity of TOI-5205b in \S\ref{sec:interiormodeling}. We present our interpretation of the observed transmission spectra and discuss the implications for the formation of TOI-5205b in \S\ref{sec:discussion}.

\section{Survey Design}\label{sec:surveydesign}
Our GEMS JWST survey has two main components. First, we aim to measure atmospheric abundances of GEMS and directly compare them with the population of hot Jupiters orbiting FGK dwarfs and Solar System gas giants. Our goal is to explore whether the atmospheric mass-metallicity trends investigated by the community for planets around FGK dwarfs \citep[e.g.,][]{welbanks_mass-metallicity_2019,edwards_characterising_2023} are applicable to the population of GEMS (\S\ref{sec:atm_massmetallicity}). Second, we seek to perform the first study of bulk metallicities for GEMS to investigate whether there are any trends that may be a result of the formation processes involved (see \S\ref{sec:bulk_massmetallicity}). The results of our survey will be compared with constraints derived from JWST programs observing the atmospheres of Jovian planets orbiting FGK dwarfs, such as HD 149026b \citep[e.g.,][]{2023Natur.618...43B}, WASP-39b \citep[e.g.,][]{2023Natur.614..649J}, WASP-77 Ab \citep[e.g.,][]{2023ApJ...953L..24A}, WASP-80b \citep[e.g.,][]{2023Natur.623..709B}, HD 209458b \citep[e.g.,][]{2024ApJ...963L...5X}, HD 189733b \citep[e.g.,][]{2024Natur.632..752F}, and WASP-17b \citep[e.g.,][]{2025AJ....169...57G}.

\begin{deluxetable*}{ccccccccccc}
    \tabletypesize{\footnotesize}
    \tablecaption{Targets for the GEMS JWST survey (sorted by increasing planet mass) \label{tab:InputSample}}
    \tablehead{
    \colhead{Name} & \colhead{Pl. Mass} & \colhead{Pl. Radius} & \colhead{Period} & \colhead{Pl. $T_{\mathrm{eq}}$} & \colhead{\teff} & \colhead{St. Mass} & \colhead{J mag}  &  \colhead{References} \\
    \colhead{} & \colhead{$\mathrm{M_J}$} & \colhead{$\mathrm{R_J}$} & \colhead{d} & \colhead{K}  & \colhead{K} & \colhead{\solmass}  & \colhead{}}
    \startdata
TOI-3984 Ab    &  $0.14\pm0.03$      &  $0.71\pm0.02$     &  4.3533       &  567     &  3476 &  0.49 $\pm$ 0.02       &  11.93 $\pm$ 0.02  & \cite{canas_toi-3984_2023}\\
TOI-3757b      &  $0.27\pm0.03$ &  $1.07\pm0.04$  &  3.4388       &  758    &  3913  &  0.64 $\pm$ 0.02       &  12.00 $\pm$ 0.03 & \cite{kanodia_toi-3757_2022} \\
HATS-6b        &  $0.32\pm0.07$   &  $1.00\pm0.02$ &  3.3253       &  713     &  3724  &  0.57$^{+0.02}_{-0.03}$        &  12.05 $\pm$ 0.02 & \cite{hartman_hats-6b_2015} \\
HATS-75b       &  $0.49\pm0.04$   &  $ 0.88\pm0.02$     &  2.7887       &  770     &  3790  &  0.60 $\pm$ 0.01       &  12.48 $\pm$ 0.02 & \cite{jordan_hats-74ab_2022} \\
TOI-5293 Ab    &  $0.54\pm0.07$   &  $1.06\pm0.04$   &  2.9303       &  690     &  3586  &  0.54 $\pm$ 0.02       &  12.46 $\pm$ 0.03 & \cite{canas_toi-3984_2023} \\
TOI-3714b      &  $0.70 \pm 0.03$    &  $1.01 \pm 0.03$  &  2.1548       &  775     &  3660  &  0.53 $\pm$ 0.02       &  11.74 $\pm$ 0.02 & \cite{canas_toi-3714_2022} \\
TOI-5205b      &  $1.08^{+0.06}_{-0.05}$  &  $1.03\pm0.03^*$     &  1.6308       &  733     &  3430   &  0.39 $\pm$ 0.01       &  11.93 $\pm$ 0.03 & \cite{kanodia_toi-5205b_2023} \\
    \enddata    
    \tablenotetext{*}{See refined parameters in \autoref{tab:5205par}, as well as \autoref{app:depthdiff} for discussion regarding the radius discrepancy for TOI-5205b.}
\end{deluxetable*}

\subsection{Target selection}
This survey was designed for JWST operating in NIRSpec PRISM mode to provide high efficiency sampling across $0.6-5.3$ \textmu{}m to facilitate (i) the simultaneous measurement of water and methane abundances, if present in the atmosphere, and (ii) the constraint of any aerosols or stellar contamination that may be present across the optical and near infrared. To select our sample of GEMS, we queried the NASA Exoplanet Archive \citep{Akeson2014} along with then-recent publications (in January 2023) for transiting gas giant planets satisfying the following requirements:
\begin{enumerate*}[label=(\roman*)]
\item $T_\mathrm{eff}<4000~K$,
\item $R_p\gtrsim8~\mathrm{R_\oplus}$, 
\item a mass and radius precision $>3\sigma$, and
\item $J$ mag $> 11.5$ to ensure no saturation in the NIRSpec PRISM mode.
\end{enumerate*} 
These constraints produced nine planets from which we excluded (i) HATS-74~Ab \citep{jordan_hats-74ab_2022} due to the presence of a bright nearby stellar companion ($\Delta J = $ 2.6 mag at a separation of 0.844\arcsec) and (ii) Kepler-45b \citep{Johnson2010} as it was within $1\sigma$ in planetary mass of two other planets (HATS-75b and TOI-5293b) albeit with a much fainter host star ($J=13.75\pm0.02$). This left our remaining sample of seven giant planets that spanned $T_{eq}$ = 570 -- 850 K with host spectral types M0 -- M4 (see \autoref{tab:InputSample}). We observed all seven planets using the JWST Near Infrared Spectrograph (NIRSpec) PRISM mode to obtain low-resolution spectra spanning $0.6-5.3$ \textmu{}m across 18 transits.

\subsection{Mass - atmospheric metallicity trend}\label{sec:atm_massmetallicity}
The atmospheric metallicity controls the efficiency of radiative cooling for a nascent gas giant planet and may impact the minimum core mass necessary to trigger runaway gas accretion, leading to predictions of a linear mass-metallicity trend \citep[e.g.,][]{Ikoma2000,Movshovitz2010,Mordasini2014,atreya_origin_2018}. For the Solar System ice and gas giants, measurements of the atmospheric methane abundance demonstrate that these planets (i) are significantly enriched in heavy elements compared to the Sun and (ii) have decreasing enrichment with increasing planetary mass \citep[e.g.,][]{Fortney2013,Movshovitz2010,Mordasini2014,Guillot2015529}. The metallicities of the Solar System ice and gas giants are inferred from their methane abundances because of their low equilibrium temperatures ($T_{eq}\lesssim120$ K).

Measurements of atmospheric methane for hot Jupiters are difficult because the dominant carbon-bearing molecule is expected to be CO instead of methane due to their high temperatures \citep[$T\gtrsim1000$~K;][]{Goukenleuque2000,Fortney2020}. As such, the atmospheric metallicity of hot Jupiters is typically measured through the water abundance \citep[e.g.,][]{Sing2016}. GEMS have typical equilibrium temperatures $T\lesssim800$ K and are cooler than hot Jupiters, which increases the likelihood of detecting methane and water \citep{baxter_evidence_2021}. A key goal of our GEMS JWST survey is to compare our population of GEMS with both hot Jupiters (using water) and Solar System giant planets (with methane) to investigate the presence of trends between planetary mass and atmospheric metallicity \citep[e.g.,][]{welbanks_mass-metallicity_2019,Pinhas2019,Sun2024}.

\subsection{Mass - bulk metallicity trend}\label{sec:bulk_massmetallicity}
In the core accretion theory of planet formation, the planetary mass is expected to be inversely proportional to the bulk metallicity, and \textit{may} serve as a probe of the formation processes involved \citep[e.g.,][]{Miller2011,thorngren_mass-metallicity_2016,Guillot2023,Swain2024}. The bulk metallicity cannot be directly measured and therefore planetary evolution models are used to determine this value for non-inflated giant planets \citep[e.g.,][]{fortney_planetary_2007, thorngren_mass-metallicity_2016}. These estimates of bulk metallicity typically rely on various model assumptions (e.g., atmospheric composition, equations of state) that result in several possible solutions for a set of measured planetary parameters \citep[e.g.,][]{muller_synthetic_2021}. Measurements of the atmospheric metallicity reduce the reliance on theoretical models and can improve the accuracy of the inferred bulk composition \citep[e.g.,][]{thorngren_connecting_2019, muller_synthetic_2021,Bloot2023,acuna_gastli_2024,Sing2024} and facilitate the study of any trends with planetary mass.

\section{Observations and Data Reduction}\label{sec:observation}
In our proposal, we requested three transits of TOI-5205 to ensure that we obtained the precision necessary for our survey goal ($\sim0.3$ dex from simulated measurements of water and methane abundances, assuming equilibrium chemistry). Multiple visits proved to be a fortuitous decision given the degree of stellar contamination described below. We obtained three consecutive transits of TOI-5205b using the JWST NIRSpec PRISM in Bright Object Time Series (BOTS) mode \citep{Birkmann2022,Espinoza2023} on 2023 October 10, October 11, and October 13 (GO 3171; observation numbers 16, 17, and 18, respectively). The host star was used directly for target acquisition with the NRSRAPID readout pattern and the SUB32 array. All observations used the NRS1 detector with the SUB512 subarray ($32\times512$ pixels) and consisted of 15784 integrations with 4 groups per integration, providing a median observational cadence of 1.15 s. Each observation was 5.05 hours long and contained the full planet transit ($\sim1.4$ hours) with a minimum of one hour baseline before and after the transit. 

We analyzed the counts for all columns in row 15 (corresponding to the trace center) for each observation and determined that the data reached a maximum value of $\sim43,300$ ADU (see \hyperref[app:rawcounts]{Appendix \ref*{app:rawcounts}}). Previous works analyzing NIRSpec/PRISM data \citep[e.g.,][]{Carter2024} demonstrated that non-linearity may impact ramp fitting if the counts exceed $\gtrsim70-75\%$ the well depth ($\sim45875-49152$ ADU). None of the observations saturated and the effects of non-linearity are not observed given the counts for all observations (maximal counts are $<70\%$ the saturation threshold of $2^{16}$ or 65,536 ADU)\footnote{\url{https://jwst-docs.stsci.edu/jwst-near-infrared-spectrograph/nirspec-instrumentation/nirspec-detectors/nirspec-detector-performance}}.

We reduced the observations with two pipelines designed for JWST exoplanet time-series observations: \texttt{ExoTiC-JEDI}\footnote{\url{https://github.com/Exo-TiC/ExoTiC-JEDI}} \citep{Alderson2022} and \texttt{Eureka!}\footnote{\url{https://github.com/kevin218/Eureka}} \citep{Bell2022}. A detailed description of the reduction for each pipeline is presented in \hyperref[app:exotic]{Appendices \ref*{app:exotic}} -- \hyperref[app:wlcgen]{\ref*{app:wlcgen}}. The second visit (observation 17) with JWST showed a discontinuity in the light curve due to a tip-tilt event, and we excluded the first 2070 data points from further analysis (see \hyperref[app:tiptilt]{Appendix \ref*{app:tiptilt}}).

For both the \texttt{ExoTiC-JEDI} and \texttt{Eureka!} data sets, we generated white and spectroscopic light curves at native resolution from our 1D spectra using Stage 4 of the \texttt{Eureka!} pipeline (see \hyperref[app:spotconfig]{Appendix \ref*{app:spotconfig}}). The light curves were extracted for columns $30-491$ ($\sim0.520-5.579$~\textmu{}m). The white light curves were generated by integrating over all wavelengths and both pipelines reveal a $\sim6\%$ transit with spot-crossing events in each transit (see \autoref{fig:wlc}). The recovered transit depth was $\sim1\%$ shallower than the value derived using ground-based photometry from \cite{kanodia_toi-5205b_2023} and we attributed this discrepancy to uncorrected trends in the ground-based photometry given the limited baseline and suboptimal observing conditions on those nights (see \autoref{app:depthdiff} for more details).

\subsection{White light curve fits}\label{sec:wlcfits}
We performed a white light curve fit only to the \texttt{ExoTiC-JEDI} reduction. We fit the white light curves separately using a modified version of \texttt{juliet} \citep{juliet2019} that models spot crossing events. To mitigate the computational time, we binned the white light curves to a cadence of 5 s and only fit the data within 1.5 transit durations ($\pm1.4$ hours) from mid-transit. In this work, all fits to the white light curves and spectroscopic light curves employed a quadratic limb darkening law \citep{quad1,quad2,quad3} in which the coefficients ($u_1$ and $u_2$) were sampled using the $q_1-q_2$ parametrization presented in \cite{kipping_efficient_2013}. We did not use limb darkening grids to apply priors (or fix values) because of the limited temperature coverage \citep{wakeford_2022_7874921} for these grids\footnote{\url{https://zenodo.org/records/7874921}} (\teff{} $\sim$ 3430 K, see a comparison of limb darkening coefficients for the quadratic limb darkening law in \autoref{app:limbdark}). We modeled starspots using a semi-analytical model calculated with the \texttt{spotrod}\footnote{\url{https://github.com/bencebeky/spotrod}} package \citep[see details on the spot model in \autoref{app:spotconfig};][]{spotrod2014}. 

\begin{figure*}[tt]
\epsscale{1.17}
\plotone{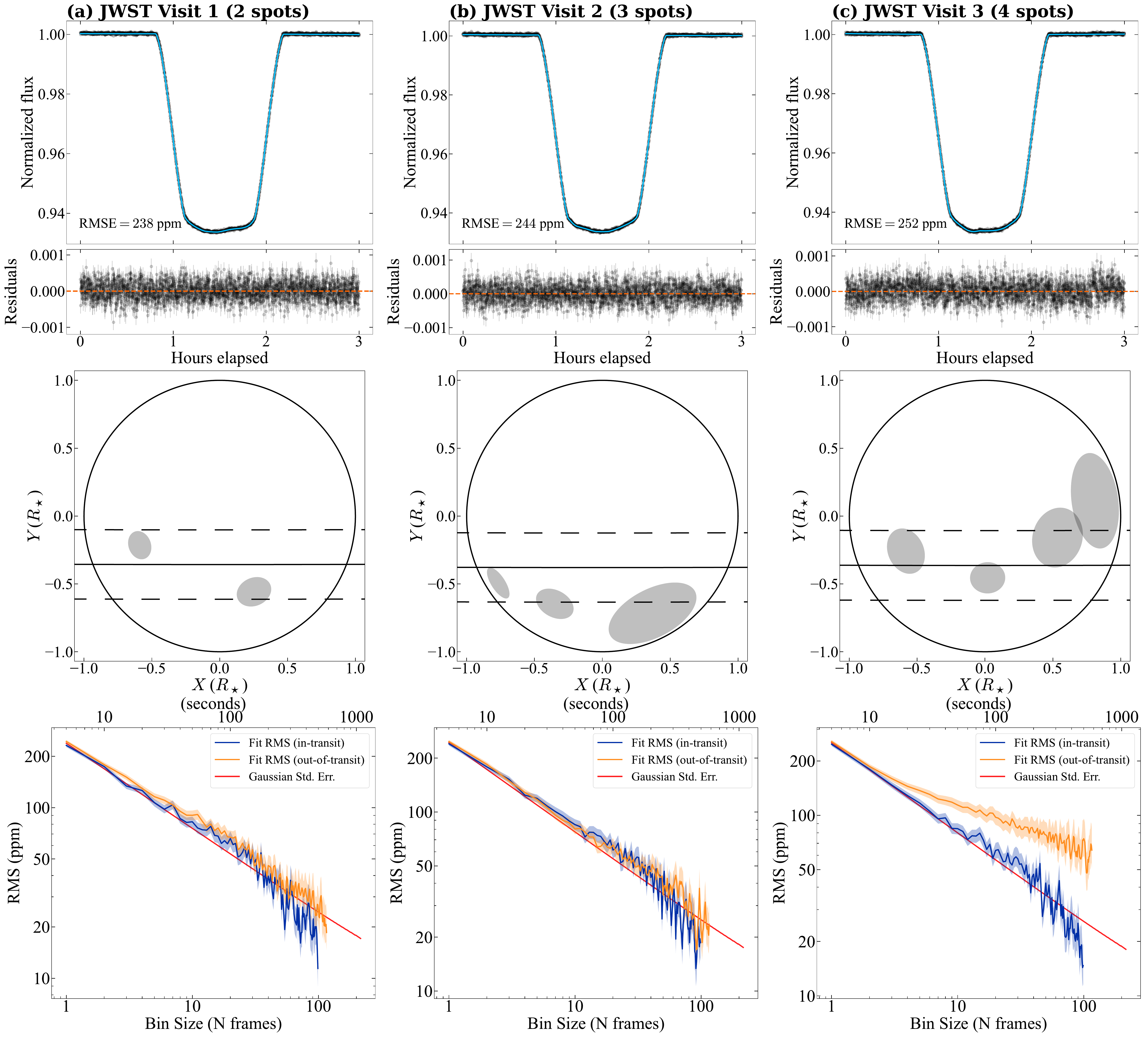}
\caption{Panels (a)-(c) JWST NIRSpec PRISM white light curves produced using \texttt{ExoTiC-JEDI} after binning to a cadence of 5 s. \textbf{Top row}: The data along with the best-fitting model (solid line) along with the residuals to the fit below. \textbf{Middle row}: The stellar surface and the adopted spot configuration.\footnote{The transparency of the spots is arbitrary and \textit{does not} reflect the spot flux ratio.} The solid line indicates the position of the transit chord (center of the planet) and the dashed lines mark the $\pm~R_p$ from the center of the transit chord. \textbf{Bottom row}: The RMS for each visit for in-transit (blue) and out-of-transit (orange) data. The prediction for Gaussian white noise is shown as a red solid line. The residuals to the model fits demonstrate there is no significant time correlated noise in-transit after modeling the spots.}
\label{fig:wlc}
\end{figure*}

We placed priors on the orbit parameters and adopted a circular orbit based on the results from \cite{kanodia_toi-5205b_2023}. We used a two-degree polynomial to model the trend known to exist in the NRS1 detector \citep[e.g.,][]{Espinoza2023,Moran2023}. We investigated various combinations of spot configurations by fitting the white light curves with a varying number of spots ($0-4$) for each transit (see more details in \autoref{app:spotconfig}). We adopted the spot configuration of 2 spots in visit 1 (observation 16), 3 spots in visit 2 (observation 17), and 4 spots in visit 3 (observation 18). The derived parameters are presented in \autoref{tab:5205par} while the spot parameters are listed in \autoref{tab:5205spotpar}. The best-fit is presented in \autoref{fig:wlc}. 

\startlongtable
\begin{deluxetable*}{lccccccc}
\tablecaption{System parameters for TOI-5205 \label{tab:5205par}}
\tablehead{\colhead{Name} &
\colhead{Units} &
\colhead{Prior$^a$} &
\colhead{Value} &
\colhead{Source} &
}
\startdata
\sidehead{Stellar parameters:}
~~~Stellar Mass ($M_\star$) & $\mathrm{M_\odot}$ & \nodata & $0.394\pm0.011$ & 1\\
~~~Stellar Radius ($R_\star$) & $\mathrm{R_\odot}$ & \nodata & $0.392\pm0.015$ & 1\\
~~~Effective Temperature ($T_{\mathrm{eff}}$) & K & \nodata & $3430\pm54$ & 1\\
~~~Surface Gravity ($\log g_\star$) & dex & \nodata & $4.84\pm0.03$ & 1\\
~~~Metallicity ($\mathrm{[Fe/H]}$) & dex & \nodata & $0.56\pm0.10$ & This work\\
\hline
\sidehead{Fitted transit parameters:}
~~~Period ($P$) & day & $\mathcal{N}(1.63,0.01)$ & $1.630731 \pm 0.000003$ & This work\\
~~~Time of mid-transit ($T_0$) & $\mathrm{BJD_{TDB}}$ & $\mathcal{N}(2460227.86,0.01)$ & $2460227.864967 \pm 0.000004$ & This work\\
~~~Eccentricity ($e$) & \nodata & Fixed & 0 & This work\\
~~~Argument of periastron ($\omega_\star$) & deg & Fixed & 90 & This work\\
~~~Scaled radius ($R_p/R_\star$) & \nodata & $\mathcal{U}(0,1)$ & $0.2475_{-0.0002}^{+0.0003}$ & This work\\
~~~Scaled semi-major axis ($a/R_\star$) & \nodata & $\mathcal{N}(10.94,0.22)$ & $10.695_{-0.005}^{+0.006}$ & This work\\
~~~Impact parameter ($b$) & \nodata & $\mathcal{U}(0,1)$ & $0.402 \pm 0.001$ & This work\\
\hline
\sidehead{Derived system parameters:}
~~~Inclination ($i$) & deg & \nodata & $87.847_{-0.006}^{+0.007}$ & This work\\
~~~Semi-major axis ($a$) & au & \nodata & $0.0195 \pm 0.0007$ & This work\\
~~~Planet Radius ($R_p$) & $\mathrm{R_J}$ & \nodata & $0.94 \pm 0.04$ & This work\\
~~~Planet Mass ($M_p$) & $\mathrm{M_J}$ & \nodata & $1.08\pm0.06$ & 1\\
~~~Equilibrium temperature ($T_\mathrm{eq}$) & K & \nodata & $742\pm12$ & This work\\
\enddata
\tablerefs{1) \cite{kanodia_toi-5205b_2023}}
\tablenotetext{a}{Normal priors with a mean of $X$ and standard deviation of $Y$ are denoted as $\mathcal{N}(X,Y)$. \\ Uniform priors between a lower limit of $X$ and upper limit of $Y$ are denoted as $\mathcal{U}(X,Y)$.}
\end{deluxetable*}

\subsection{Transmission spectra}
To derive the transmission spectra, we modeled the spectroscopic light curves of both the \texttt{ExoTiC-JEDI} and \texttt{Eureka!} reductions using stage 5 of the \texttt{Eureka!} pipeline modified to include spot modeling with \texttt{spotrod} (see \autoref{app:spotconfig} for a detailed description). We used the best-fitting spot configuration derived from the white light curve fits. The resulting spectra are presented in \hyperref[fig:spectracomp]{Fig. Set 2}. Both reductions (and all three transits) showed an increase in the transit depth towards bluer wavelengths, which we attributed to the effect of stellar contamination by unocculted spots \citep[the transit light source or TLS; e.g.,][]{Rackham2017}. At the pixel level, the differences between the two pipelines were minimal and most of the data agreed within $1\sigma$. Given the similarities between the spectra, we use the \texttt{ExoTiC-JEDI} reduction for all subsequent analysis in this work. In addition to the individual visit spectra, we created a co-added spectrum by taking the weighted average for each channel. 


\section{Atmospheric forward modeling}\label{sec:forwardmodelling}

\begin{deluxetable*}{lc|cc|cccccc}
\tablecaption{Parameters for the best-fitting models derived from \texttt{PICASO} (RCTE) and \texttt{VULCAN} (disequilibrium chemistry). \label{tab:picasovulcan} }
\tablehead{
\colhead{Parameter} &
\multicolumn{1}{c|}{Units} & 
\colhead{RCTE$^\dagger$} &
\multicolumn{1}{c|}{RCTE + Spots$^\dagger$} & 
\multicolumn{6}{c}{Disequilibrium Chemistry + Spots} 
}
\startdata
~~Atmospheric metallicity ($\mathrm{[M/H]}$) & dex & $-0.3$ & $-1.0$ & $-2.0$ & $-2.0$ & $-2.0$ & $-2.0$ & $-2.0$ & $-2.0$ \\
~~Atmospheric carbon-to-oxygen ratio (C/O)$^\ddag$ & \nodata   & 2.5    & 0.25   & 0.25   & 0.25   & 0.25   & 0.25   & 0.25   & 0.25 \\
~~Intrinsic Temperature ($T_\mathrm{int}$) & $K$   & 100    & 500    & 100    & 100    & 100    & 100    & 100    & 100\\
~~Heat redistribution factor ($r_{st}$) & \nodata  & 0.25   & 0.75   & 0.25   & 0.5    & 0.5    & 0.5    & 0.5    & 0.5\\
~~Spot coverage fraction ($f_\mathrm{spot}$) & \nodata & \nodata     & 0.3    & 0.3    & 0.3    & 0.3    & 0.3    & 0.3 & 0.3\\
~~Spot temperature ($T_\mathrm{spot}$) & K       & \nodata  & 3230   & 3330   & 3330   & 3330   & 3330   & 3330   & 3330\\
~~Eddy diffusive coefficient ($K_{zz}$) & dex & \nodata & \nodata    & 6      & 6      & 6      & 9      & 9      & 9\\
~~Sulfur enhancement (S/H) & \nodata & \nodata & \nodata             & 1      & 10     & 100    & 1      & 10     & 100\\
\hline
~~ Reduced chi-squared ($\chi^2_\nu$) & \nodata   & 21.592 & 4.223   & 4.080  & 4.065  & 4.064  & 4.083  & 4.066  & 4.059\\
~~ Bayesian Information Criterion (BIC) & \nodata & 3461.670 & $-3961.677$ & $-3974.521$ & $-3991.512$ & $-3965.990$ & $-3966.098$ & $-3975.400$ & $-3996.751$ \\
~~ Akaike Information Criterion (AIC) & \nodata & 3445.635 & $-3985.730$ & $-3994.565$ & $-4011.556$ & $-3986.034$ & $-3986.142$ & $-3995.444$ & $-4016.795$\\
\enddata
\tablenotetext{\dagger}{Parameters that are empty in a given column were not applicable to the specific model.}
\tablenotetext{\ddag}{The C/O in this table is the value reported by the Sonora Bobcat models \citep{Sonora2021} and represents multiples of $\mathrm{[C/O]}_\odot=0.458$ based on the protosolar abundances from \cite{Lodders2010}.}
\tablecomments{The \texttt{PICASO} grid parameters are described in \S\ref{sec:picaso} while the \texttt{VULCAN} grid parameters are described in \S\ref{sec:vulcan}. There are two columns for the RCTE models to reflect fits with and without spots. Each column in the disequilibrium chemistry section represents a unique combination of the listed sulfur enhancement (S/H) and eddy diffusive coefficient ($K_{zz}$). A sulfur enhancement of unity is equivalent to the Solar sulfur abundance from \cite{Lodders2009}. We note that the same models that minimized the $\chi_\nu^2$ also minimized the BIC and AIC.}
\end{deluxetable*}

\subsection{Chemical equilibrium atmospheric modeling with \texttt{PICASO}}\label{sec:picaso}
We used the open-source modeling package \texttt{PICASO}\footnote{\url{https://github.com/natashabatalha/picaso}} \citep[v3.3;][]{Batalha2019,Mukherjee2022} to compare the transmission spectrum between $0.6$ \textmu{}m $\le\lambda\le$5.3 \textmu{}m\footnote{This corresponds to the wavelengths included in the NIRSpec PRISM throughput \citep[see][]{2016SPIE.9910E..16P}.} to models of a gas giant with an atmosphere assumed to be in thermochemical equilibrium. We computed radiative-convective thermochemical equilibrium (RCTE) atmospheric models for TOI-5205b using the one-dimensional climate code implemented in \texttt{PICASO}. For each climate model, we adopted the \cite{Guillot2010} pressure-temperature profile as an initial guess. \texttt{PICASO} uses the publicly available correlated-k opacities\footnote{\url{https://zenodo.org/records/7542068}} \citep[see][]{Lupu2023} from the Sonora Bobcat models of brown dwarfs and gas giants \citep{Sonora2021} as part of its climate solver. The model atmospheres were created on a grid that spanned
\begin{enumerate*}[label=(\roman*)]
\item 13 atmospheric metallicities ($\log \mathrm{[M/H]}= -1.0$, $-0.7$, $-0.5$, $-0.3$,~0.0,~0.3,~0.5,~0.7,~1.0,~1.3,~1.5,~1.7,~2.0 dex),
\item 6 carbon-to-oxygen ratios ($\mathrm{C/O}=0.25$, 0.5, 1.0, 1.5, 2.0, 2.5 relative to solar $\mathrm{[C/O]_\odot=0.458}$),
\item 5 intrinsic temperatures ($T_{\mathrm{int}}=100$, 200, 300, 400, 500 K), and
\item 3 heat redistribution factors\footnote{The fractional contribution of the stellar radiation to the net flux per atmospheric layer (see Equation 20 in \cite{Mukherjee2022}). A value of $r_{st}=0.5$ corresponds to efficient day-night energy redistribution.} ($r_{st}=0.25,~0.50,~0.75$), resulting in 1170 models.
\end{enumerate*}
The model transmission spectra were generated from the climate models using \texttt{PICASO} with an opacity database that was resampled to $R=60,000$ \citep{picasodb2020} from the original $R\sim10^6$ line-by-line calculations presented in \cite{Freedman2008} and \texttt{EXOPLINES} \citep{exoplines2021}. There were a total of 1170 model transmission spectra representing the different combinations of the RCTE grid parameters. 

\setcounter{figure}{2}
\begin{figure*}[!ht]
\epsscale{1.17}
\plotone{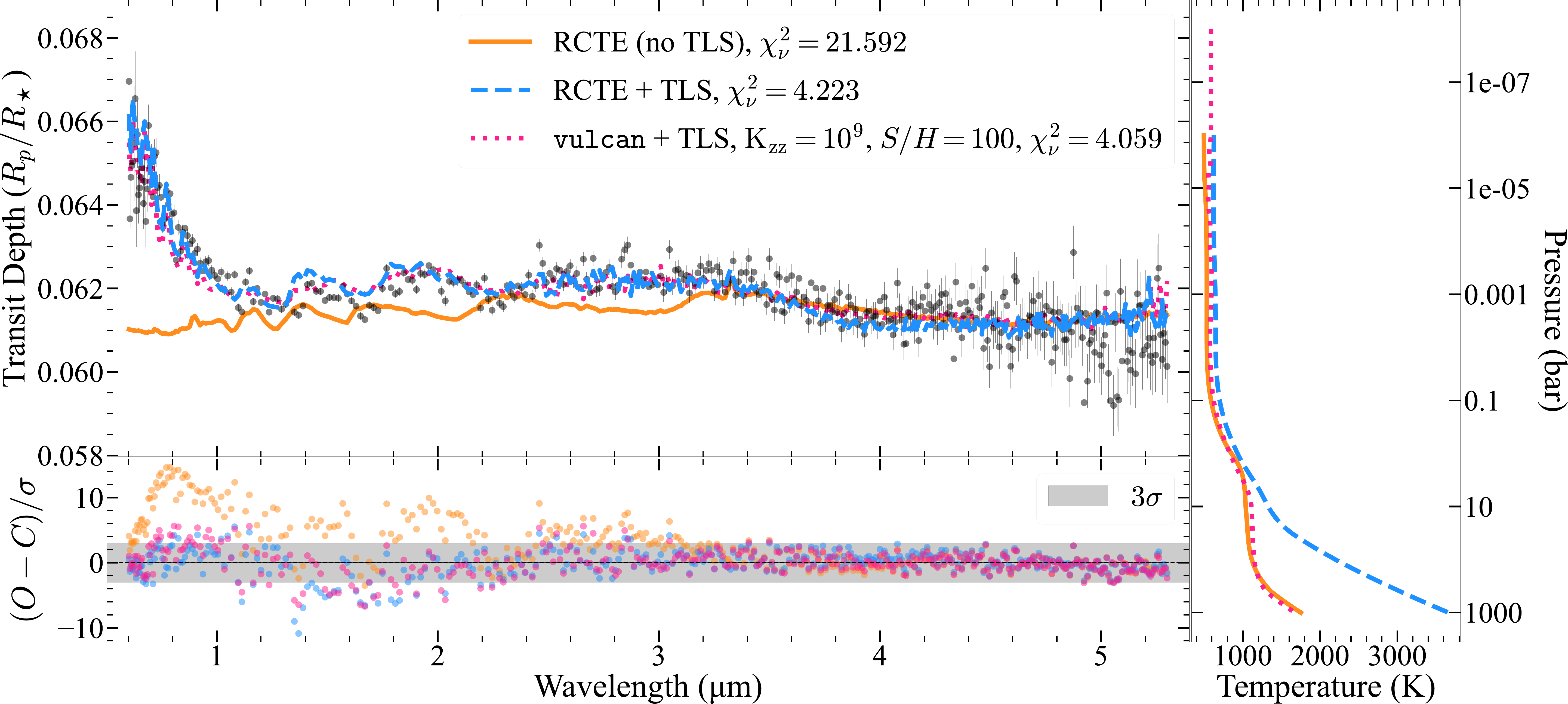}
\caption{\textbf{Top Left.} Co-added TOI-5205b \texttt{ExoTiC-JEDI} transmission spectrum along with the best-fitting grid-based models assuming equilibrium chemistry (orange line, obtained from \texttt{PICASO}), equilibrium chemistry with TLS (blue dotted line), or disequilibrium chemistry with TLS (pink dotted line, obtained from \texttt{VULCAN}). Models with TLS cannot fully replicate the slope at the blue end of the spectrum. \textbf{Bottom Left.} The difference between the data and model, scaled by the errors of the data. The $\pm3\sigma$ region is shaded for reference. \textbf{Right.} The pressure-temperature profile for each grid shown in the top left panel.}
\label{fig:picasovulcan}
\end{figure*}

Rapidly evolving spot-crossing events are observed in all JWST observations (see \autoref{fig:wlc}), and given the observed evolution of the occulted spots, we also expected the presence of unocculted spots. To model the observed contamination in the transmission spectrum by the TLS effect, we adopted a few configurations for a spot-contaminated stellar photosphere by post-processing the spectra output by \texttt{PICASO}. The disk-integrated spectral energy distribution (SED) of the host star was calculated as a weighted combination of two components: (i) a spotted region that was modeled using a PHOENIX spectrum \citep{Husser2013} with a cooler effective temperature but identical surface gravity and metallicity to the host star and a spot coverage fraction of $f_{\mathrm{spot}}$ and (ii) a quiet photosphere that was modeled using a PHOENIX spectrum that matched the known parameters of the host star (see \autoref{tab:5205par}) with a weight of $1-f_{\mathrm{spot}}$.  We followed the methodology of \cite{Rackham2018} to determine the contamination spectrum\footnote{$\epsilon_{\lambda,\mathrm{het}}$ in Equation 2 of \cite{Rackham2018}, or the multiplicative factor that is a function of wavelength and represents the ratio of the homogeneous stellar SED to the contaminated SED.} for two different spot temperatures ($T_\textrm{spot}=3230$ or 3330 K) and a range of spot coverage fractions ($f_\textrm{spot}=0.01$,~0.05,~0.1,~0.2,~0.3). Adding these additional parameters to account for the TLS effect resulted in 11700 unique model spectra. We did not consider a grid with both spots and faculae in part due to the large grid size, resulting in the inclusion of only one heterogeneity (spots) on the stellar surface. We only considered spots in this work because the shape of the features observed in the region $<1$~\textmu{}m (increasing transit depths toward bluer wavelengths) was inconsistent with faculae \citep[see][]{Rackham2018}. 

We fit the grids to the co-added (derived using a weighted average) \texttt{ExoTiC-JEDI} spectrum using the $\chi^2_\nu$ grid search method implemented in \texttt{PICASO} \citep{picasogrid}. The RCTE models which do not include the TLS effect were highly discrepant from the observed spectrum ($\chi^2_\nu>20$), but the models including the TLS effect were able to fit the observations much better with a $\chi^2_\nu<4.3$ (see \autoref{fig:picasovulcan}). The best-fitting model from the RCTE grid that included the TLS effect had $\mathrm{[M/H]}=-1.0$ dex, $\mathrm{C/O}=0.25$, $T_{\mathrm{int}}=500$ K, $r_{st}=0.25$, $T_\textrm{spot}=3230$ K and $f_\textrm{spot}=0.3$. This fit reached the lower limit of the metallicities in the correlated k-tables \citep{Lupu2023} that were used to generate the \texttt{PICASO} grid, such that we could not exclude an even lower atmospheric metallicity for TOI-5205b. Although the $\chi^2_\nu$ (see \autoref{tab:picasovulcan}) improved by five-fold when accounting for TLS, the RCTE models still showed residuals $>3\sigma$ in the region $<3.5$ \textmu{}m. \textbf{This suggested that limitations in the assumption of equilibrium chemistry, the lack of aerosol opacity in the model (e.g., clouds or hazes), or the adopted spot configuration may have impacted the recovered atmospheric parameters.}

\subsection{Investigating disequilibrium processes with \texttt{VULCAN}}\label{sec:vulcan}
To investigate the atmospheric composition when incorporating disequilibrium processes, we employed the photochemical model \texttt{VULCAN}\footnote{\url{https://github.com/exoclime/VULCAN}} \citep{Tsai2017,Tsai2021} with the temperature profiles computed by \texttt{PICASO} as input. These temperature profiles remained fixed without performing iteration to account for the radiative feedback of the disequilibrium abundances. The initial composition was determined by the equilibrium abundance for the given elemental abundances, calculated by \texttt{FastChem} \citep{Stock2018} and incorporated in \texttt{VULCAN}. We used the most recent S–N–C–H–O photochemical network\footnote{\url{https://github.com/exoclime/VULCAN/blob/master/thermo/SNCHO_photo_network_2024.txt}} for these calculations. TOI-5205 is an M4 dwarf with an effective temperature of 3430 K and we adopted the UV spectrum of GJ 436 \citep[M3;][]{muscles1,muscles2,muscles3} as an analogue for the stellar UV spectrum. For all parameters other than the metallicity, we explored an atmospheric parameter space identical to that of the RCTE models. 

As the best-fitting RCTE models reached the lower limit of the \texttt{PICASO} metallicity grid, we extended the grid to lower metallicities using \texttt{VULCAN} to investigate potential biases in atmospheric parameters due to grid limits. We did not calculate new pressure-temperature profiles for the lower metallicity models, but instead adopted the pressure-temperature profile from the nearest model in the RCTE grid. In addition, motivated by the strong detection of \ce{H2S} (see \S\ref{sec:poseidonresult}), we explored a range of sulfur elemental abundances that deviated from the solar elemental ratio. Specifically, we (i) used two different strengths of vertical mixing ($K_{zz}=10^6$ and 10$^9$ cm$^{-2}$ s$^{-1}$), (ii) expanded the grid to lower metallicities down to $\mathrm{[M/H]}=-2.0$, and (iii) included different sulfur enhancements ($\mathrm{S/H}=$ 1, 10, 100), where S/H is the sulfur enhancement factor relative to the solar value from \cite{Lodders2009}. We assumed that the lower metallicity grid points had the same pressure-temperature profiles as the grid points generated using \texttt{PICASO} for $\mathrm{[M/H]}=-1$ and $\mathrm{S/H}=1$. For reference, the adopted solar elemental abundances for the \texttt{VULCAN} analysis are $[\mathrm{C/H}]=-3.54$, $[\mathrm{O/H}]=-3.31$, and $[\mathrm{S/H}]=-4.88$.

Similar to the \texttt{PICASO} modeling in \S\ref{sec:picaso}, we post-processed the spectra to account for stellar contamination (with identical spot parameters) and fit the grid to the data using the same $\chi^2_\nu$ method. We did not consider VULCAN models without the TLS effect given the results from the RCTE grid. The modeling results are summarized in \autoref{fig:picasovulcan} and \autoref{tab:picasovulcan}. The models derived using \texttt{VULCAN} were improved fits when compared to the RCTE models ($\chi^2_\nu=4.223$ for the best-fitting RCTE model compared to $\chi^2_\nu=4.059$ for the best-fitting model processed with \texttt{VULCAN}). The eddy diffusive coefficient, $K_{zz}$, and sulfur enhancement factor, S/H, remained unconstrained as there was a minimal difference in $\chi^2_\nu$ between any of the gridded values. The best-fitting model that considered disequilibrium processes had $\mathrm{[M/H]}=-2.0$ dex, $\mathrm{C/O}=0.25$, $T_{\mathrm{int}}=100$ K, $r_{st}=0.5$, $T_\textrm{spot}=3330$ K, $f_\textrm{spot}=0.3$, $K_{zz}=10^9$ cm$^{-2}$ s$^{-1}$, and $\mathrm{S/H}=100$ ($100\times$ solar). Similar to the RCTE models, the recovered atmospheric metallicity also approached the lower limit imposed by the chemistry grids used by \texttt{VULCAN} ($\mathrm{[M/H]}\to-2$). We note that the fits with the RCTE or disequilibrium chemistry grids were still poor and showed a significant residual structure that was more than $3\sigma$ discrepant with the data at wavelengths $<3$ \textmu{}m, suggesting that the choice of grid parameters may have limited the flexibility of the model to replicate the data. Qualitatively, the RCTE and \texttt{VULCAN} grids both suggested that the planetary atmosphere was potentially metal-poor (albeit at the edge of each grid) with a possibly significant stellar contamination component (with a spot coverage fraction $\gtrsim10\%$).\footnote{For reference, the best-fitting model in \autoref{fig:wlc} suggests coverage fractions along the transit chord of 2.2\%, 8.4\%, and 13.9\% for visits 1, 2, and 3, respectively.} 

\section{Atmospheric retrievals}\label{sec:retrievals}
To aid interpretation of the transmission spectrum beyond grid-based models, we investigated the atmospheric properties of TOI-5205b by performing Bayesian atmospheric retrievals \citep[e.g.,][]{Madhusudhan2018} assuming both equilibrium and free chemistry. For completeness, we performed equilibrium chemistry retrievals and describe the results in \hyperref[app:eqchem]{Appendix \ref*{app:eqchem}}. However, given the diverse and often non-physical nature of these results, the assumption of equilibrium chemistry is shown to be flawed and demonstrates biases despite results that statistically converged (with tight uncertainties). Furthermore, the results were often discrepant with those of forward modeling (\S\ref{sec:forwardmodelling}) and free retrieval (\S\ref{sec:freechemret}) and were therefore not adopted or discussed further. Below, we provide details for Bayesian retrievals assuming free chemistry with the \texttt{ExoTiC-JEDI} reductions and note that ancillary plots and tables are found in \autoref{app:retrievalsupplement}. We discuss results from retrievals for the pixel-level \texttt{Eureka!} reductions and binned \texttt{ExoTiC-JEDI} reductions in \autoref{app:eurposeidon} and note here the general agreement with results derived using the pixel-level \texttt{ExoTiC-JEDI} reduction.

\subsection{Free Chemistry Retrievals}\label{sec:freechemret}
We used \poseidon\footnote{https://github.com/MartianColonist/POSEIDON} to investigate the atmospheric properties of TOI-5205b with retrievals allowing for freely varying individual chemical abundances to test for non-equilibrium chemical abundances. We performed this investigation using the co-added transmission spectrum, with opacities sampled to $R=20,000$ in the wavelength range of $0.6 - 5.3$ \textmu{}m. All retrievals assumed an isothermal pressure-temperature profile. We included opacities for the following species: \ce{H2O}, \ce{CH4}, \ce{CO2}, \ce{CO}, \ce{SO2}, and \ce{H2S}, as well as collision-induced absorption (CIA) from \ce{H2-H2}, \ce{H2-He}, \ce{H2-CH4}, \ce{CO2-H2}, \ce{CO2-CO2}, and \ce{CO2-CH4}. Our use of ``free chemistry'' indicates that the logarithmic volume mixing ratio (VMR) of each molecule is included as a free parameter, assuming evenly mixed vertical gas abundances, except for \ce{H2} and \ce{He} which are assumed to constitute the bulk composition of the atmosphere and are fractionated according to their relative cosmochemical abundances ($N_{\rm He}/N_{{\rm H}_2}=0.17$). 

\poseidon retrievals also included a stellar contamination model to account for the TLS effect. Our fiducial stellar contamination model included three components: a quiescent photosphere, a cooler region due to spots, and a hotter region due to faculae. For each component, we fit for an effective temperature ($T$) and a surface gravity ($\log g$). For spots and faculae, we fit for the fractional area ($f_{\mathrm{spot}}$ and $f_{\mathrm{fac}}$) covered by each component and assume that the remaining fractional stellar area ($1-f_{\mathrm{spot}}-f_{\mathrm{fac}}$) is the quiescent photosphere. The TLS effect was modeled with the PHOENIX model library \citep{Husser2013} using the interpolation scheme from the \texttt{pyMSG}\footnote{\url{https://github.com/rhdtownsend/msg}} software package \citep[v1.3;][]{2023JOSS....8.4602T}. 
The posteriors for all \poseidon{} retrievals were sampled using the nested sampling Bayesian parameter estimation code \texttt{MultiNest} \citep{Feroz2009}, which is accessed within \poseidon via the \texttt{PyMultiNest} Python wrapper \citep{Buchner2014}. Our retrievals used 2000 live points to sample the parameter space with a convergence criterion of $\Delta\ln Z=1$. 

Our baseline model included 16 free parameters to describe the planet and heterogeneous star (with both spots and faculae), but we investigated sets of nested models with larger and smaller parameter volumes to identify the most critical model components for fitting the co-added spectrum of TOI-5205b. A summary of the free parameters used in our retrievals and their associated priors is provided in the first two columns of \autoref{tab:retrievals_clear}. 

\begin{deluxetable*}{r|l|l||l|l|l||l}
\tablewidth{0.98\textwidth}
\tabletypesize{\scriptsize}
\tablecaption{Atmospheric retrieval priors and posteriors for fit to the \texttt{ExoTiC-JEDI} reduction using model M3.1 (as listed in \autoref{tab:retrieval_models}), which assumes a clear atmosphere and a TLS effect component. The analogous table for (i) the cloudy atmosphere can be found in \hyperref[tab:retrievals_clouds]{Appendix F Table \ref*{tab:retrievals_clouds}}, (ii) the \texttt{Eureka!} reduction can be found in \hyperref[tab:eureka_retrievals]{Appendix G Table \ref*{tab:eureka_retrievals}}, or (iii) the binned \texttt{ExoTiC-JEDI} reduction can be found in \hyperref[tab:binned_exotic_retrievals]{Appendix G Table \ref*{tab:binned_exotic_retrievals}}. \label{tab:retrievals_clear}}
\tablehead{
\colhead{Parameters} & \colhead{Units} & \multicolumn{1}{c||}{Priors} & \colhead{Visit 1} & \colhead{Visit 2} & \multicolumn{1}{c||}{Visit 3} & \colhead{Visits co-added}}
\startdata
$\mathrm{R}_{\mathrm{p, \, ref}}$ & $\mathrm{R_J}$ & $\mathcal{U}(0.80, 1.08)$ & $0.933\pm0.001$ & $0.933^{+0.001}_{-0.002}$ & $0.936\pm0.001$ & $0.933\pm0.001$ \\
$\mathrm{T}$ & K & $\mathcal{U}(200.00, 1500.00)$ & $787^{+54}_{-61}$ & $788^{+64}_{-68}$ & $780^{+74}_{-93}$ & $773^{+51}_{-54}$ \\
$\log \, \mathrm{CH_4}$ & dex & $\mathcal{U}(-12.00, -1.00)$ & $-5.78^{+0.20}_{-0.19}$ & $-5.54^{+0.23}_{-0.22}$ & $-6.54^{+0.29}_{-0.31}$ & $-5.89^{+0.16}_{-0.15}$ \\
$\log \, \mathrm{H_2 O}$ & dex & $\mathcal{U}(-12.00, -1.00)$ & $-9.7^{+1.6}_{-1.5}$ & $-9.8^{+1.5}_{-1.4}$ & $-9.7^{+1.6}_{-1.5}$ & $-10.0^{+1.4}_{-1.3}$ \\
$\log \, \mathrm{H_2 S}$ & dex & $\mathcal{U}(-12.00, -1.00)$ & $-6.8^{+1.7}_{-3.5}$ & $-4.33\pm0.29$ & $-5.1^{+0.6}_{-3.4}$ & $-4.59^{+0.18}_{-0.19}$ \\
$\log \, \mathrm{CO_2}$ & dex & $\mathcal{U}(-12.00, -1.00)$ & $-9.8^{+1.3}_{-1.4}$ & $-7.44^{+0.65}_{-0.89}$ & $-10.3^{+1.2}_{-1.1}$ & $-8.6^{+0.7}_{-1.6}$ \\
$\log \, \mathrm{CO}$ & dex & $\mathcal{U}(-12.00, -1.00)$ & $-9.7^{+1.8}_{-1.5}$ & $-9.7^{+1.7}_{-1.4}$ & $-9.5^{+1.9}_{-1.6}$ & $-10.0^{+1.5}_{-1.3}$ \\
$\log \, \mathrm{SO_2}$ & dex & $\mathcal{U}(-12.00, -1.00)$ & $-9.3\pm1.8$ & $-9.1^{+2.0}_{-1.9}$ & $-9.3\pm1.7$ & $-9.3\pm1.8$ \\
$\mathrm{f}_{\mathrm{spot}}$ & \nodata & $\mathcal{U}(0.00, 1.00)$ & $0.92\pm0.04$ & $0.30\pm0.09$ & $0.44^{+0.16}_{-0.09}$ & $0.33\pm0.05$ \\
$\mathrm{f}_{\mathrm{fac}}$ & \nodata & $\mathcal{U}(0.00, 0.50)$ & $0.48^{+0.01}_{-0.03}$ & $0.06^{+0.03}_{-0.02}$ & $0.013^{+0.006}_{-0.004}$ & $0.06^{+0.03}_{-0.02}$ \\
$\mathrm{T}_{\mathrm{spot}}$ & K & $\mathcal{U}(2300, 3430)$ & $3335^{+14}_{-21}$ & $3372^{+35}_{-76}$ & $3413^{+11}_{-19}$ & $3385^{+20}_{-16}$ \\
$\mathrm{T}_{\mathrm{fac}}$ & K & $\mathcal{U}(3430, 4802)$ & $3516^{+21}_{-22}$ & $3971^{+112}_{-76}$ & $4452^{+182}_{-188}$ & $3973^{+87}_{-75}$ \\
$\mathrm{T}_{\mathrm{phot}}$ & K & $\mathcal{N}(3430, 54)$ & $3427^{+8}_{-22}$ & $3607^{+32}_{-31}$ & $3529^{+31}_{-32}$ & $3600^{+24}_{-25}$ \\
$\log \, \mathrm{g}_{\mathrm{spot}}$ & dex & $\mathcal{U}(4.34, 5.34)$ & $4.37\pm0.02$ & $4.58\pm0.15$ & $4.42^{+0.07}_{-0.06}$ &$4.40^{+0.07}_{-0.05}$ \\
$\log \, \mathrm{g}_{\mathrm{fac}}$ & dex & $\mathcal{U}(4.34, 5.34)$ & $5.31\pm0.02$ & $4.43^{+0.10}_{-0.06}$ & $4.73^{+0.33}_{-0.25}$ & $4.37^{+0.04}_{-0.02}$ \\
$\log \, \mathrm{g}_{\mathrm{phot}}$ & dex & $\mathcal{U}(4.34, 5.34)$ & $4.74\pm0.03$ & $4.83^{+0.15}_{-0.12}$ & $4.71^{+0.13}_{-0.10}$ & $4.64^{+0.08}_{-0.08}$ \\
$\log~\mathrm{C/O}$ & dex & \nodata & $1.9_{-0.9}^{+1.0},~2\sigma>0.46$ & $1.3_{-0.5}^{+0.7},~2\sigma>0.28$ & $1.3_{-0.9}^{+1.1},~2\sigma>-0.28$ & $1.8_{-0.7}^{+0.8},~2\sigma>0.48$ \\
$\log~[\mathrm{M/H}]$ & dex & \nodata & $-2.9_{-0.3}^{+0.6}$ & $-1.6\pm0.3$ & $-2.4_{-1.2}^{+0.5}$ & $-1.9\pm0.2$ \\
$\log~\mathrm{C/H}$ & dex & \nodata & $-2.5\pm0.2$ & $-2.2\pm0.2$ & $-3.2\pm0.3$ & $-2.6\pm0.2$\\
$\log~\mathrm{O/H}$ & dex & \nodata & $-4.7_{-1.0}^{+0.9},~2\sigma<-3.14$ & $-3.8_{-0.7}^{+0.6},~2\sigma<-2.63$ & $-4.8_{-1.1}^{+1.0},~2\sigma<-2.97$ & $-4.6_{-0.8}^{+0.7},~2\sigma<-3.30$ \\
$\log~\mathrm{S/H}$ & dex & \nodata & $-2.0_{-2.3}^{+1.6}$ & $0.3\pm0.3$ & $-0.5_{-2.6}^{+0.6}$ & $0.0\pm0.2$\\
\hline
$\chi^2_{\nu}$ & \nodata & \nodata & 1.66 & 1.46 & 1.37 & 2.40 \\
$\ln Z$ & \nodata & \nodata & 2385.36 & 2410.83 & 2327.21 & 2420.51 \\
\enddata 
\tablecomments{The radius is reported at a pressure level of 10 bar. The abundances are the volume mixing ratios for each trace molecular species. The atmospheric metallicity ([M/H]) and elemental abundances (C/H, O/H, and S/H) are reported with respect to the solar values from \cite{Lodders2019}. The adopted solar values are $[\mathrm{M/H}]_\odot=Z_\odot=0.0011$, $\log~[\mathrm{C/H}]_\odot=-3.53$, $\log~[\mathrm{O/H}]_\odot=-3.27$, and $\log~[\mathrm{S/H}]_\odot=-4.85$.}
\end{deluxetable*}

\subsubsection{Model comparison and selection} \label{sec:retrievals:models}
Our retrievals considered (i) stellar contamination from the TLS effect, (ii) atmospheric absorption due to gas opacity, and (iii) absorption and scattering due to an opaque cloud-deck and hazes \citep[using the generalized cloud and haze prescription from][]{MacDonald2017}. We constructed sets of nested models that contained different combinations of these components with their respective free parameters, and quantified the goodness of fit ($\chi^2_{\nu}$) and Bayesian evidence ($\ln Z$). The Bayes Factor \citep[$B$;][]{Trotta2008}, derived from the likelihood ($Z$) of different models, allowed models to be statistically compared and certain models to be rejected. Detailed analysis of individual visits is presented in \S\ref{sec:retrievals:individual}. 

\autoref{tab:retrieval_models} lists the retrieval models that we considered in our initial transmission spectrum analysis and the statistical results of the model comparison. The models include a series where the spectrum is modeled (i) only assuming stellar contamination (M1.X), (ii) only assuming a planetary atmosphere (M2.X), or (iii) assuming both a planetary atmosphere with stellar contamination. Although model selection and rejection are based solely on the odds using the evidence in \autoref{tab:retrieval_models}, we also report an equivalent significance following the equations in \cite{Sellke2001}. As noted in \cite{Benneke2013} and \cite{Kipping2025}, the quoted significance represents an optimistic interpretation of the Bayes factor and the true significance will be less than the reported sigma. 

\begin{deluxetable}{c|l|l|l|l|l}
\tablecaption{Atmospheric retrieval models \& model selection results for the co-added \texttt{ExoTiC-JEDI} spectrum \label{tab:retrieval_models}}
\tablehead{
\colhead{Name} & \colhead{Model Description} & \colhead{$\chi^2_{\nu}$} & \colhead{$\ln Z$} & \colhead{$\ln B_{3.3}$} & \colhead{Rejected}}
\startdata
M1.1 & TLS only (1 spot) & 4.15 & 2072.64 & 361.81 & 27$\sigma$ \\
M1.2 & TLS (1 spot, free $\log(g)$) & 3.73 & 2160.76 & 273.69 & 24$\sigma$ \\
M1.3 & TLS (2 spots) & 4.14 & 2079.32 & 355.13 & 27$\sigma$ \\
M1.4 & TLS (2 spots, free $\log(g)$) & 2.87 & 2319.56 & 114.89 & 15$\sigma$ \\
\hline
M2.1 & Atm only & 8.97 & 1125.16 & 1309.29 & $>30\sigma$ \\
M2.2 & Atm + Cloud & 8.99 & 1124.16 & 1310.29 & $>30\sigma$ \\
M2.3 & Atm + Cloud + Haze & 3.33 & 2241.42 & 193.03 & 20$\sigma$ \\
\hline
\textbf{M3.1} & \textbf{TLS$^\dagger$ + Atm} & \textbf{2.40} & \textbf{2420.51} & \textbf{13.94} & $\mathbf{5.6\sigma}$ \\
M3.2 & TLS$^\dagger$ + Atm + Cloud & 2.38 & 2422.58 & 11.87 & 5.2$\sigma$ \\
M3.3 & TLS$^\dagger$ + Atm + Cloud + Haze & 2.32 & 2434.45 & 0.00 & --- \\
\enddata 
\tablenotetext{\dagger}{Model M1.4 is the stellar contamination model for the M3.X series (with a photosphere with two heterogeneities, each with free temperatures and surface gravities) such that M3.X is a combination of models M1.4 and M2.X.}
\tablecomments{The $\ln B$ and $\ln Z$ columns report results for the co-added \texttt{ExoTiC-JEDI} transmission spectrum with the ``2-3-4'' spot configuration. Nominally, ``Atm'' refers to a \ce{H2}/\ce{He} dominated atmosphere with trace amounts of \ce{H2O}, \ce{CH4}, \ce{CO2}, \ce{CO}, \ce{SO2}, and \ce{H2S}. The Bayes factors listed for each case are with respect to the model with the highest Bayes evidence (M3.3) and are converted to an equivalent (upper limit) significance using the equations in \cite{Sellke2001}. The adopted model, M3.1, is in bold (see \S\ref{sec:retrievals:models}).}
\end{deluxetable}

Our model complexity investigation demonstrates that both stellar contamination and the planetary atmosphere are detected with high confidence ($\ln B>100$ when comparing model M3.1 with models M2.1 and M1.1), while evidence for or against clouds is minimal ($\ln B<3$ between models M3.1 and M3.2). \autoref{tab:retrieval_models} shows that the selected model includes components for the TLS effect and a planetary atmosphere (M3.1 in \autoref{tab:retrieval_models}). Although the model with the highest evidence contains a planetary atmosphere, stellar contamination, clouds, and hazes, we select a simpler model with an atmosphere and stellar contamination as the canonical model because it provides a more physical value for the planetary temperature. 

\subsubsection{Atmospheric retrievals on individual visits} \label{sec:retrievals:individual}

Given the varying structure in the spectrum of each visit (see \hyperref[fig:spectracomp]{Fig. Set 2}), we performed retrievals on individual visits to assess the consistency of the model parameters. We performed retrievals using the M3.1 (TLS + Atm) and M3.2 (TLS + Atm + Cloud) model configurations.   

\hyperref[fig:retrieval_corner_visits]{Fig. Set 4.1} compares the posterior distributions obtained by separately fitting the three visits and the co-added transmission spectrum using retrievals with and without clouds. Figures showing individual results with marginalized 1D posterior constraints labeled are provided as \hyperref[fig:retrieval_corner_visits]{Fig. Set 4} in the online journal. For a quantitative comparison, \autoref{tab:retrievals_clear} and \hyperref[tab:retrievals_clouds]{Appendix F Table \ref*{tab:retrievals_clouds}} provide the retrieval constraints for clear and cloudy retrievals for each transmission spectrum. We note that the inclusion of clouds to the model was not significantly favored ($\ln B<3$) in retrievals for the individual visits or the co-added spectrum.


\hyperref[fig:retrieval_corner_visits]{Fig. Set 4.1} enables a comparison between the retrieval model behavior in the presence of both strong stellar contamination and an atmospheric signal for the individual visit spectra and the co-added spectrum. When clouds were included in the model, significant visit-to-visit variability was observed among the retrieved atmospheric parameters, particularly the atmospheric temperature, cloud-top pressure, and \ce{CH4} (and other gas) abundance. 
For example, the $\log \, \mathrm{CH_4} - \log \, \mathrm{P}_{\mathrm{cloud}}$ covariance was anti-correlated for cloud-top pressures lower than ${\sim}1$ bar, requiring high \ce{CH4} abundances when the clouds are high in the atmosphere (low pressure) and lower \ce{CH4} abundances when the clouds are deep (higher pressure). We note that this degeneracy between cloud-top pressure and chemical composition has been reported in previous atmospheric studies \citep[e.g.,][]{Iyer2016,MacDonald2017,Pinhas2019}. In addition to correlations with gas abundances, the cloud-top pressure was also correlated with the 10 bar planet reference radius and the atmospheric temperature. When the retrieval found a preference for the clouds at relatively low pressures ($P\lesssim1$ bar), it also found that the atmosphere was hot ($T\gtrsim1000$ K), significantly exceeding the equilibrium temperature of the planet ($T_{eq}=742$ K). 

\setcounter{figure}{4}
\begin{figure*}
    \centering
    \includegraphics[width=0.97\linewidth]{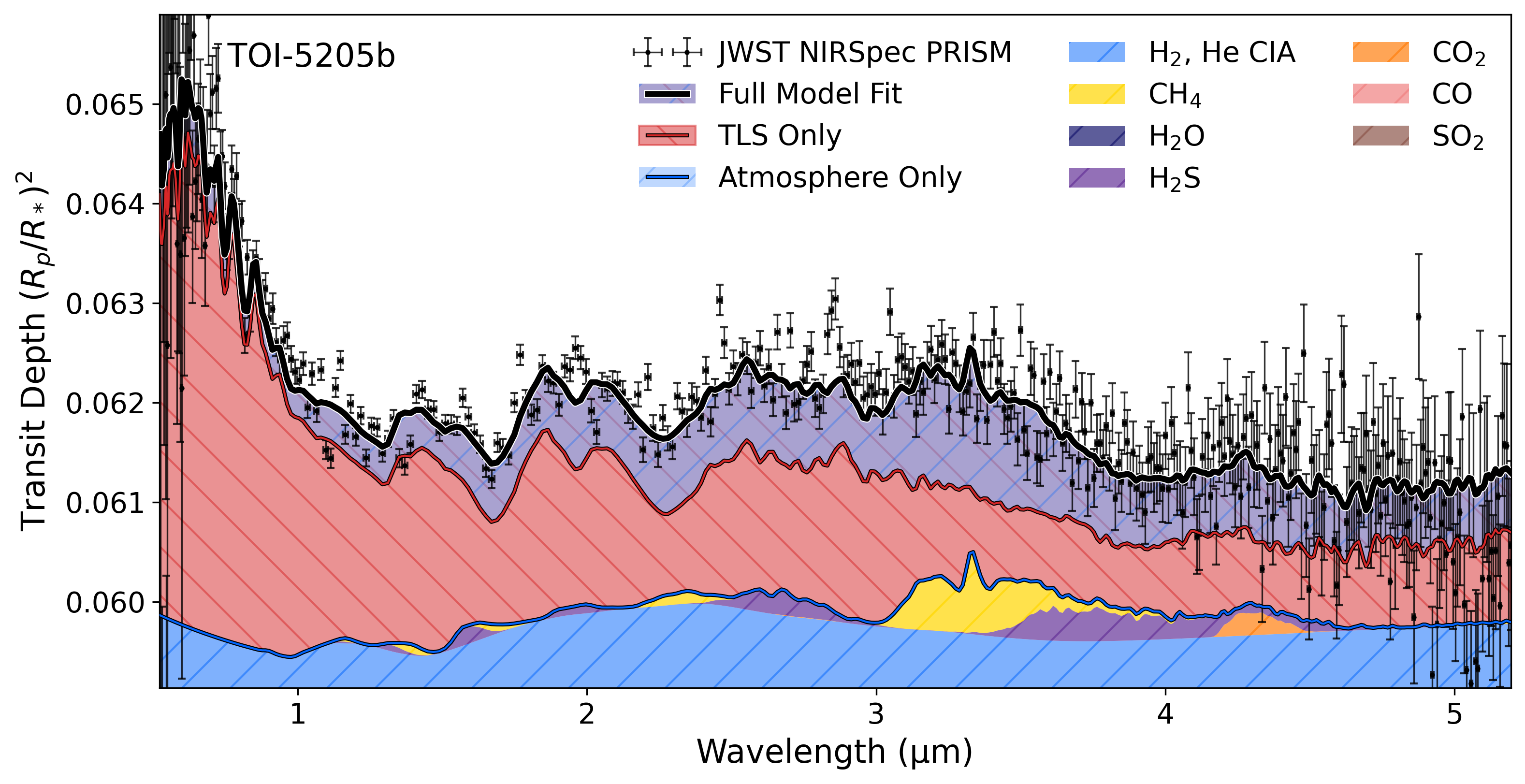}
    \caption{Maximum \textit{a posteriori} retrieved transmission spectrum (black line) with the contributions, derived using \texttt{POSEIDON}, shown from stellar contamination (red line) and atmospheric opacity (blue line). The contributions from individual atmospheric species are shown with various colors (see legend).
    The transmission spectrum of TOI-5205b is characterized by strong stellar contamination throughout much of the $0.6 - 3.5$ \textmu{}m wavelength range, with evident absorption from the planetary atmosphere due to \ce{CH4} and \ce{H2S}. No other molecular species were significantly detected in our retrievals (see \autoref{tab:retrievals_clear}).}
    \label{fig:retrieval_spectrum_breakdown}
\end{figure*}

\hyperref[fig:retrieval_corner_visits]{Fig. Set 4.1} also shows a positive correlation between the reference radius and the starspot temperature. Given the significant stellar contamination present in the transmission spectrum, it is not surprising that slight vertical offsets in transit depths between visits could be explained by either slight offsets to the planet (reference) radius or slightly different characteristics in the star, as probed by the starspots and their effective temperature and area covering fraction. Thus, we propose that slight differences in the TLS contamination and noise properties intrinsic to each visit propagate through the planet reference radius and cloud-top pressure covariances into the other planetary parameters, such as the atmospheric temperature and gas abundances. In this way, retrieval degeneracies between the planet and stellar parameters obscure atmospheric characteristics from being accurately inferred. \textbf{We emphasize that without multiple visits these retrieval degeneracies would not have been as simple to diagnose.} 

Considering the full covariances explored by the three visits, some atmospheric characteristics are much more physically plausible than others. When clouds are near $P_\mathrm{cloud}\sim1$ bar, we observe degeneracies in both the planet radius and temperature, in which the spectrum can be equally modeled with a hotter and smaller planet. The atmospheric temperature reaches $>1000$ K in this region of the solution space (particularly for Visit 1 with M3.2), such that these solutions are discarded as non-physical based on the orbit of TOI-5205b and the properties of the host star. The hazy M3.3 models in \autoref{app:hazyret} further illustrate these erroneous hot solutions. Based on this analysis, it appears that the cloud and haze parameters may over-fit the spectrum and lead to non-physical solutions, and we choose to consider cloud-free models for the visit-to-visit analyses. \autoref{tab:retrieval_models} shows that including clouds in the model does not improve the goodness-of-fit and that clouds are only weakly favored over the cloud-free model. Therefore, we next consider visit-to-visit analyses using the M3.1 ``TLS + Atm'' model. 

When clouds and hazes are omitted from the retrieval, we find that the three visits converge to a more consistent atmospheric interpretation. The gas abundances not only agree more closely between the three visits, but they are consistent with the cloudy models when the clouds were found to be deep in the atmosphere. The isothermal temperature (see \autoref{tab:retrievals_clear}) agrees between visits to within $1\sigma$ and does not significantly exceed the planet's equilibrium temperature ($T_{\mathrm{eq}}=742$ K), as was found with several cloudy and hazy models. \textbf{Therefore, we favor the cloud-free results (M3.1), which suggests that clouds, if present in an isothermal atmosphere, are likely deep at pressures ${\gtrsim} 0.5$ bar (see \hyperref[fig:retrieval_cornervisit3]{Fig. Set 4.4}) where the spectrum is not significantly affected by them. }

\subsubsection{Atmospheric retrievals on the co-added transmission spectrum} \label{sec:retrievals:combined}

\autoref{fig:retrieval_spectrum_breakdown} compares the median model posteriors to the co-added transmission spectrum, along with the contributions from the stellar and planetary components to visualize their individual impact on the observations. The stellar contamination component (red line) is calculated to show how the transmission spectrum would appear if the planet possessed a gray transmission spectrum with a radius aligned with the y-axis lower limit (the retrieved 10 bar reference radius). Similarly, the planetary component (blue line) is calculated to visualize how the transmission spectrum would appear if TOI-5205 were inactive and did not contaminate the spectrum. 

The transmission spectrum of TOI-5205b is highly contaminated by a heterogeneous stellar disk. Stellar contamination increases the transit depth by nearly 0.005 (${\sim}8\%$ increase) in the optical and ${\sim}0.001$ at 5 \textmu{}m (${\sim}2\%$ increase). Despite stellar contamination, substantial atmospheric features remain identifiable in the spectrum, primarily due to \ce{CH4} and \ce{H2S} which are not found in stellar spots. In the co-added transmission spectrum, we significantly detect \ce{CH4} ($\ln B=47.8$ or $10\sigma$ between model M3.1 and model M3.1 without \ce{CH4}) and \ce{H2S} ($\ln B=10.0$ or $4.8\sigma$ between model M3.1 and model M3.1 without \ce{H2S}). We also found independent evidence for \ce{CH4} in each visit and \ce{H2S} in two visits where \ce{CH4} was detected with $\ln B=25.7$ ($7.5\sigma$) in Visit 1, $\ln B=30.2$ ($8.1 \sigma$) in Visit 2, and $\ln B=5.0$ ($3.6 \sigma$) in Visit 3, and \ce{H2S} was detected with $\ln B=8.6$ ($4.5\sigma$) in Visit 1, $\ln B=4.9$ ($3.6\sigma$) in Visit 2, and $\ln B=1.0$ ($2\sigma$) in Visit 3. 
An additional case with \ce{NH3} (not shown) added to the \poseidon free retrieval produced no evidence favoring the inclusion of \ce{NH3} ($\ln B=-0.4$ between model M3.1 and model M3.1 with \ce{NH3}) (see more details in \ref{app:exploreotherspecies}). 

\subsubsection{Summary of retrieval results}
Below, we highlight a few insights into the preferred model based on these retrievals.

\textbf{TLS alone cannot replicate the spectrum:}
Previous works investigated the TLS effect as a plausible source to generate a large spectral slope in transmission spectra \cite[e.g.,][]{Rackham2018,Espinoza2019}. The recovered spot and faculae fractions from M3.1 ($f_\mathrm{spot}=0.3\pm0.05$ and $f_\mathrm{fac}=0.06_{-0.02}^{+0.03}$) suggested that the unocculted surface of TOI-5205 has a small, but non-negligible, percentage of spots and a negligible percentage of faculae. For visits 2 and 3 and the co-added spectrum, the faculae coverage fraction may be small ($\lesssim6\%$) but the faculae temperature is often $\gtrsim400$ K hotter than the photosphere and thus contributes significantly to the TLS model. We note that although the robustness of the TLS model and the constraints on surface inhomogeneities are limited by the ability of 1D stellar models to replicate the spectra for starspots and plages \citep{Smitha2024}, these values are consistent with the measured spot coverage fractions for other mid-M dwarfs \citep[e.g.,][]{Almenara2022,Mori2024,Mori2025}. Therefore, the observed features (particularly the spectral slope at bluer wavelengths) are most likely dominated by spots. 

The M1 series of models was designed to investigate whether the spectra were completely dominated by the TLS effect (e.g., only TLS contamination and no planetary atmosphere). Retrievals with these models are comparatively poor fits to the data and are rejected at ${\ge}16\sigma$ (i.e., strong detection of the atmosphere). The most complex TLS model (M1.4), with two heterogeneous components and free surface gravity for the photosphere, spot, and faculae components, is strongly preferred over the simpler TLS models (e.g. M1.4 preferred over M1.3 with an odds of $\ln~B=240$ or at most $\sim22\sigma$). 

For models including both the TLS effect and atmosphere (M3 series), all cases are modifications to the M1.4 TLS model. In these cases, we found that including clouds in the model marginally improved the Bayesian evidence ($\ln B\sim2$) over the cloud-free case, while incorporating a haze component increased the evidence significantly. 

However, all models that include the TLS effect and atmosphere provided similar $\chi^2_{\nu}$ values, indicating similar quality fits to the data. \textbf{Moreover, the cloudy and hazy M3.3 case tended to favor high atmospheric temperatures ($T=1447^{+34}_{-49}$ K), which raised significant concerns that the retrieval was using the haze model to fit the TLS contamination slope, thus compromising the validity of retrieved parameters when the haze model was used.} We provide additional details about the hazy M3.3 model in \hyperref[app:hazyret]{Appendix \ref*{app:hazyret}} and further discuss the invalidity of that case in the following section. Ultimately, given the tendency for the haze scattering slope to erroneously fit the TLS slope that is instead induced by cool starspots (which are known to be present from our observation of spot crossing events), we favor the M3.1 model as the clearest indicator of the atmospheric composition of TOI-5205b.   

\textbf{Hazes and clouds alone cannot replicate the blue slope:}
By comparison, all retrievals without TLS contamination (M2 series) were very discrepant with the co-added spectrum and were rejected with $\ln B>190$ (at most $19\sigma$, i.e., strong detection of TLS). The model with clouds and hazes (M2.3) was the best model among those without TLS contamination because the hazes produced a slope at short wavelengths, which was similar, but not identical, to the contamination from cool spots. Moreover, the M2.3 case resulted in non-physical planet atmospheric temperatures ($\mathrm{T} = 1447^{+34}_{-43}$ K), large Rayleigh enhancement factors ($\log a = 7.84^{+0.12}_{-0.21}$), and large negative scattering slopes ($\gamma =-16.54^{+0.46}_{-0.28}$) in an attempt to fit $\sim30$ scale heights worth of stellar contamination.  If the spectral slope observed in the spectrum of TOI-5205b were solely due to hazes, this would require enhanced scattering \citep[``super-Rayleigh'' scattering in][]{Ohno2020}, which is much larger than the values observed for hot Jupiters. For comparison, $\gamma=-4$ is the nominal value of Rayleigh scattering caused by small aerosol particles \citep{Lecavelier2008}, while $-7\lesssim\gamma\lesssim-5$ is the median value derived from a population of hot Jupiters observed with HST \citep[e.g.,][]{Pinhas2019,welbanks_mass-metallicity_2019}. Due to the non-physical values for the haze parameters and the detection of spot crossing events in the photometry, we conclude that a TLS component is required to model the observed slope.

\section{Stellar Metallicity}\label{sec:stellarmetallicity}
Knowledge of the stellar metallicity enables comparisons between other exoplanetary systems and the Solar System \citep[e.g.,][]{welbanks_mass-metallicity_2019,Sun2024}. The metallicity of TOI-5205 was assumed to be solar ($\mathrm{[M/H]}=0$) in \cite{kanodia_toi-5205b_2023} based on predictions from various photometric relationships. To verify the stellar metallicity, we analyzed an archival $0.8-2.4$ \textmu{}m SpeX spectrum \citep{Rayner2003} of TOI-5205 from the NASA Infrared Telescope Facility Data Archive at IRSA (Program ID 2022B-049, PI: B. Rackham). The data were processed using the IDL \texttt{Spextool} pipeline \citep{Cushing2004} and corrected for telluric absorption using the \texttt{xtellcor} routine \citep{Vacca2003}. The spectrum was then shifted to the vacuum rest frame of the star via cross-correlation with a template star spectrum HD 36395. We measured the K-band metallicity using empirical relationships between line indices and metallicity for M dwarfs in wide-binary systems with an FGK companion with a well-measured metallicity \citep{Mann2013,Rojas-Ayala2012}. This yielded an [Fe/H] of $0.56\pm0.10$ which was consistent (within $1\sigma$) to the measurement of $\mathrm{[Fe/H]}=0.68\pm0.08$ by \cite{GanMetal}. 

\begin{figure}[!ht]
\epsscale{1.15}
\plotone{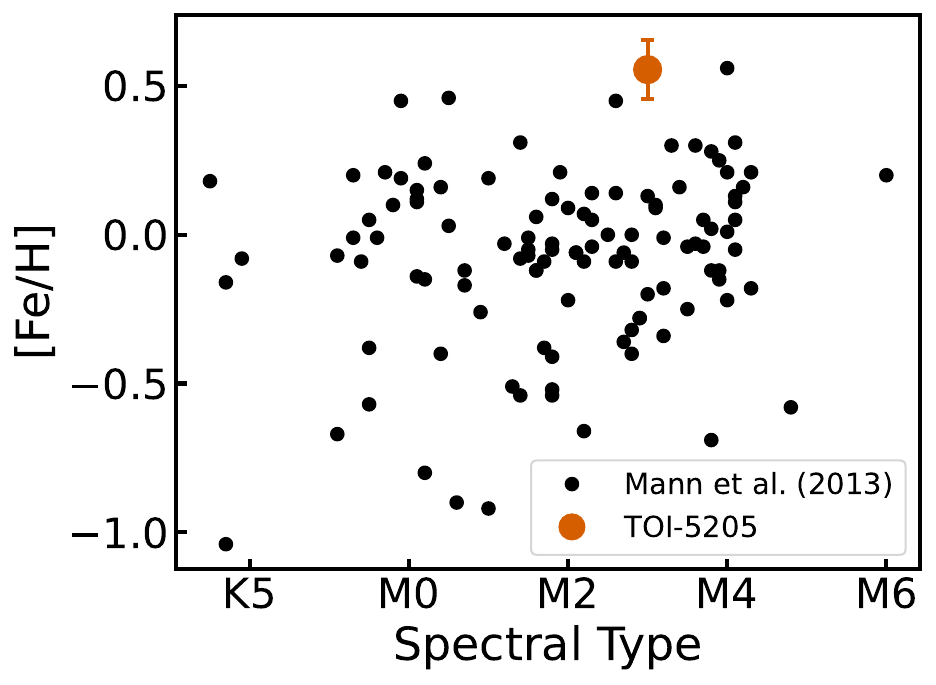}
\caption{Comparison of the metallicity calibration sample from \citet{Mann2013} (black points) to our measured metallicity of TOI-5205 (red point). The metallicities are shown plotted against host star spectral type. Our measured metallicity is at the upper limit of the calibration sample.}
\label{fig:met-spt}
\end{figure}

We caution that our measured metallicity is at the upper limit of the metallicity range in the calibration sample used in \citet{Mann2013} (see \autoref{fig:met-spt}). We chose to adopt the empirical metallicity relationships derived from moderate-resolution K-band infrared spectra over both photometric relationships \citep[e.g.,][]{Bonfils2005,Schlaufman2010,Neves2012} and those from high-resolution red-wavelength spectra \citep[e.g.,][]{Yee2017}. Photometric metallicities are calibrated on spectroscopically derived metallicities and may be susceptible to training set biases \citep[e.g.,][]{Dittmann2016,Hardegree-Ullman2020,Anderson2021}. Determining the stellar metallicity from high-resolution spectra with template matching methods \citep[e.g., \texttt{SpecMatch-Emp};][]{Yee2017} 
relies on comparison with a grid of spectra in $T_{\mathrm{eff}}$, $\log g$, and [Fe/H] space. This may be unreliable because of degeneracies between stellar parameters, particularly for M dwarfs with many atomic and molecular lines. The empirical relationships we adopt are derived from measured companion star metallicities and calibrated on several line indices to a precision of $\sim$0.1 dex. As such, despite the uncertainties associated with our relatively high measurement for TOI-5205, it likely does have a super-solar metallicity. We adopt this metallicity for the star and note that this measurement could be refined in the future with a higher S/N spectrum, as the archival IRTF spectrum measured a S/N $\approx$ 60 at 2.2 \textmu{}m.

\section{Bulk metallicity}\label{sec:interiormodeling}

Characterizing the interiors of giant planets is a crucial step in improving our understanding of their formation history \citep{2006A&A...453L..21G,2017AREPS..45..359J,2018ApJ...865...32H,2024arXiv240906670H}. A key property of a giant planet is the bulk metallicity (or bulk heavy element mass). This property, however, must be inferred indirectly from mass-radius measurements \citep{Miller2011,thorngren_mass-metallicity_2016}. The measured atmospheric metallicity of TOI-5205b is a crucial additional constraint for the planet's interior and lifts part of the degeneracy concerning its composition and the distribution of heavy elements within the planet \citep[e.g.,][]{thorngren_connecting_2019,2023A&A...669A..24M,2007ApJ...661..502B,2020ApJ...903..147M}.

To infer the heavy element content of TOI-5205b, we used the \texttt{planetsynth}\footnote{\url{https://github.com/tiny-hippo/planetsynth}} evolution models that account for stellar irradiation and atmospheric composition \citep{muller_synthetic_2021}. The inputs that generate a model with \texttt{planetsynth} are the planet mass, planet radius, system age, and incident flux. The evolution models used a $50-50$ \ce{H2O}-\ce{SiO2} (water-rock) mixture for the heavy elements \citep{1988PhFl...31.3059M}, and a proto-solar hydrogen-helium mixture \citep{2019ApJ...872...51C}. We employed a common Monte-Carlo approach \citep[e.g.,][]{2023FrASS..1079000M} in which the observations were used as priors, sample planets were drawn, and their thermal evolution was modeled to find the heavy element mass that fit the observed radius. Gaussian priors were used for the mass, radius, and stellar irradiation. Since the host-star age is largely unknown \citep{kanodia_toi-5205b_2023}, we used a uniform prior between 2 and 10 Gyr. For the atmospheric metallicity, we used a uniform prior for $\log$ [M/H] from -1.0 to -0.3 to ensure that the samples were within the limits of the models used by \texttt{planetsynth}.  

\begin{figure}
    \centering
    \includegraphics[width=0.8\columnwidth]{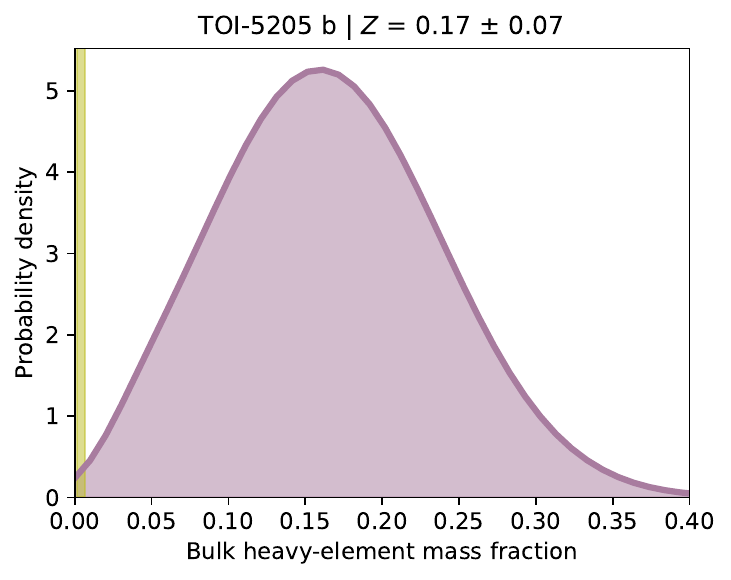}
    \caption{Posterior distribution of the bulk metallicity ($Z_\mathrm{planet}$) inferred with thermal evolution models. The shaded olive region denotes the atmospheric metallicity ($\log\mathrm{[M/H]}=-1.9$).
    The inferred bulk metallicity, much higher than the atmospheric metallicity, is incompatible with a homogeneously mixed planet.}
    \label{fig:toi5205b_kde}\label{fig:mztoi5205}
\end{figure}

The posterior distribution of the bulk metallicity estimated from $10^6$ planet samples is shown in \autoref{fig:toi5205b_kde}. We also highlight the region where the bulk metallicity equals the atmospheric metallicity. The inferred bulk metallicity and heavy element mass were $Z_{\mathrm{planet}} = 0.17 \pm 0.07$ and $M_z = 57 \pm 25$ M$_\oplus$, respectively. The large errors were mainly driven by the radius and age uncertainties. For comparison, interior models of Jupiter infer a heavy element mass of $\sim20 - 30$ M$_\oplus$ \citep[see][for a review]{2024arXiv240705853H}. Interestingly, the high bulk metallicity of TOI-5205b challenges the finding that giant planets around M dwarf stars are metal-poor compared to those around FGK stars \citep{2024arXiv241116197M}.

\begin{figure}
    \centering
    \includegraphics[width=0.9\columnwidth]{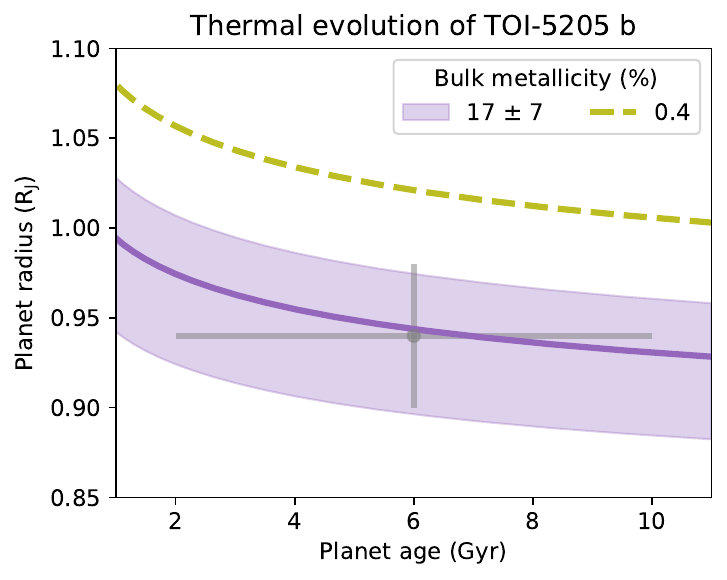}
    \caption{Thermal evolution of TOI-5205b assuming either that the bulk metallicity equals the inferred value and its uncertainty (purple) or the median atmospheric metallicity (dashed olive). The gray error bars show the measurement values and their uncertainties. Note that, unlike the radius, we consider each planetary age shown as equally likely. 
    The cooling curves demonstrate that the planet must be metal-rich and inhomogeneous.}
    \label{fig:toi5205b_evolution}
\end{figure}

\autoref{fig:toi5205b_evolution} shows the cooling curves of TOI-5205b for two scenarios: (i) models that use the range of inferred bulk metallicities (purple line and shaded region) and (ii) one that has a bulk metallicity equal to the median atmospheric metallicity ($\log\mathrm{[M/H]}=-1.9$, dashed olive line). The cooling curve for the latter case clearly shows that the observed radius ($R_p=0.94\pm0.04~\mathrm{R_J}$ derived from the white light curve (see \autoref{tab:5205par}) cannot be matched if the planet had a sub-stellar bulk metallicity. This implies that TOI-5205b is not fully mixed and that the heavy element mass fraction is higher in the interior. Recent observations of Jupiter with Juno and Galileo have called into question the distribution of heavy elements in its interior \citep{2019ApJ...872..100D, 2023A&A...680L...2H,2024ApJ...967....7M}, but similar inferences regarding the extent to which atmospheric metallicity is linked to the planetary bulk composition for exoplanets have not yet been determined.  

\begin{figure*}
    \centering
    \includegraphics[width=\textwidth]{figures/f9_vulcan_profile.pdf}
    \caption{The volume mixing ratios of the most select carbon-bearing species (\ce{CH4}, \ce{C2H4}, \ce{C2H6}, and \ce{CS2}), \ce{H2O}, and \ce{H2S} from the best-fitting models with a TLS component in \S\ref{sec:vulcan} for \textbf{(a)} $\log K_{zz}=9$ \textbf{(b)} $\log K_{zz}=6$, and \textbf{(c)} RCTE. The $\pm1\sigma-3\sigma$ VMRs retrieved for model M3.1, assuming free chemistry, in \S\ref{sec:retrievals:combined} using \texttt{POSEIDON} for the co-added \texttt{ExoTiC-JEDI} data are the shaded regions for \ce{H2O} (blue), \ce{CH4} (green), and \ce{H2S} (orange).}
    \label{fig:vulcanprofiles}
\end{figure*}

We note that in addition to the formal error from the Monte-Carlo inference, there are other uncertainties related to the model details, such as equations of state, atmospheric models, energy transport, and composition gradients \citep{2008A&A...482..315B,2015ApJ...815...78K,2019Atmos..10..664P,2023A&A...672L...1H}. In particular, the inferred bulk composition can be influenced by the assumed chemical composition of the heavy elements. The \texttt{planetsynth} models used a 50-50 water-rock mixture for the heavy elements and, to test the sensitivity of the inferred bulk metallicity to our choice of heavy element materials, we calculated additional evolution models using the modified version of the stellar evolution code Modules for Experiments in Stellar Astrophysics \citep[\texttt{MESA};][]{Paxton2011,Paxton2013,Paxton2015,Paxton2018,Paxton2019,Jermyn2023} from \cite{2024ApJ...967....7M}. For these models, we adopted a pure rock equation of state for the heavy elements and attempted to fit the mean observed radius at the median stellar age. The inferred bulk metallicity was $Z_{\mathrm{planet}} \sim 0.12$ and remained larger than the observed atmospheric metallicity ($\lesssim$ 1\%). Overall, we conclude that the model uncertainties due to the choice of heavy element materials are unlikely to be responsible for this super-solar bulk metallicity.

\section{Discussion}\label{sec:discussion}

\subsection{The TOI-5205b transmission spectrum is not shaped by clouds and/or hazes, but by stellar contamination}
 
We have found that the TLS effect is the primary source of the observed spectral slopes in the transmission spectra of TOI-5205b based on the presence of spot-crossing events and results from our atmosphere models (see \hyperref[fig:spectracomp]{Fig. Set 2}). However, we acknowledge that scattering from photochemical hazes and/or clouds could also introduce a spectral slope where the observed transit depth increases towards bluer wavelengths \citep[e.g.,][]{Sing2016,Sedaghati2017,May2020}. We disfavor the models containing aerosols for the following reasons. Our retrievals with haze component and no TLS (M2.3) are able to roughly reproduce the blue slope in the TOI-5205b transmission spectrum, demonstrating this potential ambiguity. However, this case required an unrealistically hot atmosphere ($\mathrm{T} \sim 1450$ K) with extreme haze parameters to produce the nearly 30 atmospheric scale heights of slope amplitude observed, indicating that TLS must play a dominant role in generating the observed spectral features. 

We found that the presence of retrieval degeneracies between a cloudy/hazy atmosphere and stellar contamination was the primary source of the discrepancies between visits. These visit discrepancies were almost entirely eliminated for our clear sky atmospheric models (M3.1; \autoref{tab:retrievals_clear}) and were instead attributable to modest changes in the unocculted starspot covering fraction. Although some combination of aerosols and TLS could produce the observed transmission spectrum (e.g., M3.3), inconsistencies in the planetary atmospheric state between the three separate visits (see \autoref{fig:retrieval_corner_hazy_visits} and \autoref{tab:retrievals_clouds}) led us to disfavor such cases and prefer a clear atmosphere. It remains possible that optically thin clouds and/or hazes exist in the atmosphere of TOI-5205b, obscured in our analysis by the massive contaminating TLS signal, however, we lack evidence for or against such scenarios. 

\cite{Ohno2020} investigated microphysical models of hazes and demonstrated that, under certain conditions (preferentially when $T_{\mathrm{eq}}=1000-1500$ K or for efficient eddy mixing when the diffusion coefficient is $\log K_{zz}\gtrsim9$), photochemical hazes can enhance the spectral slope by factors of $2-4$.  \cite{Pinhas2017} investigated the impact of mineral clouds (e.g., sulfide and chloride clouds) on the transmission spectra. \ce{KCl}, \ce{Na2S}, and \ce{ZnS} are the only known major mineral cloud species with condensation curves that cross the equilibrium temperature of TOI-5205b \citep[based on \texttt{Virga};][]{Batalha2020,Rooney2022}. These cloud species are unable to produce a scattering slope $\gamma<-8$ even if the cloud scale height is assumed to be identical to the atmospheric scale height. This further supports the conclusion that neither mineral clouds nor hazes alone can reproduce the observed spectral slope in the transmission spectrum of TOI-5205b.

We note that hydrocarbon aerosols, as opposed to silicate clouds and aerosols, are expected to be the dominant aerosol composition for giant exoplanets with $T_{\mathrm{eq}}<950$ K \citep{Gao2020} in part because the primary carbon reservoir changes from \ce{CO} to \ce{CH4} \citep[e.g.,][]{Lodders2002,Liang2004,Fortney2020}, which can photodissociate and is a known source of hydrocarbon hazes in the Solar System \citep[e.g.,][]{Horst2017,Mandt2023,Zhang2024,Gao2024}. In the context of warm gas giants, such as TOI-5205b, various studies \citep{Line2010,Line2011,Venot2015,Molaverdikhani2019} have shown that photochemistry could destroy simple species (such as \ce{CH4}) and produce more complex hydrocarbons, such as acetylene (\ce{C2H2}) and benzene (\ce{C6H6}), which may be important precursor molecules for haze \citep[e.g.,][]{Morley2015,Venot2020,Tsai2021}. 

We use the best-fitting \texttt{VULCAN} models (see \S\ref{sec:vulcan}) to investigate the inventory of hydrocarbon species that may be present in the atmosphere and could form photochemical hazes. In the \texttt{VULCAN} model, all hydrocarbons, except \ce{CH4}, are present at VMRs that may be too low to robustly recover in the presence of \ce{CH4} when using NIRSpec PRISM \citep[e.g., like the warm Jupiter Kepler-30 c in][]{Gasman2022}. \textbf{Furthermore, such low VMRs ($\log X<-8$ for all hydrocarbons except \ce{CH4} and \ce{C2H2}, see \autoref{fig:vulcanprofiles}) could limit the formation rate of polycyclic aromatic hydrocarbons, such as benzene (\ce{C6H6}), which are major precursors of more complex haze species \citep[e.g.,][]{Tsai2021}, suggesting that extensive hydrocarbon haze formation may not occur on TOI-5205b.}

\subsection{TOI-5205b appears to have an atmosphere with a low metallicity and high C/O ratio}\label{sec:poseidonresult}
\begin{figure}[!t]
\epsscale{1.15}
\plotone{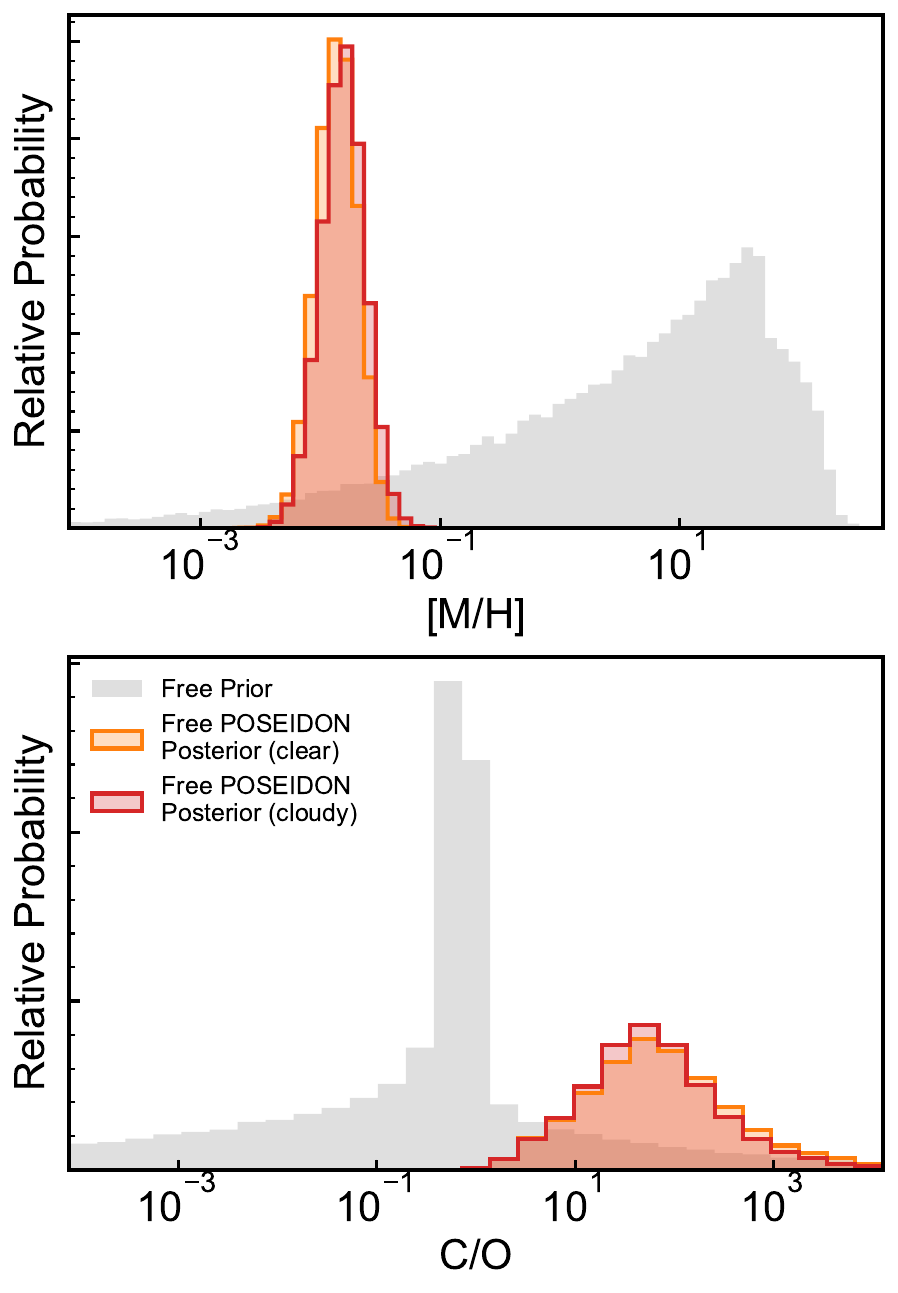}
\caption{The implicit prior (gray) and posterior distributions using results from the fits to the co-added \texttt{ExoTiC-JEDI} spectrum from retrievals with \texttt{POSEIDON}, assuming free chemistry and a clear (orange) or cloudy (red) atmosphere. \textbf{Top}: The posteriors for the atmospheric metallicity, [M/H]. \textbf{Bottom}: The posterior distributions for the C/O ratio. 
Chemical equilibrium retrievals tended to push the limits of their respective grids, but the free retrievals showed that the metallicity, while likely sub-solar, remained dependent on the unconstrained presence of clouds.  The C/O ratio is consistently super-solar and insensitive to the presence of clouds.}
\label{fig:posterioratmprop}
\end{figure}

The retrievals indicate a sub-solar atmospheric metallicity and a super-solar C/O ratio. \autoref{fig:posterioratmprop} displays the posteriors of all retrievals on the co-added spectrum (assuming free chemistry) along with the implicit priors calculated for our free chemistry retrievals (see \autoref{tab:retrievals_clear}). We derived the atmospheric [M/H] and C/O using the VMRs from the \texttt{POSEIDON} free retrievals for all molecular species in the models \citep[e.g.,][]{MacDonald2019}. Briefly, we determined the metallicity as the contribution from all elements except hydrogen or helium such that $\mathrm{[M/H]}=Z_\mathrm{atm}=(1-X-Y)/Z_\odot$.\footnote{$X$ and $Y$ are the elemental abundances of hydrogen and helium, respectively. $Z_\odot=0.0011$ is the solar metal abundance for the present day Sun from \cite{Lodders2019}. For these calculations, we make use of the convenience function \texttt{volume\_mixing\_ratios2metallicity} in the \texttt{petitRADTRANS} software package \citep{prt2019}.} Similarly, the C/O is the ratio of the elemental mix between carbon ($[\mathrm{C}]=[\mathrm{CH_4}]+[\mathrm{CO_2}]+[\mathrm{CO}]$) and oxygen ($[\mathrm{O}]=[\mathrm{H_2 O}]+2[\mathrm{CO_2}]+[\mathrm{CO}]+2[\mathrm{SO_2}]$).\footnote{[X] refers to the volume mixing ratio for a given molecular species.} The posteriors of the cloud-free model (M3.1) favored $\log\mathrm{[M/H]} = -1.9\pm0.2$ and $\log\mathrm{C/O} = 1.8^{+0.8}_{-0.7}$. Our retrievals including clouds (M3.2) gave approximately the same constraints ($\log\mathrm{[M/H]} = -1.8\pm0.2$ and $\log\mathrm{C/O} = 1.7^{+0.7}_{-0.6}$).

In the free chemistry retrieval, only \ce{CH4} and \ce{H2S} were significantly detected (see M3.1 in \S\ref{sec:retrievals:combined}). The abundances for these molecules from the model without clouds were comparable to the VMRs from the best-fitting \texttt{VULCAN} models with enhanced S abundance, regardless of $K_{zz}$. For comparison, the retrieved VMRs for M3.1 using \texttt{POSEIDON} are $\log \mathrm{[CH_4]}=-5.89^{+0.16}_{-0.15}$ and $\log \mathrm{[H_2S]}=-4.59^{+0.18}_{-0.19}$ while the VMRs from the best-fitting \texttt{VULCAN} model that accounts for disequilibrium processes (with $\mathrm{[M/H]}=-2.0$ and $\mathrm{S/H}=100$) reach values of $\log \mathrm{[CH_4]}=-5.31$ and $\log \mathrm{[H_2S]}=-4.57$. These VMRs are consistent (within $3\sigma$). 

We also note that VMRs assuming equilibrium chemistry, especially for water, are very discrepant with the values from our free retrievals (see \autoref{fig:vulcanprofiles}). Water (\ce{H2O}) is expected to be readily detected in the atmosphere of warm gas giants \citep[e.g.,][]{Fortney2020}. In the case of TOI-5205b, the low retrieved VMR of \ce{H2O} ($\log \mathrm{[H_2O]}=-10.0^{+1.4}_{-1.3}$) may be due in part to modeling uncertainties associated with the TLS effect. \cite{Moran2023} caution that heterogeneities on the surface of M dwarfs may significantly impact the transmission spectrum and the retrieved water abundance when there is stellar contamination. \autoref{fig:retrieval_spectrum_breakdown} displays the contamination spectrum derived with \texttt{POSEIDON}, which contributes the most to the spectral features between $1-3.5$ \textmu{}m. This is the same region that is most sensitive to water features in the NIRSpec/PRISM bandpass \citep[e.g.,][]{Fortney2020}. This overlap between $\log \mathrm{[H_2O]}$ and stellar contamination suggests that the recovered water abundance should be interpreted with caution, as some spectral features associated with water may be attributed to stellar contamination. Although no individual retrievals smoothly traversed a degeneracy between planetary atmospheric water and stellar parameters, fits to the $\lambda > 1.5$ \textmu{}m spectrum  (\autoref{fig:retrieval_corner_wavelength}) showed a well-constrained water abundance, suggesting that portions of the spectrum can be fitted by water opacity. However, the existence of data $\lambda < 1.5$ \textmu{}m precluded this scenario by tightly constraining the stellar contamination. Ultimately, the free chemistry retrieval did not robustly detect any oxygen-bearing molecule (\ce{H2O}, \ce{CO}, \ce{CO2}, or \ce{SO2}). 

The detection of \ce{H2S} and not \ce{SO2} is consistent with a high C/O, low-metallicity atmosphere. \ce{H2S} may be a primary sulfur reservoir for cool gas giants and its detectability is strongly dependent on the abundance of sulfur and mostly independent of the temperature or the C/O ratio \citep[e.g.,][]{Zahnle2009,Wang2017,Polman2023,Mukherjee2024}. \ce{SO2}, however, is preferentially detected in atmospheres that are oxygen-rich (low C/O) with high-metallicity because it is primarily formed by the photodissociation of \ce{H2O} \citep{Tsai2023}. For a high C/O ratio atmosphere of a cool gas giant, the detectability of \ce{H2S} with NIRSpec/PRISM will depend on the presence of other molecules \citep[due to its lower absorption cross-section, see][]{Polman2023}. The contribution to the transmission spectrum from \ce{H2S} in regions $\le3$ \textmu{}m overlaps with the stellar contamination spectrum and so any abundances derived from these regions could be degenerate with stellar surface inhomogeneities (similar to \ce{H2O}). However, \ce{H2S} also absorbs at 3.8 \textmu{}m \citep{Polman2023}, which can explain the shape of the significant contribution between $3.5-4.5$ \textmu{}m in  \autoref{fig:retrieval_spectrum_breakdown} observed for TOI-5205b. 

\subsection{The discrepancy between bulk and atmospheric metallicity reveal TOI-5205b is not fully mixed}
Our results demonstrate that TOI-5205b is rather metal-rich in its bulk and is not fully mixed.
The planetary metallicity is $Z_{\mathrm{planet}} = 0.17 \pm 0.07$ ($\sim 17\%$), which is intriguing because the host star is an M dwarf that is less massive than the Sun ($M_\star=0.394\pm0.011~\mathrm{M_\odot}$) and the low-mass of the star should be associated with a less massive protoplanetary disk and less available solids \citep[e.g.,][]{laughlin_core_2004}. While formation and evolution models clearly demonstrate that the atmosphere could be enriched due to heavy element accretion  \citep{2009ApJ...696.1348M,2019MNRAS.487.4510S,2020A&A...633A..33S,2024ApJ...967....7M}, our measured low atmospheric metallicity requires either little pollution or for most of it to have settled. In addition, primordial composition gradients are expected to become (at least partially) eroded, polluting the envelope from below \citep{2018A&A...610L..14V,2020A&A...638A.121M,2024arXiv240709341K}. Our inferred bulk metallicity being inconsistent with a fully mixed planet could hint that there was no significant enrichment with material from the deep interior. This would require the planet to form very cold or mixing to be inhibited.

Although the apparent discrepancy between the atmospheric and bulk metallicity appears surprising, we note that the region where the atmospheric metallicity is measured is only at very low pressures and therefore unlikely to represent the bulk or even the full atmospheric metallicity. We also emphasize that the transmission spectrum of TOI-5205b has significant stellar contamination (see a more detailed discussion in \S\ref{sec:limitations}), which may contribute to the apparent low metallicity and high super-solar C/O ratio due to a non-detection of water. Future measurements of the atmospheric metallicities of giant planets around low-mass stars, including from this survey, will provide a more statistical view of the metallicities and heavy element distributions of this enigmatic population \citep{kanodia_searching_2024}.

\subsection{Implications for formation}
The sub-solar atmospheric metallicity of TOI-5205b combined with its high atmospheric C/O ratio currently appears to be uncommon for hot Jupiters orbiting FGK dwarfs \citep[e.g.,][]{Kempton2024}. The atmosphere of a giant planet is predicted to have a sub-solar metallicity if, during runaway gas accretion, heavy elements in the form of pebbles or planetesimals are not accreted or if the accreted gas is metal-poor. This depends on the planet's formation process, its location in the disk, and its migration history \citep{2020A&A...633A..33S,2021ApJ...909...40T,2022ApJ...937...36P}. We note that TOI-5205b has a bulk heavy element mass of about $55 M_\oplus$, which is rather high given that it orbits an M dwarf star where less heavy elements should be available during the early stages of planet formation \citep{2013ApJ...771..129A,pascucci_steeper_2016}. The inferred high bulk metallicity and low atmospheric metallicity of TOI-5205b hint that most (if not all) of the available heavy elements accreted by the planet were deposited in the deep interior during its formation.

The high C/O ratio measured in the atmosphere could be explained if the planet accreted its gas in a region where carbon was abundant due to the evaporation of methane-rich pebbles \citep{2021A&A...654A..71S,2023A&A...677L...7M}. Recent JWST/MIRI observations of the inner regions of mid-to-late M dwarf protoplanetary disks \citep{tabone_rich_2023, arabhavi_abundant_2024, 2024arXiv241205535L} are finding carbon-rich disks with abundant hydrocarbons in the gas phase, weak water emission, and high C/O ratios.\footnote{Though exceptions like the molecule-rich Sz 114 disk exist \citep{2023ApJ...959L..25X}.} It is important to note, however, that the C/O ratio can be influenced by factors beyond the planet's initial formation location. Late-stage accretion of carbon-rich planetesimals can increase the atmospheric C/O ratio, even if it initially formed in a region with a lower gas-phase C/O ratio \citep{2011ApJ...743L..16O}. Similarly, processes like atmospheric escape or dredge-up of core material due to convective mixing can alter the observable C/O ratio after the planet has formed, albeit such processes would also increase the atmospheric metallicity \citep{2020A&A...638A.121M,Kempton2024}.

As a result, using the atmospheric C/O ratio as a tracer for formation should be done cautiously, since pebble evaporation, disk chemistry, planetary migration, accretion and mixing history, and late pollution of the atmosphere can affect the atmospheric C/O ratio  \citep{2021A&A...654A..71S,2021ApJ...909...40T}. Since most of the sulfur in a protoplanetary disk is locked in solids, the enhanced sulfur in the atmosphere of TOI-5205b may suggest that the planet had undergone extensive migration and accreted some planetesimals into its envelope \citep{2021ApJ...909...40T}. However, to better constrain the formation history of a giant planet, measurements of other elemental ratios (e.g., C/N, N/O, S/N) combined with a reliable measurement of the host-star abundances are required \citep{2021ApJ...909...40T,2022ApJ...937...36P}. Finally, additional measurements of the C/O ratios and atmospheric metallicity in giant planets orbiting M dwarfs would put the atmospheric composition of TOI-5205b in perspective.  

\subsection{Current limitations \& future improvements}\label{sec:limitations}
The treatment of stellar contamination in both the light curves and the transmission spectra is a limitation of studies of exoplanet atmospheres. The data show strong evidence for stellar activity (see \autoref{fig:wlc}) that is evolving with time (e.g., different spot configurations, spectral slopes, and spectral baselines). The TLS effect is a recurring problem for recent M dwarf atmospheric studies \citep[e.g.,][]{Lim2023,Moran2023,May2023,Forunier2024} and the observations of TOI-5205b, along with other planets in our survey, are a benchmark data set to investigate further improvements for data analysis in the presence of stellar contamination. We attempted to correct for stellar activity by (i) deriving transmission spectra using a spot-crossing model and (ii) including a stellar contamination model in our atmospheric retrievals \citep[e.g.,][]{Rackham2018}. However, we made certain assumptions in our modeling that can be improved upon in future work, such as (i) accounting for rotational effects in the light curve model \citep[e.g.,][]{macula2012,starry2021,fleck2022} or (ii) attempting a multi-epoch transmission spectra retrieval with a fixed planetary atmosphere model and variable stellar contamination spectra unique to each observation. 

A persistent issue with atmospheric retrievals for planets orbiting M dwarfs is that the photospheric models for these stars currently impose limitations on the reliability of the calculated contamination spectra \cite[e.g.,][]{Lim2023,Radica2024,TRAPPISTPath2024}. The contamination spectra calculated in atmospheric retrievals are based on 1D spectral models (PHOENIX in this work) and, given the lack of model fidelity for M dwarfs \citep[e.g.,][]{Rajpurohit2018}, this imposes a limit to the accuracy of these corrections. These interpolated stellar models are in stark contrast to the relatively complex, high-dimensional 1D atmospheric models used for the planet, despite the fact that the contaminating TLS signal exceeded that of the planet atmosphere for TOI-5205. Our TLS modeling also requires assumptions about the surface inhomogeneities (namely that starspots or plages can be modeled with a 1D stellar model) and their shape and position on the stellar surface, which impose further limitations on the accuracy of the TLS correction. \cite{spotexep2023} note that reliably correcting stellar contamination, in both photometry and transmission spectra, requires various theoretical improvements to model features that may impact the spectra (e.g., small-scale and spot-like magnetic heterogeneities). Our work here further motivates such improvements. 

Recent efforts to model surface inhomogeneities have emphasized the need to account for magnetohydrodynamic effects to reliably model spots and faculae \citep[e.g.,][]{Witzke2022,Rackham2023}. One-dimensional radiative-equilibrium models are unable to reproduce the spectral features and profiles that are predicted through comprehensive three-dimensional radiative magnetohydrodynamic simulations of spotted stars, and this can result in significant error for the contamination spectra of M dwarfs \citep[e.g.,][]{Norris2023,Smitha2024}. The limb darkening derived from 1D spectra may also impose limitations on the modeling of the photometry (see \autoref{app:limbdark}) and there are active efforts to account for magnetohydronamic effects in the generation of stellar atmosphere models \citep[e.g.,][]{Kostogryz2024,Witzke2024}. While there may be empirical improvements to the stellar contamination model using higher-resolution observations of the host star that are contemporaneous with the transmission spectra \citep[e.g.,][]{Berardo2024,Waalkes2024}, such a dataset does not exist for the GEMS JWST survey. The faintness of our targets also necessitates the use of large aperture telescopes to obtain the S/N necessary for such empirical improvements. 

We also adopted an isothermal pressure-temperature profile for its simplicity (i.e., limited number of additional free parameters) in our complex atmospheric models. We note that previous studies have shown that the simple assumption of an isothermal pressure-temperature profile may bias the retrieved chemical abundances to lower values \citep[e.g.,][]{Welbanks2019isotherm}. In addition, there may be potential degeneracies between the stellar contamination model and atmospheric chemistry as a result of imperfect TLS modeling (e.g., using PHOENIX model spectra) that could affect the retrieved abundances.

Observing TOI-5205b at redder wavelengths would also mitigate the impact of TLS on the spectra \citep{Seager2024}. Observations of a warm Jupiter with MIRI may provide valuable constraints on the presence of more complex hydrocarbons \citep[e.g.,][]{Gasman2022}, microphysical cloud models \citep[e.g.,][]{Kiefer2024}, additional nitrogen chemistry \citep{Welbanks2024}, and additional haze precursors \citep[e.g.,][]{Mukherjee2024}. We note that the data obtained from this GEMS JWST survey are exclusively transmission spectra. TOI-5205b is a favorable target for emission spectroscopy \citep{kanodia_toi-5205b_2023} which may provide additional constraints on the atmospheric chemistry \citep{Mukherjee2024} while significantly decreasing the impact of stellar activity \citep[e.g.,][]{Seager2024}. As part of cycle 4, JWST \citep[GO 7683;][]{2025jwst.prop.7683K} will perform emission spectroscopy of TOI-5205b using MIRI-LRS to confirm or refute (i) the fidelity of the TLS treatment in this work, (ii) the water abundance, and (iii) the apparent low atmospheric metallicity and super-solar C/O ratio. GO 7683 will serve to empirically calibrate the suitability of various assumptions for the stellar contamination and surface heterogeneities in this work. 

\section{Summary}
We have presented the first transmission spectra of TOI-5205b, a warm Jupiter orbiting an M4 dwarf. We reduced the data with two reduction pipelines (\texttt{Eureka!} and \texttt{ExoTiC-JEDI}) and derived consistent transmission spectra using a model that included spot-crossing events. Although there is significant stellar contamination in the transmission spectra, free chemistry retrievals reveal that the atmosphere of TOI-5205b has a sub-solar metallicity (2$\sigma$ upper limit of $\log \mathrm{[M/H]}\lesssim-1.5$) and a super-solar carbon-to-oxygen ratio ($2\sigma$ lower limit of $\log \mathrm{[C/O]}\gtrsim0.5$) because there were no robust detections of oxygen-bearing molecules in the atmosphere. The retrievals provide a significant ($\ln B>10$ or at best $4.8\sigma$) detection of both methane ($\log \mathrm{[CH_4]}=-5.89^{+0.16}_{-0.15}$) and hydrogen sulfide ($\log \mathrm{[H_2S]}=-4.59^{+0.18}_{-0.19}$) in the atmosphere of TOI-5205b at abundances that are within $3\sigma$ from the expected values when accounting for disequilibrium processes and sulfur enhancement. We do not robustly detect water with the NIRSpec/PRISM data, most likely due to significant stellar contamination. From the planet mass and radius, we estimate a bulk metallicity of $Z_{\mathrm{planet}} = 0.17 \pm 0.07$, which is in stark contrast to the metal-poor atmosphere observed. The high bulk and low atmospheric metallicity may suggest that TOI-5205b is not fully mixed, similar to the giant planets in the solar system. We caution that extensive stellar contamination and the non-detection of water may bias the atmospheric metallicity to lower values, and future observations through GO 7683 will help corroborate or refute these findings. TOI-5205b is one of seven planets in our GEMS JWST survey (GO 3171) --- \textit{Red Dwarfs and the Seven Giants: First Insights Into the Atmospheres of Giant Exoplanets around M dwarf Stars} \citep{2023jwst.prop.3171K} --- that will provide a sample of well-characterized warm Jupiter atmospheres that will (i) provide atmospheric and bulk metallicities to place TOI-5205b in greater context and (ii) allow for a comparison with hot Jupiters and Solar System gas giants to investigate potential constraints on the formation of GEMS.

\begin{acknowledgments}
We thank the anonymous referee for the thorough and valuable feedback that has improved the quality of this manuscript. We thank (i) Adriana Kuehnel and Zafar Rustamkulov for assistance in reducing JWST data as part of the GEMS JWST survey, (ii) Michael Zhang and Ryan MacDonald for assistance with the \texttt{PLATON} and \texttt{POSEIDON} codes, respectively, and (iii) Amanda Marrione and Marcio Melendez Hernandez, members of the STScI JWST Help Desk, for assistance in confirming the nature of the tilt event in visit 2 (Obs. 17). 

This work is based on observations made with the NASA/ESA/CSA James Webb Space Telescope. The data were obtained from MAST at STScI, which is operated by the Association of Universities for Research in Astronomy, Inc., under NASA contract NAS 5-03127 for JWST. These observations are associated with program \#3171. Support for program \#3171 was provided by NASA through a grant from the Space Telescope Science Institute, which is operated by the Association of Universities for Research in Astronomy, Inc., under NASA contract NAS 5-03127.

The JWST data presented in this paper were obtained from MAST at STScI. The specific observations analyzed can be accessed via \dataset[DOI: 10.17909/29st-dz13]{https://doi.org/10.17909/29st-dz13}. Support for MAST for non-HST data is provided by the NASA Office of Space Science via grant NNX09AF08G and by other grants and contracts.
This research has used the NASA/IPAC Infrared Science Archive, which is funded by the National Aeronautics and Space Administration and operated by the California Institute of Technology.

CIC and DRL acknowledge support by NASA Headquarters through an appointment to the NASA Postdoctoral Program at the Goddard Space Flight Center, administered by ORAU through a contract with NASA, and support from NASA under award number 80GSFC24M0006.

The Center for Exoplanets and Habitable Worlds is supported by the Pennsylvania State University and the Eberly College of Science.

Resources supporting this work were provided by the (i) NASA High-End Computing (HEC) Program through the NASA Center for Climate Simulation (NCCS) at Goddard Space Flight Center, 
(ii) Pennsylvania State University's Institute for Computational and Data Sciences' (ICDS) Roar supercomputer, and (iii) Carnegie Science Earth and Planets Laboratory. This content is solely the responsibility of the authors and does not necessarily represent the views of the NCCS, ICDS, or Carnegie Science.
\end{acknowledgments}

\vspace{5mm}
\facilities{JWST, IRSA, IRTF}

\software{
\texttt{astroquery} \citep{Ginsburg2019},
\texttt{astropy} \citep{astropy:2013,astropy:2018,astropy:2022},
\texttt{batman} \citep{Kreidberg2015},
\texttt{dynesty} \citep{Speagle2020},
\texttt{Eureka!} \citep{Bell2022},
\texttt{ExoTiC-JEDI} \citep{Alderson2022},
\texttt{ExoTiC-LD} \citep{Grant2024_exotic-ld},
\texttt{GGChem} \citep{Woitke2018},
\texttt{jwst} \citep{bushouse2023},
\texttt{juliet} \citep{juliet2019},
\texttt{lightkurve} \citep{LightkurveCollaboration2018},
\texttt{matplotlib} \citep{Hunter2007},
\texttt{numpy} \citep{vanderWalt2011},
\texttt{pandas} \citep{McKinney2010},
\texttt{petitRADTRANS} \citep{prt2019},
\texttt{PICASO} \citep{Batalha2019,Mukherjee2022},
\texttt{planetsynth} \citep{muller_synthetic_2021}, 
\texttt{PLATON} \citep{Zhang2019},
\texttt{POSEIDON} \citep{MacDonald2017},
\texttt{pyMSG} \citep{2023JOSS....8.4602T},
\texttt{PyMultiNest} \citep{Buchner2014},
\texttt{scipy} \citep{Virtanen2020},
\texttt{spotrod} \citep{spotrod2014},
\texttt{Spextool} \citep{Cushing2004},
\texttt{TauREx} \citep{Refaie2021,Refaie2022};
\texttt{VULCAN} \citep{Tsai2017,Tsai2021}
}

\bibliography{references}
\bibliographystyle{aasjournal}

\clearpage

\figsetstart
\figsetnum{2}
\figsettitle{Transmission spectra for individual visits}

\figsetgrpstart
\figsetgrpnum{2.1}
\figsetgrptitle{Overview}
\figsetplot{f2_spec_comparison_JWST1.pdf}
\figsetgrpnote{\textbf{Top}. The transmission spectra for the first observation (observation 16) of TOI-5205b on 2023 October 10. The \texttt{ExoTiC-JEDI} reduction is the blue circles and the \texttt{Eureka!} reduction is the orange squares. \textbf{Bottom}. The differences between both reductions, scaled by the errors of the \texttt{ExoTiC-JEDI} derived data. The 1, 2, and 3$\sigma$ regions are shaded for reference. The complete figure set (3 images, one for each observation) is available in the online journal. All transmission spectra are included as data behind the figure.}\label{fig:spectracomp}
\figsetgrpend

\figsetgrpstart
\figsetgrpnum{2.2}
\figsetgrptitle{Visit 2}
\figsetplot{f2_spec_comparison_JWST2.pdf}
\figsetgrpnote{Comparison of transmission spectra of TOI-5205b on 2023 October 11 (observation 17, similar to \hyperref[fig:spectracomp]{Fig. Set 2.1}).}
\figsetgrpend

\figsetgrpstart
\figsetgrpnum{2.3}
\figsetgrptitle{Visit 3}
\figsetplot{f2_spec_comparison_JWST3.pdf}
\figsetgrpnote{Comparison of transmission spectra of TOI-5205b on 2023 October 13 (observation 18, similar to \hyperref[fig:spectracomp]{Fig. Set 2.1}).}
\figsetgrpend

\figsetgrpstart
\figsetgrpnum{2.4}
\figsetgrptitle{\texttt{ExoTiC-JEDI} reductions}
\figsetplot{f2_spec_comparison_exotic.pdf}
\figsetgrpnote{Comparison of all transmission spectra of TOI-5205b derived with \texttt{ExoTiC-JEDI}.}
\figsetgrpend

\figsetgrpstart
\figsetgrpnum{2.5}
\figsetgrptitle{\texttt{Eureka!} reductions}
\figsetplot{f2_spec_comparison_eureka.pdf}
\figsetgrpnote{Comparison of all transmission spectra of TOI-5205b derived with \texttt{Eureka!}.}
\figsetgrpend

\figsetend

\clearpage
\figsetstart
\figsetnum{4}
\figsettitle{Retrieval Results for individual visits}

\figsetgrpstart
\figsetgrpnum{4.1}
\figsetgrptitle{Overview}
\figsetplot{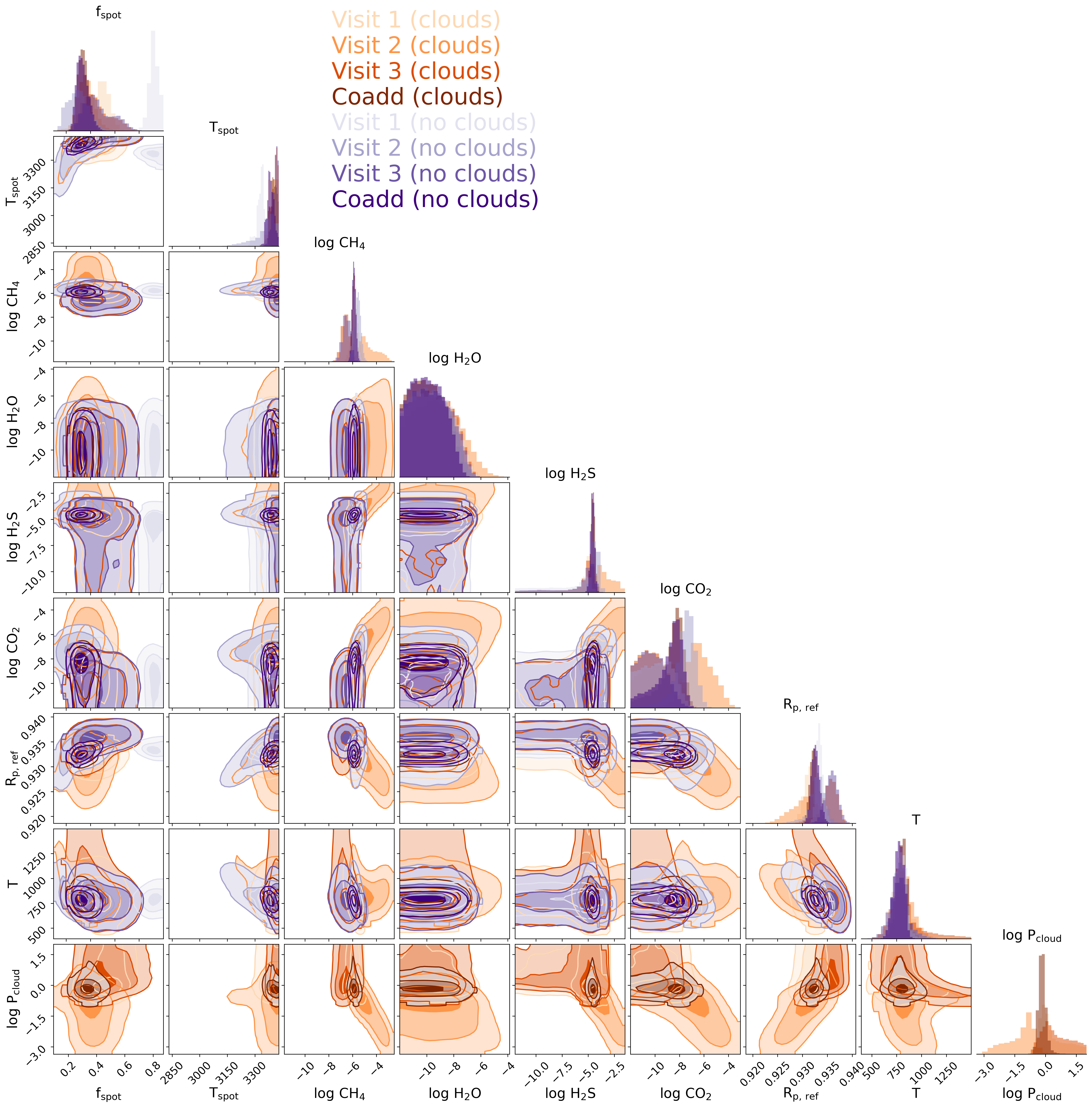}
\figsetgrpnote{Posterior probability distributions (1- and 2-D) for a subset of retrieval parameters showing constraints obtained from different visits and simulations with and without clouds (colors). When clouds are included in the retrieval, significant visit-to-visit variability is seen among the retrieved atmospheric parameters, particularly temperature, cloud-top pressure, and the \ce{CH4} abundance. When clouds are neglected, the three visits converge to a consistent atmospheric interpretation. \textit{Individual visit analyses do not reveal the full extent of the correlations that exist between retrieval parameters. Comparing multiple visits illuminates the covariances along which the visits appear discrepant.} The complete figure set (9 images, for the 8 individual figures plus one  combined figure) is available in the online journal.}\label{fig:retrieval_corner_visits}
\figsetgrpend

\figsetgrpstart
\figsetgrpnum{4.2}
\figsetgrptitle{Visit 1 (M3.2; clouds)}
\figsetplot{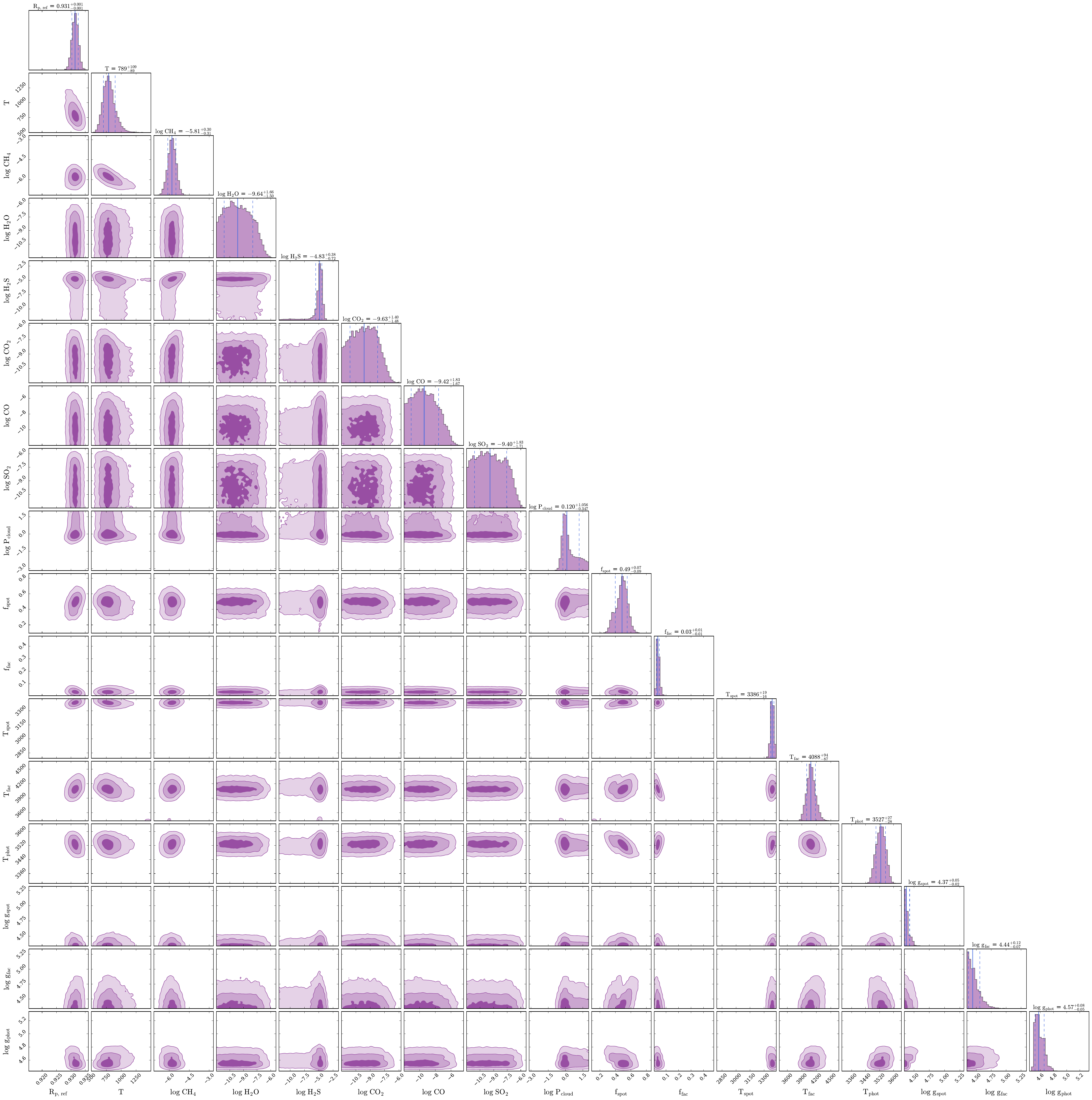}
\figsetgrpnote{Retrieval results for Visit 1 with clouds (as shown in \hyperref[fig:retrieval_corner_visits]{Fig. Set 4.1}).}
\figsetgrpend

\figsetgrpstart
\figsetgrpnum{4.3}
\figsetgrptitle{Visit 2 (M3.2; clouds)}
\figsetplot{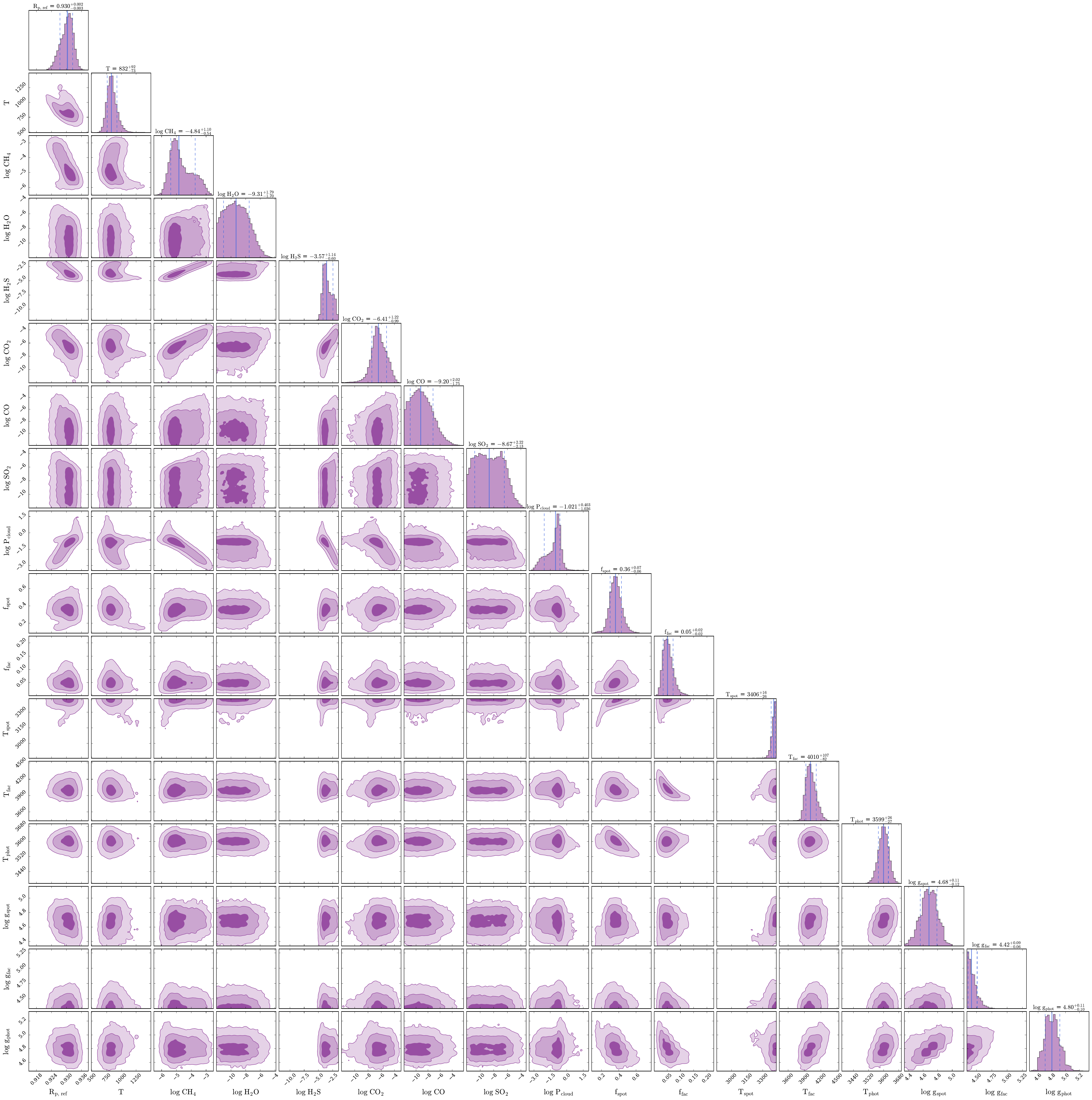}
\figsetgrpnote{Retrieval results for Visit 2 with clouds (as shown in \hyperref[fig:retrieval_corner_visits]{Fig. Set 4.1}).}
\figsetgrpend

\figsetgrpstart
\figsetgrpnum{4.4}
\figsetgrptitle{Visit 3 (M3.2; clouds)}
\figsetplot{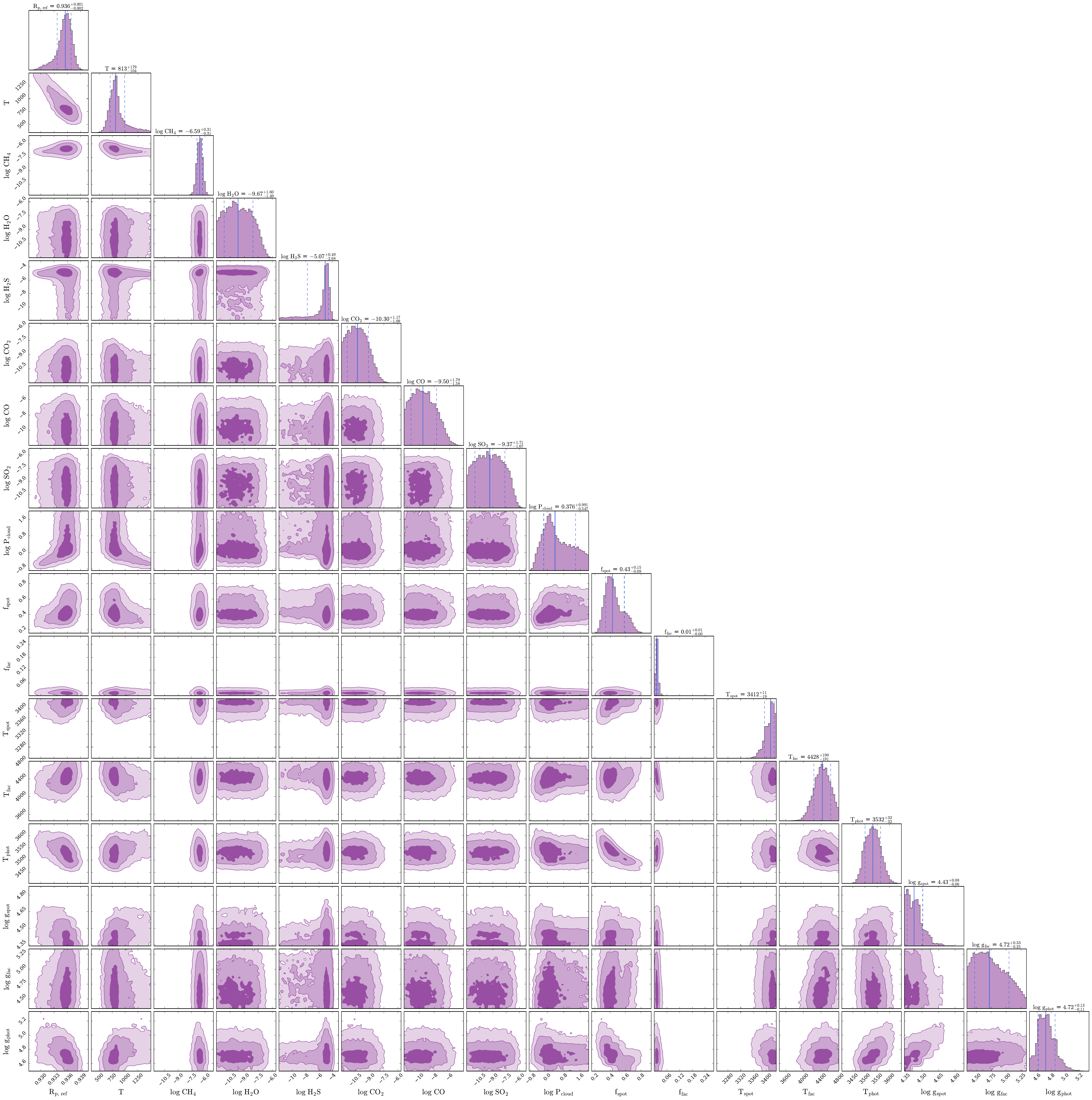}
\figsetgrpnote{Retrieval results for Visit 3 with clouds (as shown in \hyperref[fig:retrieval_corner_visits]{Fig. Set 4.1}).}\label{fig:retrieval_cornervisit3}
\figsetgrpend

\figsetgrpstart
\figsetgrpnum{4.5}
\figsetgrptitle{Co-added spectrum (M3.2; clouds)}
\figsetplot{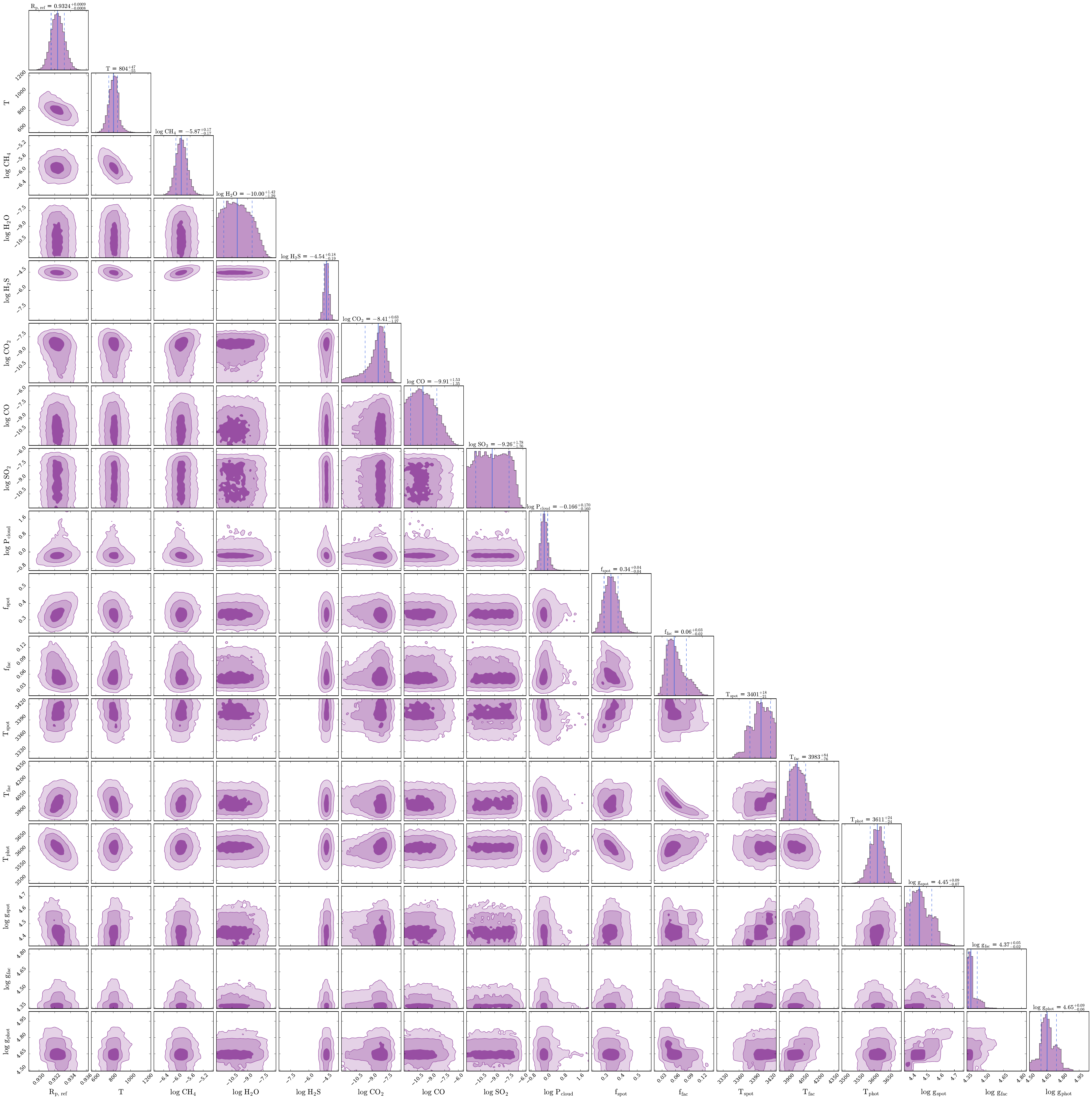}
\figsetgrpnote{Retrieval results for the Visit co-added spectrum with clouds (as shown in \hyperref[fig:retrieval_corner_visits]{Fig. Set 4.1}).}
\figsetgrpend

\figsetgrpstart
\figsetgrpnum{4.6}
\figsetgrptitle{Visit 1 (M3.1; no clouds)}
\figsetplot{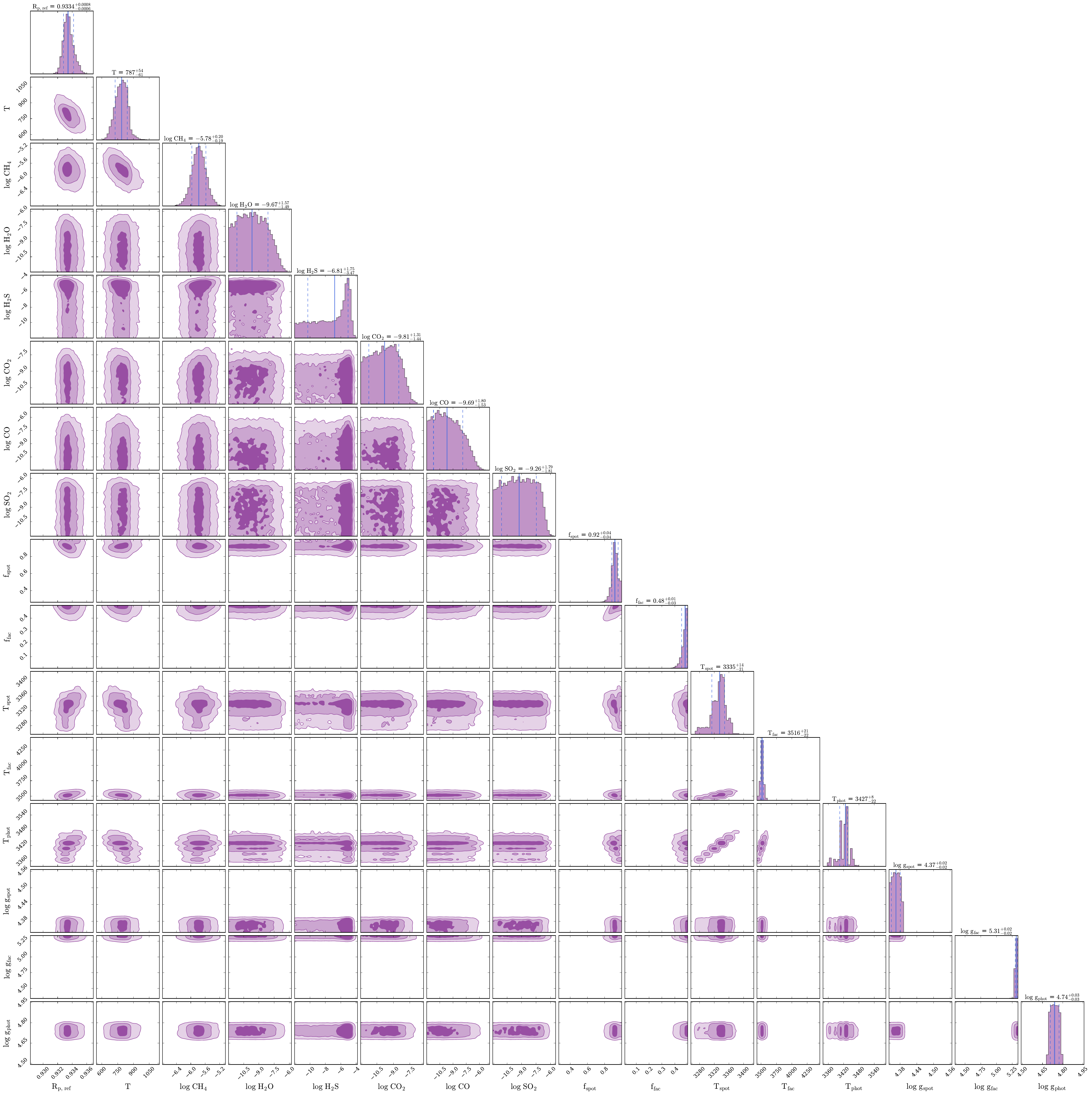}
\figsetgrpnote{Retrieval results for Visit 1 without clouds (as shown in \hyperref[fig:retrieval_corner_visits]{Fig. Set 4.1}).}
\figsetgrpend

\figsetgrpstart
\figsetgrpnum{4.7}
\figsetgrptitle{Visit 2 (M3.1; no clouds)}
\figsetplot{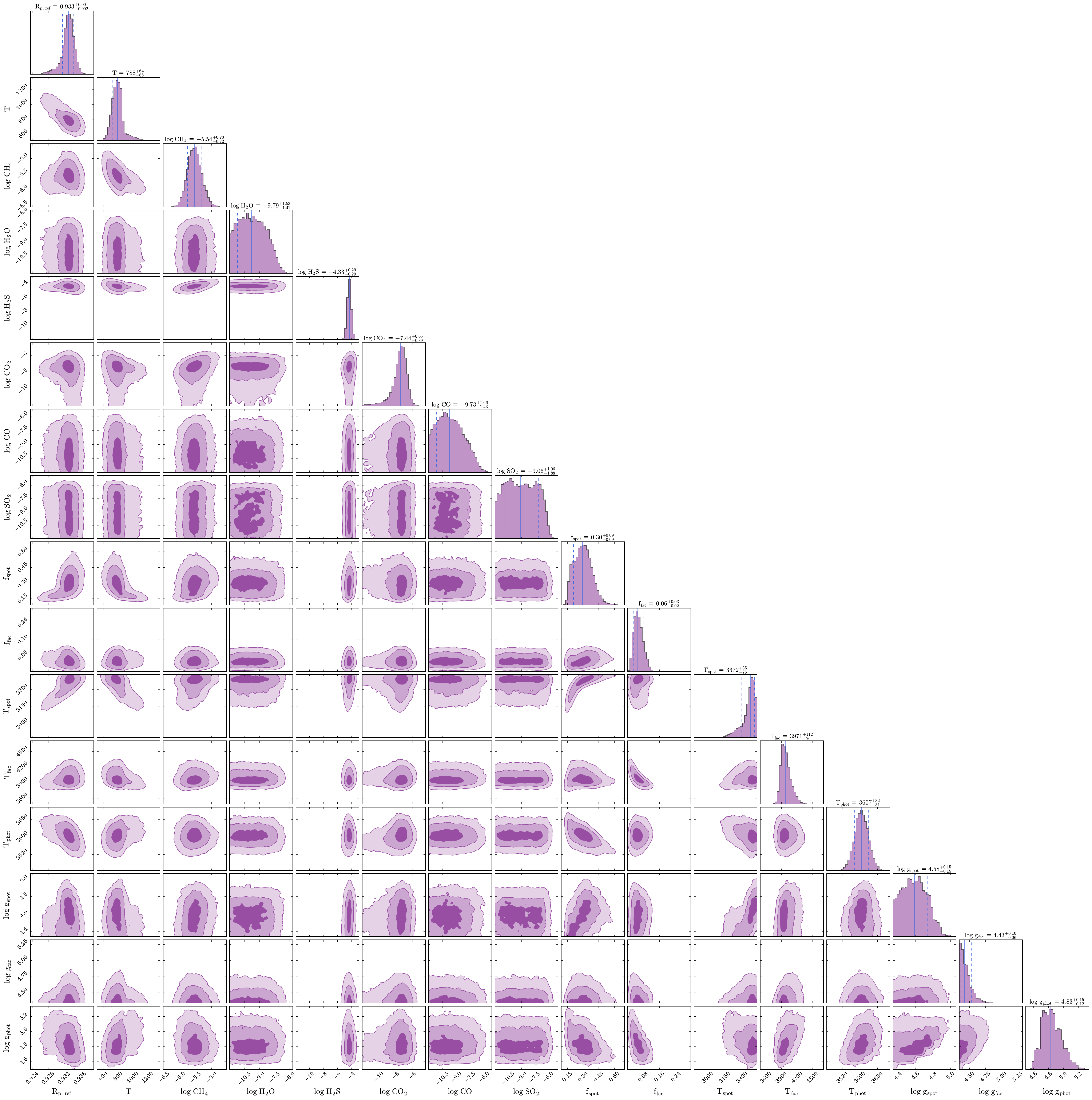}
\figsetgrpnote{Retrieval results for Visit 2 without clouds (as shown in \hyperref[fig:retrieval_corner_visits]{Fig. Set 4.1}).}
\figsetgrpend

\figsetgrpstart
\figsetgrpnum{4.8}
\figsetgrptitle{Visit 3 (M3.1; no clouds)}
\figsetplot{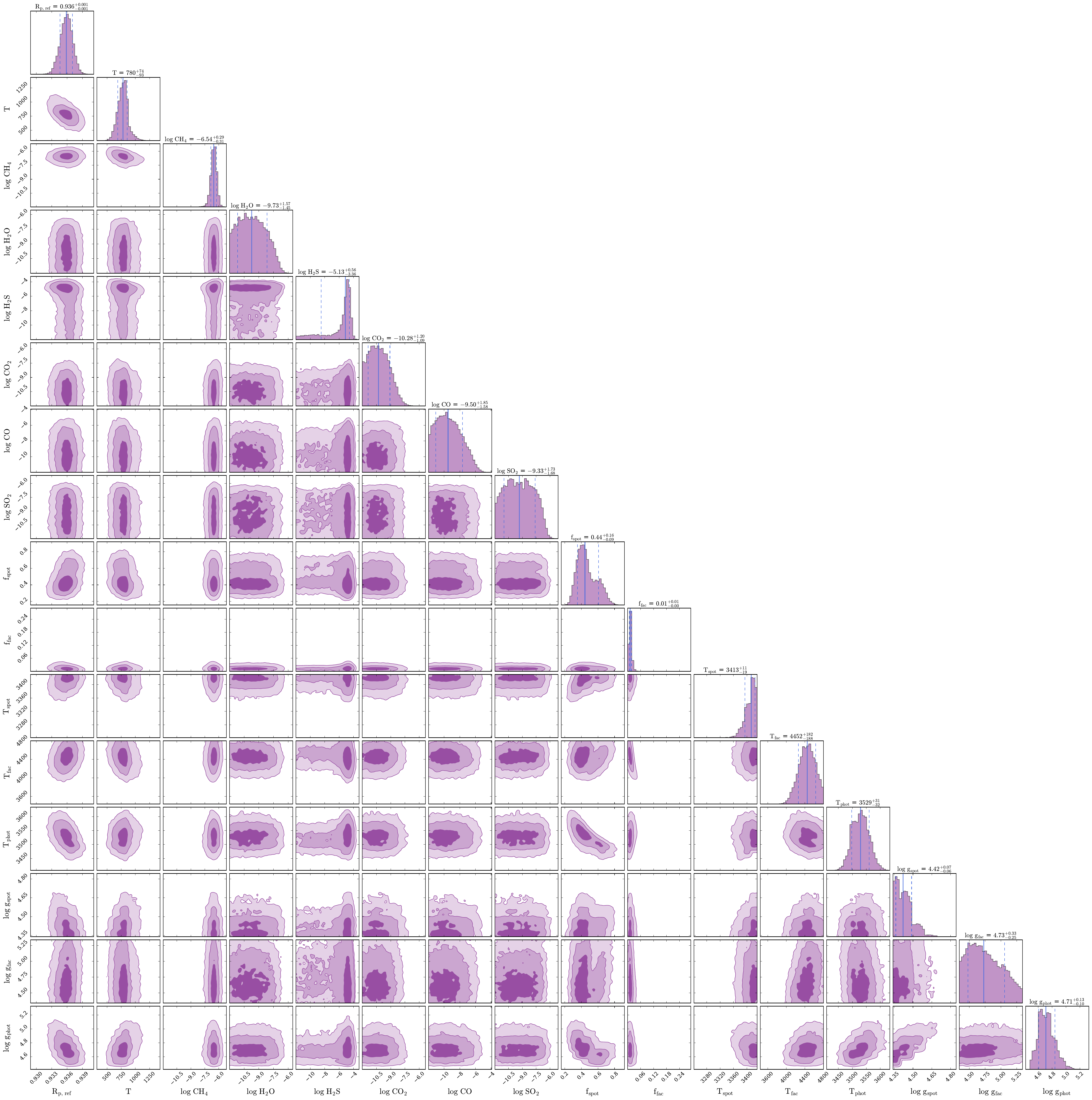}
\figsetgrpnote{Retrieval results for Visit 3 without clouds (as shown in \hyperref[fig:retrieval_corner_visits]{Fig. Set 4.1}).}
\figsetgrpend

\figsetgrpstart
\figsetgrpnum{4.9}
\figsetgrptitle{Co-added spectrum (M3.1; no clouds)}
\figsetplot{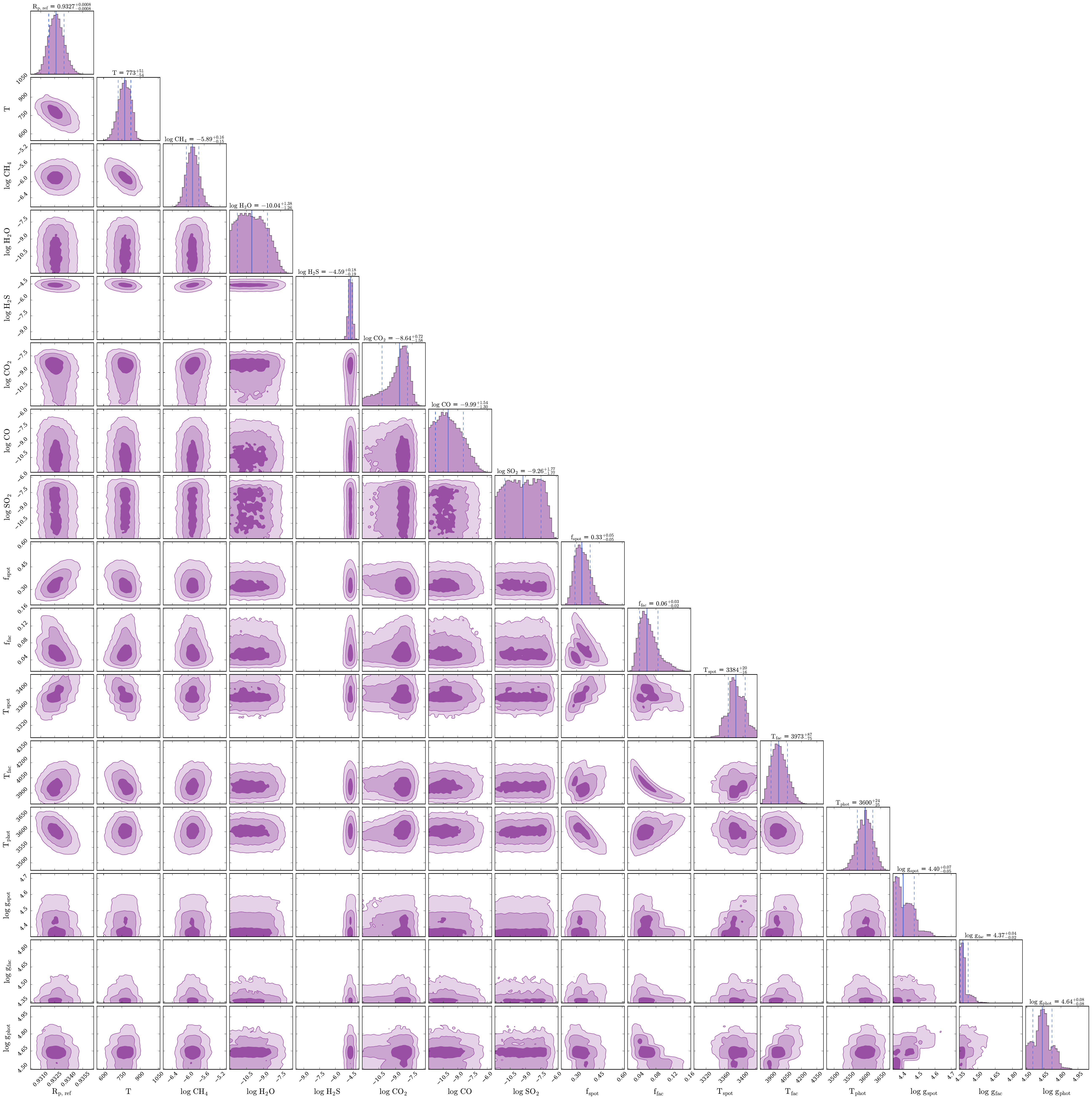}
\figsetgrpnote{Retrieval results for the Visit co-added spectrum without clouds (as shown in \hyperref[fig:retrieval_corner_visits]{Fig. Set 4.1}).}
\figsetgrpend

\figsetend

\clearpage

\appendix
\section{JWST Data Reduction}\label{app:datareduction}
\subsection{Raw ADU counts and non-linearity}\label{app:rawcounts}

\cite{Carter2024} noted that non-linearity may impact the ramp fitting step in the reduction if the counts exceed a conservative threshold of $70\%$ the well depth ($45875$ ADU). \autoref{fig:rawcounts} shows the raw counts for each column along row 15 (the peak in the trace). The maximum count is below the $70\%$ well-depth threshold such that we do not expect any effects from non-linearity on our data.

\setcounter{figure}{10}
\begin{figure*}[!ht]
\epsscale{0.9}
\plotone{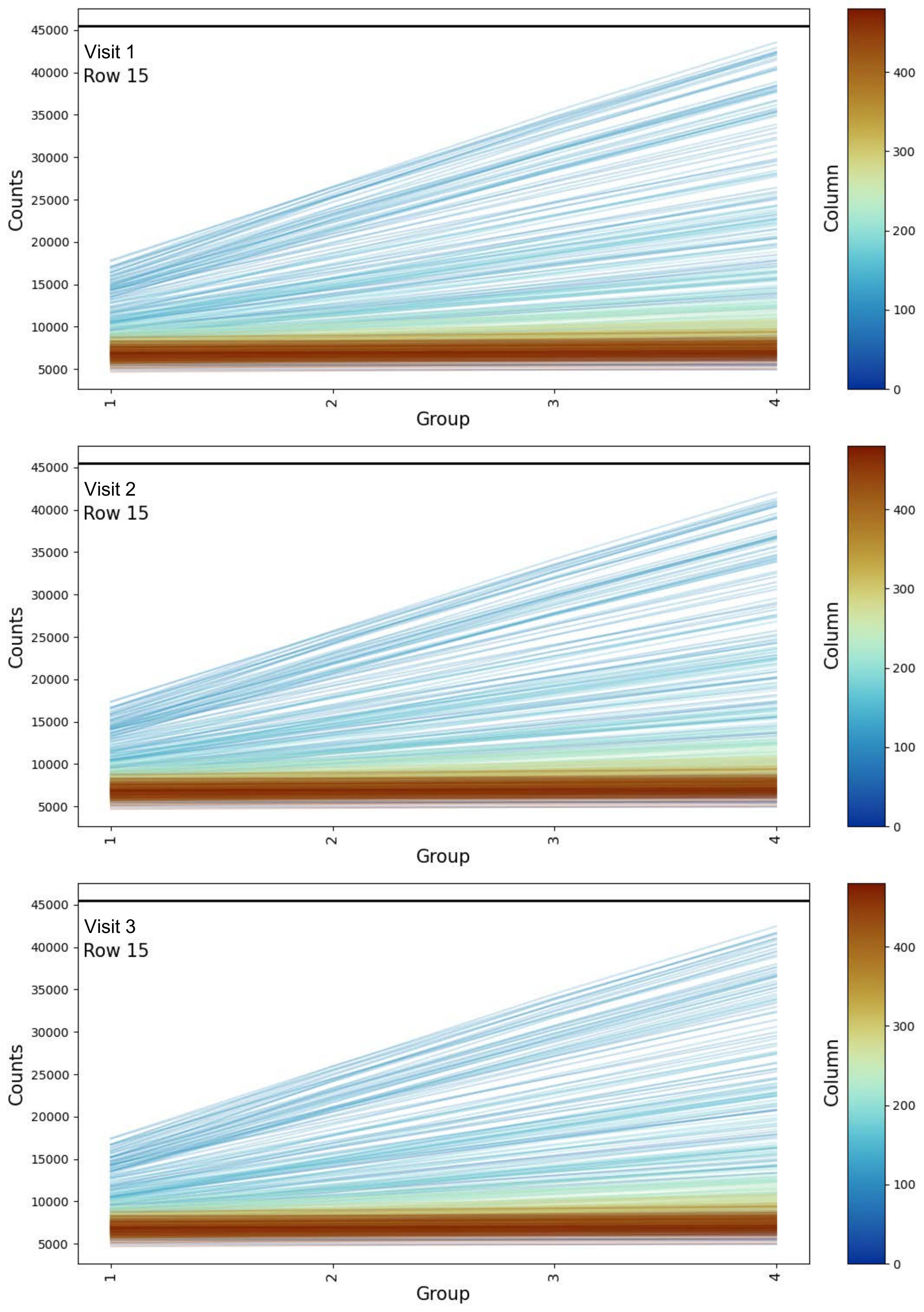}
\caption{\textbf{Top.} The ADU counts along row 15 (center of the trace and maximum flux) for all columns of visit 1 (Obs. 16). \textbf{Middle.} The ADU counts along row 15 for all columns of visit 2 (Obs. 17). \textbf{Bottom.} The ADU counts along row 15 for all columns of visit 3 (Obs. 18). In each panel, the 70\% well-depth count (45875 ADU) is noted as a black line for reference.}\label{fig:rawcounts}
\end{figure*}

\subsection{ExoTiC-JEDI}\label{app:exotic}
We reduced the uncalibrated data files (\texttt{*uncal.fits}) downloaded from the Barbara A. Mikulski Archive for Space Telescopes (MAST)\footnote{\url{https://mast.stsci.edu/portal/Mashup/Clients/Mast/Portal.html}} using the Exoplanet Timeseries Characterisation -- JWST Extraction and Diagnostics Investigator pipeline (\texttt{ExoTiC-JEDI} v0.10). The first two stages of \texttt{ExoTiC-JEDI} are wrappers to the JWST Science Calibration Pipeline (\texttt{jwst} v1.12.5 with CRDSv11.17.21 and CRDS context \texttt{jwst\_1230.pmap}) \citep{bushouse2023} and adhere to the default detector (\texttt{calwebb\_detector1})\footnote{\url{https://jwst-pipeline.readthedocs.io/en/latest/jwst/pipeline/calwebb_detector1.html}} and spectroscopic (\texttt{calwebb\_spec2})\footnote{\url{https://jwst-pipeline.readthedocs.io/en/latest/jwst/pipeline/calwebb_spec2.html}} pipeline processing steps with minor adjustments. Stage 1 of the reduction skipped both the custom \texttt{superbias} step and \texttt{jump} step. Cosmic rays were instead removed as outliers using sigma clipping when extracting spectra in Stage 3. Stage 2 of the reduction included only the following steps:
\begin{enumerate*}[label=(\roman*)]
\item \texttt{assign\_wcs},
\item \texttt{extract\_2d},
\item \texttt{srctype}, and
\item \texttt{wavecorr}.
\end{enumerate*}

We found that a hot pixel in column index 125 and row index 14 resulted in biased flux levels for the corresponding wavelength during stellar spectral extraction. To correct this problem, we created a new custom mask by flagging this pixel as ``DO\_NOT\_USE'' in a modified version of the mask file \texttt{jwst\_nirspec\_mask\_0074.fits} file. While the resulting fit to this channel had less correlated noise, the transit depth was still discrepant. As such, we ignore data from column 125 ($\lambda\sim1.58-1.60$ \textmu{}m) for all spectra in this work. Before ramp-fitting in Stage 1, we used the custom destripping step in \texttt{ExoTiC-JEDI} to ``destrip'' the group-level images and correct for any correlated noise that may be introduced by the detector readout system \citep[``1/f noise'';][]{Moseley2010,Birkmann2022,Rustamkulov2023}. This step masked all pixels within $\sim7$ pixels (3 times the full-width-at-half-maximum of $\sim1.1$ pixels padded with a 4 pixel buffer) from the trace center along the dispersion axis for all groups within an integration and subtracted the median values of non-masked pixels in each column for each group. We tested other apertures (4 and 5 $\times$ FWHM), but found that $3\times$FWHM minimized the scatter in the white light curves.

Before extracting the spectra, we used the data-quality flags to identify any pixels flagged as bad, saturated, dead, hot, low quantum efficiency or no gain value and replaced these with the median value of the surrounding 5 pixels on either side along the row. We similarly replaced pixels that were either $5\sigma$ outliers from the median of the surrounding 20 pixels in each row or $10\sigma$ outliers from the median of that pixel in the surrounding 10 integrations. 

We determined the trace position by fitting a Gaussian to each column of an integration and fitting a constant value to the trace centers and full-width-at-half-maximum (FWHMs). This simple box aperture extended to three times the FWHM of the spectral trace for a full aperture width of $\sim7$ pixels. \texttt{ExoTiC-JEDI} corrected for any remaining $1/f$ and background noise by subtracting the median of the region in each column after masking all pixels 5 FWHMs from the edge of the aperture region (the top and bottom 6 rows). The 1D spectra were extracted following the steps outlined in \cite{horne_optimal_1986} and realigned for positional shifts that were determined via cross-correlation.\footnote{Optimal spectral extraction is included in the public version of \texttt{ExoTiC-JEDI}, see \url{https://github.com/Exo-TiC/ExoTiC-JEDI}.} 

\subsection{Eureka!}\label{app:eureka}
We used the publicly available \texttt{Eureka!} pipeline (v0.1) to analyze the Stage 0 uncalibrated data (\texttt{*uncal.fits}). \texttt{Eureka!} employs the  \texttt{jwst} pipeline (v1.12.5 with CRDSv11.17.21 and CRDS context \texttt{jwst\_1293.pmap}) to reduce uncalibrated files. \texttt{Eureka!} Stages 1 and 2 are wrappers for the \texttt{jwst} pipeline modules \texttt{calwebb\_detector1} and \texttt{calwebb\_spec2}, which perform detector-level and spectroscopic processing, respectively. 

The default \texttt{jwst} Stage 1 detector processing pipeline does not perform reference pixel correction (the \texttt{clean\_flicker\_noise} module is skipped by default\footnote{\url{https://jwst-pipeline.readthedocs.io/en/latest/jwst/pipeline/calwebb_detector1.html}}). As the NRS1 SUB512 detector region used by the NIRSpec PRISM mode does not contain reference pixels, we ran \texttt{Eureka!}'s custom Row-by-row, Odd-Even By Amplifier (ROEBA) routine to correct systematic noise, using 7 rows along both the top and the bottom of the detector as reference pixels. We also employed the top and bottom 7 rows of the SUB512 subarray to perform group-level background subtraction using the median value of this region. For Stage 1, we skipped the \texttt{jump} step (designed to find and flag outliers that are usually due to cosmic rays) because it has proven problematic for observations with low numbers of groups \citep[e.g.,][]{Rustamkulov2023}.  

We followed the recommendation of the Early Release Science team in Stage 2 and skipped the \texttt{flat\_field} step because it removes regions of the spectral trace \citep[e.g.,][]{Alderson2023}. We also skipped the \texttt{photom} step that converts data from units of count rate to flux density in MegaJanskies (MJy), which is not necessary for the ensuing data analysis steps in exoplanet transit TSO observations. We used the same modified mask as described in the ExoTiC-JEDI reductions to mask the identical hot pixel, but ultimately ignored columns 125 ($\lambda\sim1.58-1.60$ \textmu{}m) for all subsequent analysis.  

\texttt{Eureka!} Stage 3 performs background subtraction and optimal spectral extraction \citep{horne_optimal_1986} on the Stage 2 outputs, producing a time series of 1D stellar spectra for use in light curve fitting (\autoref{app:spotconfig}). We adopted a spectral half-width of 3 pixels in our final analyses, since we found this half-width minimized scatter in our white light curve fits. Our Stage 3 background region included pixels along the top and bottom of the SUB512 subarray, greater than 7 pixels from the center of the 2D stellar spectrum. In our optimal extraction, we created a normalized spatial profile by using the median of all data frames, and we clipped outliers greater than 10-$\sigma$ \citep[step 7 of][]{horne_optimal_1986}. 

\subsection{Light curve generation}\label{app:wlcgen}
The outputs from both pipelines were extracted following an identical procedure. The outputs from Stage 3 of \texttt{ExoTiC-JEDI} were repackaged into a format compatible with \texttt{Eureka!} Stage 4 for light curve extraction and spectroscopic fitting.\footnote{A description of the format can be found here \url{https://eurekadocs.readthedocs.io/en/latest/tutorials/formatting_eureka_s4_compatible_data.html}} For both the reduction produced by \texttt{ExoTiC-JEDI} and \texttt{Eureka!}, we used \texttt{Eureka!} Stages 4 to generate the spectroscopic light curves at the native resolution of NIRSpec PRISM ($R<100$), producing one light curve for columns spanning $20-491$ (0.52 \textmu{}m $\lesssim\lambda\lesssim$ 5.6 \textmu{}m) for a total of 471 light curves per observation. The remaining outliers in each light curve were removed using a five-iteration $4\sigma$-outlier rejection step applied to the photometric time series after applying a rolling median filter with a width of 10 integrations.

\subsection{Tilt event in observation 17} \label{app:tiptilt}
In the second visit (observation 17) there is an offset in the photometry (see \autoref{fig:tiltevent}) of $\sim5.7$ ppt in the white light curve. We observed a change in the position of the centroid that coincided with this offset (near integration 1917). We also examined the flux from the fine guidance sensor (FGS) instrument through the mnemonic \texttt{SA\_ZFGINSTCT} \citep{jwstengdb}, which has been shown to correlate with tilt events \citep{Schlawin2023}. The FGS flux displays a discontinuity occurring 5 minutes before\footnote{This is within the temporal calibration uncertainty of $\sim10$ minutes for the FGS instrument \url{https://jwst-docs.stsci.edu/methods-and-roadmaps/jwst-time-series-observations/jwst-time-series-observations-noise-sources\#JWSTTimeSeriesObservationsNoiseSources-MonitoringFGSguidestardata&photometry}} the observed jump in photometry. \cite{Rigby2023} demonstrated that when a primary mirror segment changes orientation rapidly ($\sim$ a few seconds), the magnitude and timing of the change may introduce an offset in the flux. We attribute the observed change in flux to a ``tilt event'' \citep{Schlawin2023,Perrin2024} where an abrupt change in the position of a mirror segment introduces a sudden discontinuity in the time series and the position of the source. Furthermore, we confirmed with the STScI JWST Help Desk\footnote{\url{https://stsci.service-now.com/jwst}} that a tilt event did occur at 2023-10-11T21:20:01.873 based on the FGS flux (compared to 2023-10-11T21:25:00.894 in the time series). We conservatively excluded the first 2070 integrations from the data to remove the region impacted by this event. After excluding these points, observation 17 still had a pre-transit baseline longer than one hour. 

\section{Light curve fitting}\label{app:spotconfig}

\subsection{Modeling spot crossing events}
\begin{figure*}[tt]
\epsscale{1.17}
\plotone{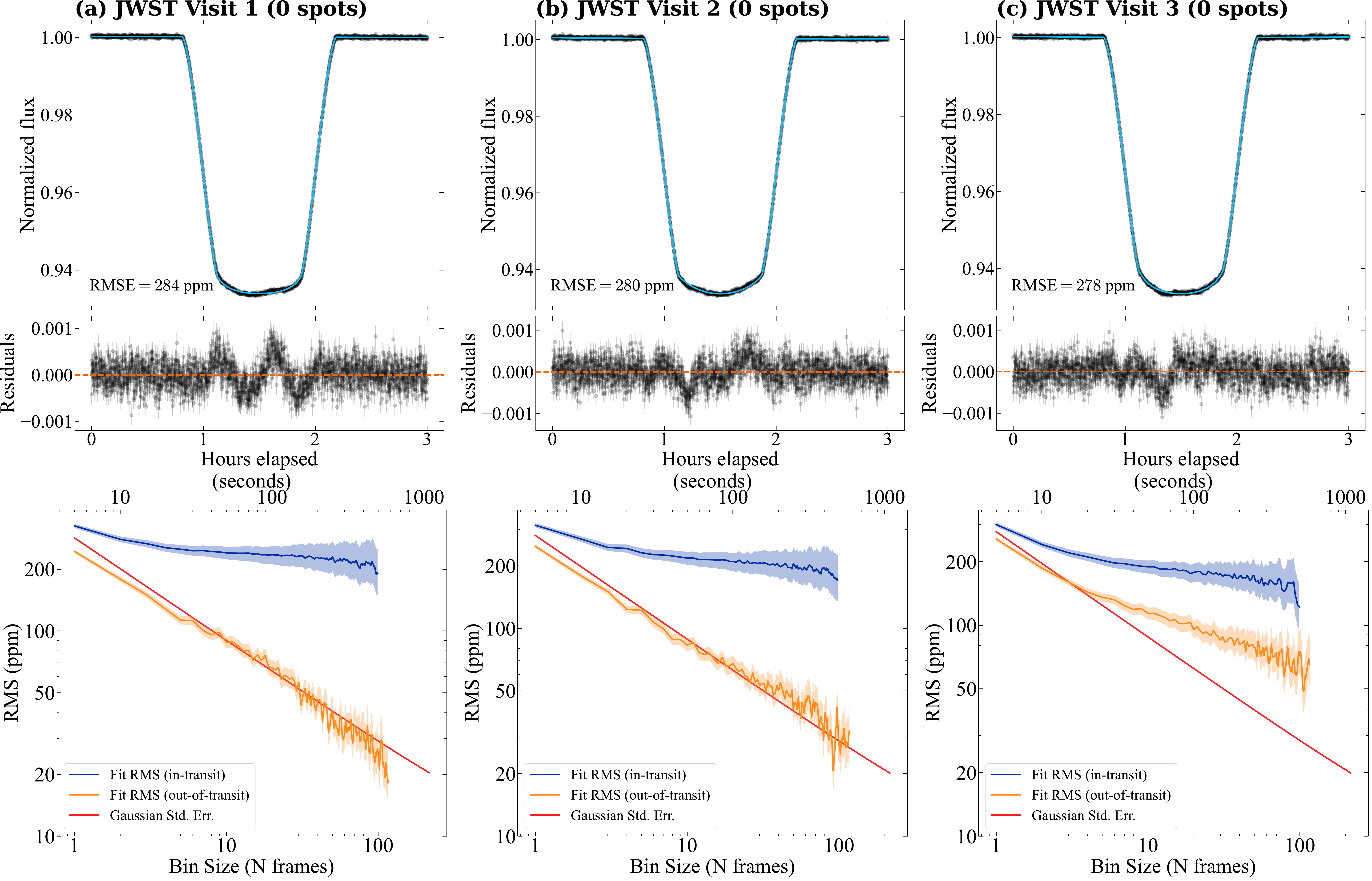}
\caption{Similar to \autoref{fig:wlc} except displaying the model that includes no spots. \textbf{Top.} For each visit, the best-fitting model is indicated by the solid line with residuals to the fit indicated in the bottom panel. \textbf{Bottom.} The RMS for each visit for in-transit (blue) and out-of-transit (orange) data. There is significant structure in-transit when the data are not modeled to account for spot crossing events.}
\label{fig:nospotwlc}
\end{figure*}

We show the best-fitting model to the white light curves with no spot crossing events in \autoref{fig:nospotwlc}. These fits displayed significant in-transit variability and motivated models that accounted for surface inhomogeneities on the star (e.g., spots and faculae).  Spots were modeled using \texttt{spotrod}\footnote{\url{https://github.com/bencebeky/spotrod}} \citep{spotrod2014} and treated as projected circles on a spherical surface that were individually parameterized with the following four variables:
\begin{enumerate*}[label=(\roman*)]
\item spot flux ratio, $f_\mathrm{spot}$, or the ratio of the integrated flux of the spot compared to the stellar photosphere ($f_\mathrm{spot}=1$ is the photosphere),
\item spot radius, $r_\mathrm{spot}$, or the size of the spot in units of stellar radii, and 
\item two positional parameters, $x_\mathrm{spot}$ and $y_\mathrm{spot}$, which describe the center of the spot in stellar radii using a projected planetary coordinate system \citep[e.g.,][]{Winn2010}.
\end{enumerate*}

\subsection{White light curve fits}
To determine an optimal configuration, we individually fit the white light curves derived from the \texttt{ExoTiC-JEDI} reduction using all combinations of spots (between $1-4$) for each visit. We binned the white light curves (5s cadence) for these fits to decrease the computational time required to compute a model. Each fit applied
\begin{enumerate*}[label=(\roman*)]
\item a uniform prior on the spot flux ratio ($0\le f_\mathrm{spot}\le1$) which was different for all visits,
\item a uniform prior on the radius ($0\le r_\mathrm{spot}\le0.5$) that was different for each spot,
\item and spot position that varied for each spot and was sampled from a unit disk.
\end{enumerate*} The remaining free parameters were the orbital parameters ($P$, $T_0$, $a/R_\star$, $b$), the radius ($R_p/R_\star$), quadratic limb darkening coefficients sampled following the parameterization in \cite{kipping_efficient_2013} ($q_1$, $q_2$), and the coefficients for a quadratic polynomial to model any out of transit systematics ($c_0$, $c_1$, $c_2$). The priors for all free parameters in the fits to the white light curves are listed in \autoref{tab:5205priors}.

\startlongtable
\begin{deluxetable*}{lccc}
\tablecaption{Priors for the white and spectroscopic light curve fits for TOI-5205\label{tab:5205priors}}
\tablehead{\colhead{Name} &
\colhead{Units} &
\colhead{Prior}
}
\startdata
\sidehead{Free parameters in white light curve fits:}
~~~Period & days & $\mathcal{N}(1.63,0.01)$\\
~~~Time of mid-transit ($T_0$) & $\mathrm{BJD_{TDB}}$ & $\mathcal{N}(2460227.86,0.01)$ \\
~~~Scaled semi-major axis ($a/R_\star$) & \nodata & $\mathcal{N}(10.94,0.22)$ \\
~~~Eccentricity ($e$) & \nodata & 0 \\
~~~Argument of periastron ($\omega_\star$) & deg & 90\\
~~~Scaled radius ($R_p/R_\star$) & \nodata & $\mathcal{U}(0,1)$ \\
~~~Impact parameter ($b$) & \nodata & $\mathcal{U}(0,1)$ \\
~~~Dilution term ($D_{\mathrm{JWST}}$) & \nodata & 1 \\
~~~Jitter term ($\sigma_{\mathrm{JWST}}$) & \nodata & 0 \\
~~~Linear limb darkening coefficient ($q_1$) & \nodata & $\mathcal{U}(0,1)$ \\
~~~Quadratic limb darkening coefficient ($q_2$) & \nodata & $\mathcal{U}(0,1)$ \\
~~~Polynomial baseline constant term ($c_0$) & \nodata & $\mathcal{U}(-3,3)$ \\
~~~Polynomial baseline linear term ($c_1$) & \nodata & $\mathcal{U}(-3,3)$ \\
~~~Polynomial baseline quadratic term ($c_2$) & \nodata & $\mathcal{U}(-3,3)$ \\
~~~Spot flux ratio ($f_\mathrm{spot}$) & \nodata & $\mathcal{U}(0,1)$\\
~~~Spot position ($U_{i}$) & \nodata & $\mathcal{U}(0,1)$ \\
~~~Spot position ($V_{i}$) & \nodata & $\mathcal{U}(0,1)$ \\
~~~Spot radius ($r_{\mathrm{spot},i}$) & \nodata & $\mathcal{U}(0,0.5)$ \\
\sidehead{Free parameters in spectroscopic light curve fits:$^\dagger$}
~~~Scaled radius ($R_p/R_\star$) & \nodata & $\mathcal{U}(0,1)$ \\
~~~Errorbar Scaling ($\sigma_{\mathrm{JWST}}$) & \nodata & $\mathcal{U}(0,5)$ \\
~~~Linear limb darkening coefficient ($q_1$) & \nodata & $\mathcal{U}(0,1)$ \\
~~~Quadratic limb darkening coefficient ($q_2$) & \nodata & $\mathcal{U}(0,1)$ \\
~~~Polynomial baseline constant term ($c_0$) & \nodata & $\mathcal{U}(-3,-3)$ \\
~~~Polynomial baseline linear term ($c_1$) & \nodata & $\mathcal{U}(-3,-3)$ \\
~~~Polynomial baseline quadratic term ($c_2$) & \nodata & $\mathcal{U}(-3,-3)$ \\
~~~Spot flux ratio ($f_\mathrm{spot}$) & \nodata & $\mathcal{U}(0,2)$\\
\enddata
\tablenotetext{\dagger}{The transit model parameters and spot size and position are fixed to the results from the white light curve fits.}
\tablecomments{For all spots, the center is sampled from a unit disk using the dimensionless parameters $U_i$ and $V_i$, where the positional inputs to \texttt{spotrod} are calculated as $X_i=\sqrt{U_i}\cos(2\pi V_i)$ and $Y_i=\sqrt{U_i}\sin(2\pi V_i)$.}
\end{deluxetable*}

We sampled the parameter space using the dynamic nested sampler \texttt{dynesty} \citep{Speagle2020} using 5000 live points and a convergence criterion of $\Delta\ln Z=0.01$. The models with different spot configurations were then compared using various model selection criteria (see \autoref{tab:spotevidence}). We adopted the configuration with 2 spots in visit 1, 3 spots in visit 2, and 4 spots in visit 3 as it maximized the AIC for each visit compared to other spot configurations. Due to the complex and multimodal nature of starspot modeling, we did not exclusively rely on the Bayesian evidence but instead used a selection criterion that penalized more complex models, such as the AIC \citep[Akaike Information Criterion;][]{Akaike1974} or BIC \citep[Bayesian Information Criterion;][]{Schwarz1978}, to prevent instances where greater complexity yielded minimal impacts on the model (as with 5 spots in visit 1). We ultimately chose the models that minimized AIC because it was the metric that consistently yielded strong support for only the selected models \citep[$\Delta \mathrm{AIC}>10$ for all other configurations; see][]{Burnham2002}. We note that the preferred model for visits 2 and 3 is different if relying on the BIC or the Bayesian evidence and this highlights the degenerate and multi-modal nature of spot modeling using only the white light curves. The complexities and degeneracies in fitting star spots have also been observed in other JWST datasets \citep[e.g.,][]{Libby-Roberts2025,Murray2025}. \hyperref[fig:spectracomp]{Fig. Set 2} shows the spectra derived from our selected model. The spot parameters, along with their priors, for the adopted configuration are listed in \autoref{tab:5205spotpar}.

\startlongtable
\begin{deluxetable*}{cccccc}
\tablecaption{Statistical results for various spot configurations\label{tab:spotevidence}}
\tablehead{\colhead{Number of starspots} &
\colhead{$\ln Z$} &
\colhead{$\ln B$} &
\colhead{$\Delta \mathrm{BIC}$} &
\colhead{$\Delta \mathrm{AIC}$} &
\colhead{RMSE (ppm)}
}
\startdata
\sidehead{Visit 1 (Obs 16):}
~~~0 & $14297.3\pm0.5$ & 543.1 & 1088.2 & 1127.93 & 284.0\\
~~~1 & $14581.1\pm0.5$ & 259.3 & 516.8 & 533.8 & 260.9\\
~~~\textbf{2} & $\mathbf{14840.4\pm0.6}$ & \nodata & \nodata & \nodata & \textbf{238.5}\\
~~~3 & $14822.7\pm0.6$ & 17.7 & 36.7 & 19.7 & 239.3\\
~~~4 & $14800.3\pm0.7$ & 40.1 & 74.7 & 40.6 & 239.7\\
~~~5 & $14838.8\pm0.6$ & 1.6 & 62 & 10.9 & 238.2\\
\sidehead{Visit 2 (Obs 17):}
~~~0 & $14356.5\pm0.5$ & 391.7 & 280.1 & 816.8 & 280.1\\
~~~1 & $14696.9\pm0.5$ & 51.3 & 252.2 & 140.4 & 252.2\\
~~~2 & $14700.6\pm0.6$ & 47.6 & 251.0 & 135.1 & 251.0\\
~~~\textbf{3} & $\mathbf{14748.2\pm0.7}$ & \nodata & \nodata & \nodata & \textbf{244.3}\\
~~~4 & $14754.0\pm0.6$ & -5.8 & 245.4 & 48.0 & 245.4\\
~~~5 & $14738.6\pm0.6$ & 9.6 & 247.2 & 115.3 & 247.2\\
\sidehead{Visit 3 (Obs 18):}
~~~0 & $14383.3\pm0.5$ & 244.6 & 548.3 & 622.0 & 277.6\\
~~~1 & $14442.8\pm0.6$ & 185.1 & 385.6 & 436.7 & 270.2\\
~~~2 & $14637.3\pm0.7$ & -9.4 & -7.9 & 26.2 & 253.7\\
~~~3 & $14649.5\pm0.6$ & -21.6 & 11.0 & 28.0 & 253.6\\
~~~\textbf{4} & $\mathbf{14627.9\pm0.7}$ & \nodata & \nodata & \nodata & \textbf{252.2}\\
~~~5 & $14634.7\pm0.7$ & -6.8 & 31.5 & 14.4 & 252.5\\
\enddata
\tablecomments{The adopted configuration, based on the AIC, for each visit is in bold. For each visit, all differences are calculated with respect to the adopted configuration (bolded rows) such that $\ln B=\ln Z_{\mathrm{bold}} - \ln Z_i$, $\Delta \mathrm{BIC} = \mathrm{BIC}_i - \mathrm{BIC}_{\mathrm{bold}}$, and $\Delta \mathrm{AIC} = \mathrm{AIC}_i - \mathrm{AIC}_{\mathrm{bold}}$, where $i$ refers to the number of spots. RMSE refers to the root mean square error of the residuals and is only included for reference.}
\end{deluxetable*}

\subsection{Spectroscopic light curve fits}\label{app:specfit}
Both the \texttt{ExoTiC-JEDI} and \texttt{Eureka!} spectroscopic light curves were fit using Stage 5 from \texttt{Eureka!} with identical priors.\footnote{We did not perform the spectroscopic fits with \texttt{juliet} out of convenience, as \texttt{Eureka!} provided a well-vetted pipeline that could use the same systematic model (quadratic polynomial), transit code (\texttt{spotrod}), and sampler (\texttt{dynesty}) while also generating the resulting spectra and various diagnostic plots. We also note that the version of \texttt{Eureka!} (v0.1) used in this work did not have the ability to perform a joint white light curve fits.} We fixed the orbital parameters $P$, $T_0$, $e$, $\omega_\star$, $a/R_\star$, and $i$\footnote{This implicitly fixes the impact parameter, $b\equiv a/R_\star \cos i$} from \autoref{tab:5205par} along with the spot configuration to the values from \autoref{tab:5205spotpar} that were derived from our white light curve fit. The priors for the free parameters in the fits to the spectroscopic light curves are listed in \autoref{tab:5205priors}. The light curve for each spectral channel had eight free parameters, including the 
\begin{enumerate*}[label=(\roman*)]
\item radius ratio ($R_p/R_\star$),
\item parameterized coefficients for a quadratic limb darkening law ($q_1,~q_2$),
\item coefficients for a second-order polynomial as a function of time to model the out-of-transit baseline,
\item spot flux ratio ($f_\textrm{spot}$), and
\item a multiplicative factor to the uncertainties for each channel.
\end{enumerate*}
We sampled the parameter space using the dynamic nested sampler \texttt{dynesty} with 1000 live points and a convergence criterion of $\Delta\ln Z=0.1$. Some channels for all visits for both the \texttt{ExoTiC-JEDI} and \texttt{Eureka!} reductions are shown in \hyperref[figset:lc]{Fig. Set 13}. The transit depths as a function of wavelength (i.e., the transmission spectra) are shown in \hyperref[fig:spectracomp]{Fig. Set 2}.

\startlongtable
\begin{deluxetable*}{lhcccccc}
\tablecaption{Spot parameters for TOI-5205\label{tab:5205spotpar}}
\tablehead{\colhead{Name} &
\nocolhead{Units} &
\colhead{Prior} &
\colhead{Value} &
\colhead{Source} &
}
\startdata
\sidehead{Visit 1 (Obs. 16) spot parameters:}
~~~Spot flux ratio ($f_\mathrm{spot}$) & \nodata & $\mathcal{U}(0,1)$ & $0.928_{-0.009}^{+0.008}$ & This work\\
~~~$U_{16,1}$ & \nodata & $\mathcal{U}(0,1)$ & $0.40 \pm 0.01$ & This work\\
~~~$V_{16,1}$ & \nodata & $\mathcal{U}(0,1)$ & $0.552_{-0.006}^{+0.003}$ & This work\\
~~~$r_{\mathrm{spot},16,1}$ & \nodata & $\mathcal{U}(0,0.5)$ & $0.12_{-0.01}^{+0.02}$ & This work\\
~~~$U_{16,2}$ & \nodata & $\mathcal{U}(0,1)$ & $0.37_{-0.02}^{+0.04}$ & This work\\
~~~$V_{16,2}$ & \nodata & $\mathcal{U}(0,1)$ & $0.818_{-0.005}^{+0.004}$ & This work\\
~~~$r_{\mathrm{spot},16,2}$ & \nodata & $\mathcal{U}(0,0.5)$ & $0.13 \pm 0.02$ & This work\\
\sidehead{Visit 2 (Obs. 17) spot parameters:}
~~~Spot flux ratio ($f_\mathrm{spot}$) & \nodata & $\mathcal{U}(0,1)$ & $0.916 \pm 0.008$ & This work\\
~~~$U_{17,1}$ & \nodata & $\mathcal{U}(0,1)$ & $0.68_{-0.03}^{+0.02}$ & This work\\
~~~$V_{17,1}$ & \nodata & $\mathcal{U}(0,1)$ & $0.835 \pm 0.004$ & This work\\
~~~$r_{\mathrm{spot},17,1}$ & \nodata & $\mathcal{U}(0,0.5)$ & $0.31 \pm 0.02$ & This work\\
~~~$U_{17,2}$ & \nodata & $\mathcal{U}(0,1)$ & $0.51 \pm 0.03$ & This work\\
~~~$V_{17,2}$ & \nodata & $\mathcal{U}(0,1)$ & $0.666 \pm 0.005$ & This work\\
~~~$r_{\mathrm{spot},17,2}$ & \nodata & $\mathcal{U}(0,0.5)$ & $0.13_{-0.02}^{+0.03}$ & This work\\
~~~$U_{17,3}$ & \nodata & $\mathcal{U}(0,1)$ & $0.87 \pm 0.02$ & This work\\
~~~$V_{17,3}$ & \nodata & $\mathcal{U}(0,1)$ & $0.594_{-0.003}^{+0.005}$ & This work\\
~~~$r_{\mathrm{spot},17,3}$ & \nodata & $\mathcal{U}(0,0.5)$ & $0.16 \pm 0.02$ & This work\\
\sidehead{Visit 3 (Obs. 18) spot parameters:}
~~~Spot flux ratio ($f_\mathrm{spot}$) & \nodata & $\mathcal{U}(0,1)$ & $0.941 \pm 0.004$ & This work\\
~~~$U_{18,1}$ & \nodata & $\mathcal{U}(0,1)$ & $0.74_{-0.10}^{+0.08}$ & This work\\
~~~$V_{18,1}$ & \nodata & $\mathcal{U}(0,1)$ & $0.10_{-0.07}^{+0.19}$ & This work\\
~~~$r_{\mathrm{spot},18,1}$ & \nodata & $\mathcal{U}(0,0.5)$ & $0.38_{-0.08}^{+0.07}$ & This work\\
~~~$U_{18,2}$ & \nodata & $\mathcal{U}(0,1)$ & $0.33_{-0.01}^{+0.02}$ & This work\\
~~~$V_{18,2}$ & \nodata & $\mathcal{U}(0,1)$ & $0.98 \pm 0.01$ & This work\\
~~~$r_{\mathrm{spot},18,2}$ & \nodata & $\mathcal{U}(0,0.5)$ & $0.27_{-0.04}^{+0.05}$ & This work\\
~~~$U_{18,3}$ & \nodata & $\mathcal{U}(0,1)$ & $0.41\pm0.01$ & This work\\
~~~$V_{18,3}$ & \nodata & $\mathcal{U}(0,1)$ & $0.562_{-0.005}^{+0.004}$ & This work\\
~~~$r_{\mathrm{spot},18,3}$ & \nodata & $\mathcal{U}(0,0.5)$ & $0.17_{-0.01}^{+0.02}$ & This work\\
~~~$U_{18,4}$ & \nodata & $\mathcal{U}(0,1)$ & $0.21_{-0.01}^{+0.02}$ & This work\\
~~~$V_{18,4}$ & \nodata & $\mathcal{U}(0,1)$ & $0.760 \pm 0.003$ & This work\\
~~~$r_{\mathrm{spot},18,4}$ & \nodata & $\mathcal{U}(0,0.5)$ & $0.139_{-0.010}^{+0.009}$ & This work\\
\enddata
\end{deluxetable*}

\figsetstart
\figsetnum{13}
\figsettitle{A few channels from the spectroscopic fits.}

\figsetgrpstart
\figsetgrpnum{13.1}
\figsetgrptitle{\texttt{ExoTiC-JEDI}}
\figsetplot{fb_chromatic_lc_exotic.pdf}
\figsetgrpnote{\textbf{Top.} Light curves from a few channels for the \texttt{ExoTiC-JEDI} reduction, after binning to 5 s for clarity. Each observation had a total of 471 light curves and 8 are shown as a reference. The best-fitting model for each channel is indicated by the solid line and adopts the spot configuration (position and size) from the white light curve fit (shown in \autoref{fig:wlc}). \textbf{Bottom.} Residuals after subtracting the model.}\label{figset:lc}
\figsetgrpend

\figsetgrpstart
\figsetgrpnum{13.2}
\figsetgrptitle{\texttt{Eureka!}}
\figsetplot{fb_chromatic_lc_eureka.pdf}
\figsetgrpnote{Similar to \hyperref[figset:lc]{Fig. Set 13.1} but showing data for \texttt{Eureka!}.}
\figsetgrpend

\subsection{Spot flux ratios for TOI-5205}
The fit to the white light curve revealed spots with a high-flux ratio ($f_{\mathrm{spot}}\sim0.9-0.94$) and a spot temperature ($T_{\mathrm{spot}}\sim3200-3400$ K), which is $\sim100-200$ K cooler than the photosphere. Although the sample of mid-M dwarfs with known spot temperatures is limited, it is expected that the spot contrast decreases with decreasing stellar temperature \citep[e.g., ][]{Berdyugina2005}. \hyperref[fig:spotcomp]{Fig. Set 14} displays the derived spot flux ratio for TOI-5205 for all 3 visits.  In the bluest orders, we have lower spot flux ratios, indicative of higher contrast between the surface and stellar inhomogeneity. At redder wavelengths, the fits routinely favor a spot flux ration centered on unity, confirming that the red channels are less impacted by stellar activity. A flux ratio of $\sim0.9$ is consistent with the temperature differential between spotted and ambient temperatures for other mid-M dwarfs \citep[e.g.,][]{Herbst2021,Almenara2022,Waalkes2024,Mori2024}.

\figsetstart
\figsetnum{14}
\figsettitle{Spot contrasts from the spectroscopic fits}

\figsetgrpstart
\figsetgrpnum{14.1}
\figsetgrptitle{Overview}
\figsetplot{fb_spotcomp_transit1.pdf}
\figsetgrpnote{\textbf{Top}. The spot flux ratio, $f_{\mathrm{spot}}$, for the first observation (observation 16) of TOI-5205 on 2023 October 10. The \texttt{ExoTiC-JEDI} reduction is the green circles and the \texttt{Eureka!} reductions is the orange squares. The dashed line at unity denotes the photosphere, such that values below represent cooler surface features (spots) and values above unity represent hotter surface features (faculae).}\label{fig:spotcomp}
\figsetgrpend

\figsetgrpstart
\figsetgrpnum{14.2}
\figsetgrptitle{Visit 2}
\figsetplot{fb_spotcomp_transit2.pdf}
\figsetgrpnote{Comparison of the spot flux ratio for visit 2 of TOI-5205 on 2023 October 11 (observation 17, similar to \hyperref[fig:spotcomp]{Fig. Set 14.1}).}
\figsetgrpend

\figsetgrpstart
\figsetgrpnum{14.3}
\figsetgrptitle{Visit 3}
\figsetplot{fb_spotcomp_transit3.pdf}
\figsetgrpnote{Comparison of the spot flux ratio for visit 3 of TOI-5205 on 2023 October 13 (observation 18, similar to \hyperref[fig:spotcomp]{Fig. Set 14.1}).}
\figsetgrpend

\figsetend

\section{Discrepancy with the published radius}\label{app:depthdiff}
\cite{kanodia_toi-5205b_2023} measured the radius ratio of TOI-5205b as $R_{\mathrm{ARCTIC}}/R_\star=0.2720^{+0.0039}_{-0.0043}$ using ground-based data obtained with the Astrophysical Research Consortium (ARC) Telescope Imaging Camera on the 3.5m telescope at the Apache Point Observatory \citep[ARCTIC;][]{ARCTIC2016}. For reference, the SDSS $i^\prime$ data were collected on 2022 April 22 and the SDSS $g^\prime$ data were collected on 2022 July 03. The ARCTIC transit depth is deeper ($\Delta (R_p/R_\star)=0.0245$, $>5\sigma$ difference) than the transit depth recovered with JWST ($R_{JWST}/R_\star=0.2475_{-0.0002}^{+0.0003}$). To investigate this discrepancy, we extracted all available TESS photometry using an effective point spread function to correct for contamination from on-sky companions with \texttt{tglc} \citep{tglc2023}. The TESS photometry included data from sectors 15 (2019 August 15 $-$ 2019 September 10), 41 (2021 July 23 $-$ 2021 August 20), 55 (2022 August 5 $-$ 2022 September 1), and 82 (2024 August 10 $-$ 2024 September 5). A comparison of the photometry is shown in \autoref{fig:groundbasecomp}, illustrating that JWST has the best precision and the smallest transit depth of all the photometry ($\sim1\%$ shallower). 

\setcounter{figure}{14}
\begin{figure*}[!ht]
\epsscale{1.15}
\plotone{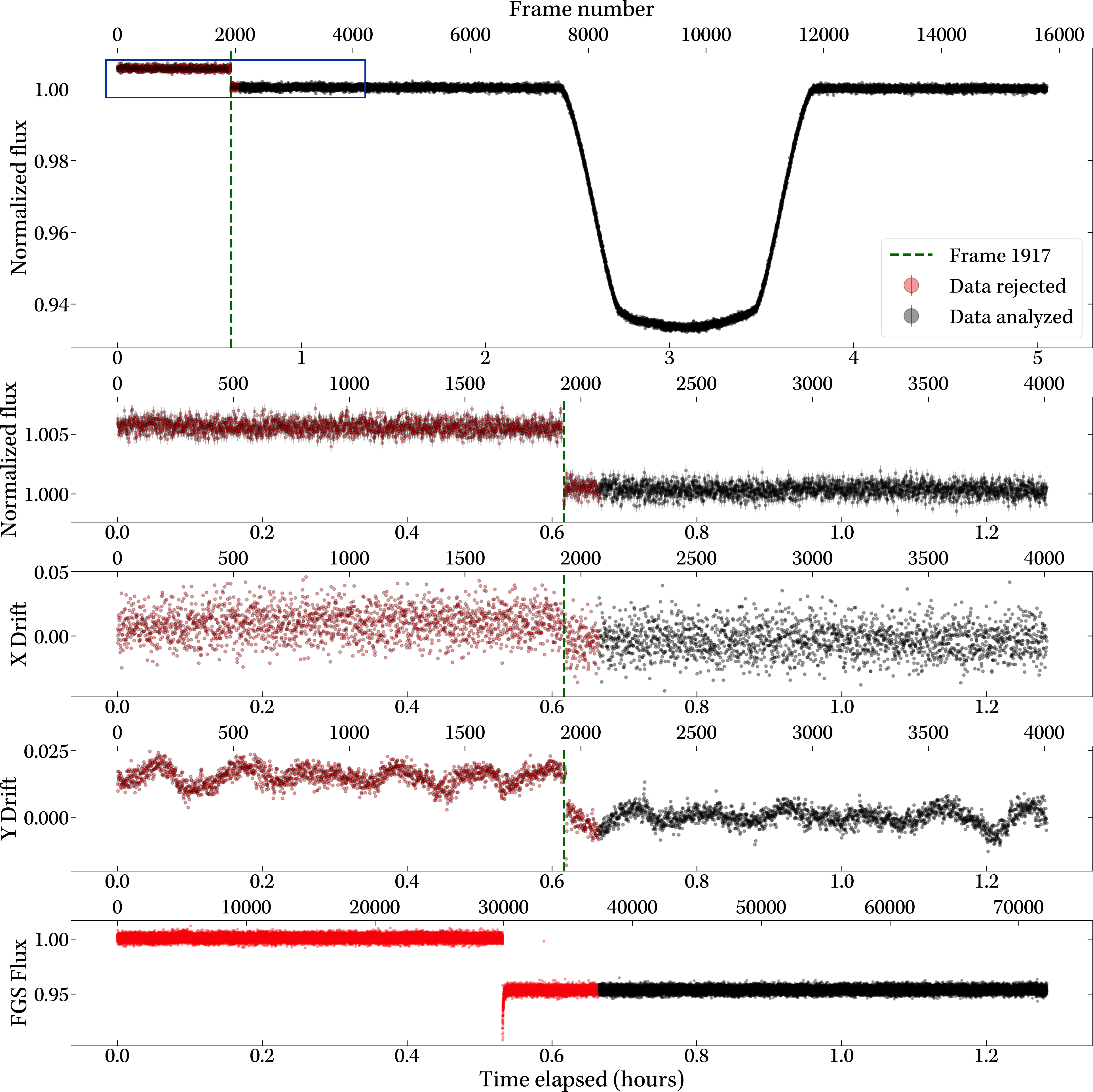}
\caption{\textbf{Top.} The white light curve for observation 17 calculated with \texttt{ExoTiC-JEDI}, the box indicates the region highlighted in the next 3 sub-panels. \textbf{Bottom four panels}: The photometry, shift in the centroid along the dispersion axis (X drift), shift in the centroid along the cross-dispersion axis (Y drift), and the counts in the fine guidance sensor (FGS flux). The discontinuity occurs simultaneously in the position of the trace and the photometry while the FGS flux shows a discontinuity 5 minutes earlier. The dashed vertical line marks integration frame 1917 (the start of the event). The red points indicate the data excised from subsequent analysis (the first 2070 frames) and the black data are the points analyzed in this work.}
\label{fig:tiltevent}
\end{figure*}

\cite{tglc2023} note there exist limitations to the dilution-correction for point-spread function photometry from TESS due to the incomplete nature of the Gaia DR3 catalog such that the PSF-derived photometry may not match the true depth of the transit, but the \texttt{tglc} photometry is still consistent with the JWST depth at the $1-2\sigma$ level. The ground-based data from ARCTIC had limited out-of-transit baselines either due to poor weather conditions or dusk, which could impact the measured transit depth. We investigated whether this depth difference could be due to unocculted spots and modeled the ground-based ARCTIC Sloan $i^\prime$ transit using the modified \texttt{juliet} code described in \autoref{app:spotconfig}. For this fit, we used the posterior values for the orbital parameters from \autoref{tab:5205par} ($P$, $T_0$, $a/R_\star$, $R_p/R_\star$, $b$) as priors and retained the uninformed spot priors. The fit requires an unocculted spot with a flux ratio of $f_\mathrm{spot,ARCTIC}=0.62\pm0.01$ (corresponding temperature of $T_{\mathrm{spot,ARCTIC}}=3170\pm50$ K). This spot would have to be present for both the ground-based transits from ARCTIC (separated by 72 days but with identical depths) while remaining outside the transit chord and subsequently disappearing before the JWST observations. This scenario is unlikely due to the significant surface evolution observed in the three consecutive JWST transits (see \autoref{fig:wlc}), which suggested that a specific spot configuration is unlikely to be long-lived. Furthermore, this high flux ratio is inconsistent with the observed spot contrasts and the predictions for other mid-M dwarfs \citep[e.g.,][]{Mori2024, Mori2025}. Even with the change in spot configuration seen in the JWST data, the transit depths across the three visits are remarkably consistent. We do not assume that this change in depth between JWST and ARTIC is astrophysical in nature, and instead attribute the discrepancy to uncorrected systematics in the ARCTIC photometry (due to weather, airmass, or other sky conditions). These systematics cannot be robustly characterized or corrected with the existing out-of-transit baseline and subsequently lead to an apparently inflated radius. The importance of a proper baseline to correct for systematic trends and derive an unbiased depth has been noted for other planetary systems \citep[e.g.,][]{Libby-Roberts2023}. For this work, we adopt the radius derived using the JWST data. 

\begin{figure*}[!ht]
\epsscale{1.15}
\plotone{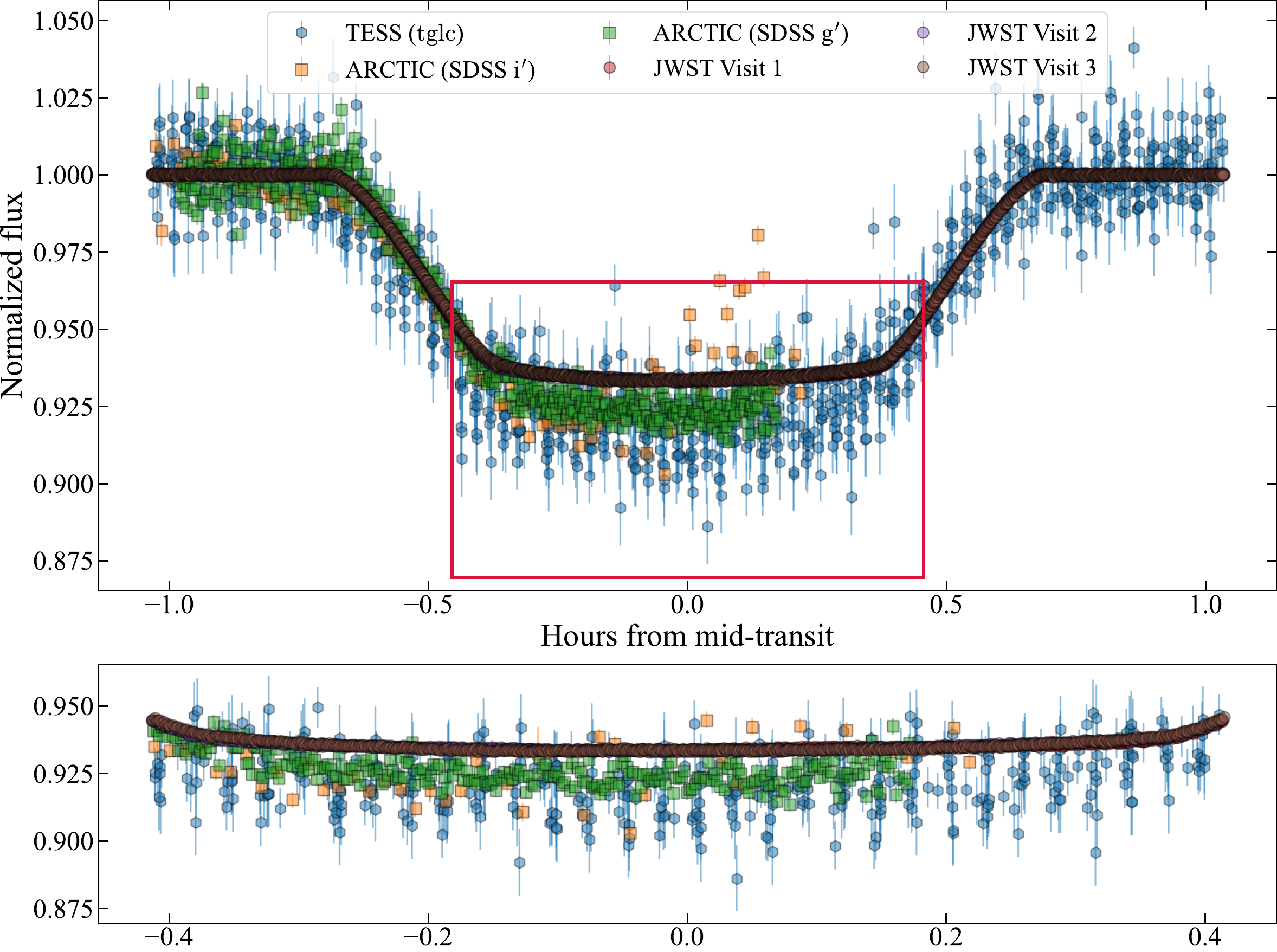}
\caption{\textbf{Top}: A comparison of published ARCTIC photometry from \cite{kanodia_toi-5205b_2023} with the photometry derived in this work. \textbf{Bottom}: A zoom into the region indicated by a rectangle on the top panel. Both panels display the phase-folded photometry for ARCTIC transits from \cite{kanodia_toi-5205b_2023}, the TESS data derived using a point-spread function to attempt to correct for dilution, and the JWST data. All ARCTIC transits and all TESS transits (before and after JWST observations) are deeper than the white light curves from JWST.}
\label{fig:groundbasecomp}
\end{figure*}

\section{A comparison of grid-based limb darkening coefficients}\label{app:limbdark}
We used \texttt{ExoTiC-LD} \citep{Grant2024_exotic-ld} to investigate the impact of different limb darkening coefficients for the quadratic limb darkening law. We investigated most options available with \texttt{ExoTiC-LD}, including limb darkening coefficients for a quadratic limb darkening law derived with the
\begin{enumerate*}[label=(\roman*)]
\item PHOENIX stellar library \citep[henceforth, PHOENIX;][]{Husser2013},
\item ATLAS9 stellar library \citep[henceforth, Kurucz;][]{Kurucz1970,Kurucz1993},
\item MPS-ATLAS stellar library \citep[henceforth, MPS1/2;][]{mpsgrid2022}, and 
\item the STAGGER grid of three-dimensional hydrodynamic model atmospheres \citep[henceforth, STAGGER;][]{magic2015stagger}.
\end{enumerate*} 
For this test, we fix the quadratic limb darkening coefficients to the values of $u_1$ and $u_2$ derived with \texttt{ExoTiC-LD} and do not rely on the parameterization of \cite{kipping_efficient_2013} (as was done in \S\ref{sec:wlcfits}).

Although the comparison includes the STAGGER grid \citep{magic2015stagger}, the closest STAGGER grid point in \texttt{ExoTiC-LD} had an effective temperature of $T_{\mathrm{eff}}=4500$ K, which is $>1000$ K hotter than the observed $T_{\mathrm{eff}}=3430$ K for TOI-5205. We also note that the MPS1/2 and Kurucz grids have lower limits of $T_{\mathrm{eff}}=3500$ K. For this comparison, we used the derived limb darkening coefficients from the spectra presented in \S\ref{sec:wlcfits}. We converted the $q_1$ and $q_2$ coefficients to $u_1$ and $u_2$ to enable a direct comparison using equations 15 and 16 presented by \cite{kipping_efficient_2013}. All spectra were produced almost identically to the procedure described in \hyperref[app:specfit]{Appendix \ref*{app:specfit}} except that the limb darkening coefficients were fixed to the predicted values. 

A comparison of the spectra for the \texttt{ExoTiC-JEDI} reductions is presented in \hyperref[figset:ld]{Fig. Set 17}. There is a significant discrepancy in transit depth in the region $\le2.5$ \textmu{}m when the limb darkening coefficients were fixed to values from the grids. The spectra derived using the PHOENIX limb darkening coefficients are most similar to the spectra where we fit the coefficients. Other stellar grids showed differences at the $1-2\sigma$ level until $\sim4$~\textmu{}m. Given the limitations in (i) temperature coverage of the MPS1/2, Kurucz, and STAGGER grids, (ii) large disagreement between grids, and (iii) mismatch between the PHOENIX spectra in the mid-infrared for mid-M dwarfs \citep[e.g.,][]{Rajpurohit2018}, we use spectra where the limb darkening coefficients are sampled following \cite{kipping_efficient_2013} for all data analyzed in the body of this manuscript. 

\figsetstart
\figsetnum{17}
\figsettitle{A comparison of limb-darkening coefficients for the quadratic limb-darkening law.}

\figsetgrpstart
\figsetgrpnum{17.1}
\figsetgrptitle{Visit 1 (Obs. 16)}
\figsetplot{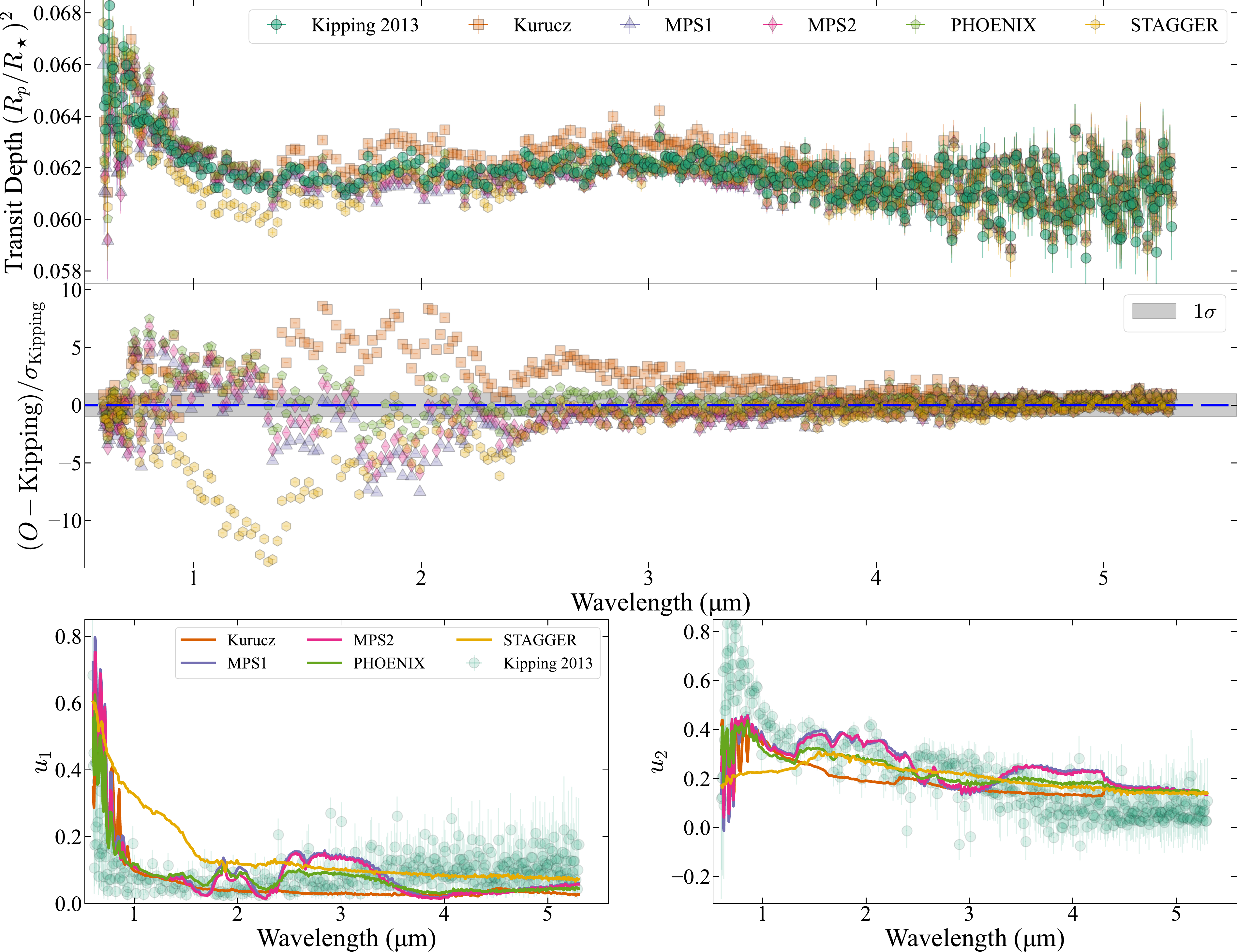}
\figsetgrpnote{\textbf{Top}: TOI-5205b transmission spectra for visit 1 (Obs. 16) derived using either fixed quadratic limb-darkening coefficients ($u_1$ and $u_2$) from \texttt{ExoTiC-LD} or fitting for the quadratic limb-darkening coefficients following the parametrization from \cite{kipping_efficient_2013} ($q_1$ and $q_2$). Spectra are labeled according to the limb-darkening coefficient grid. \textbf{Middle}: The residuals with respect to the spectrum with free limb-darkening coefficients (labeled ``Kipping 2013''). The residuals are divided by the observed error of the spectrum where limb darkening is a free parameter. The $\pm1\sigma$ region is shaded for reference. \textbf{Bottom}: The limb-darkening coefficients, $u_1$ and $u_2$, for a quadratic limb-darkening law that were used to generate the spectra.}\label{figset:ld}
\figsetgrpend

\figsetgrpstart
\figsetgrpnum{17.2}
\figsetgrptitle{Visit 2 (Obs. 17)}
\figsetplot{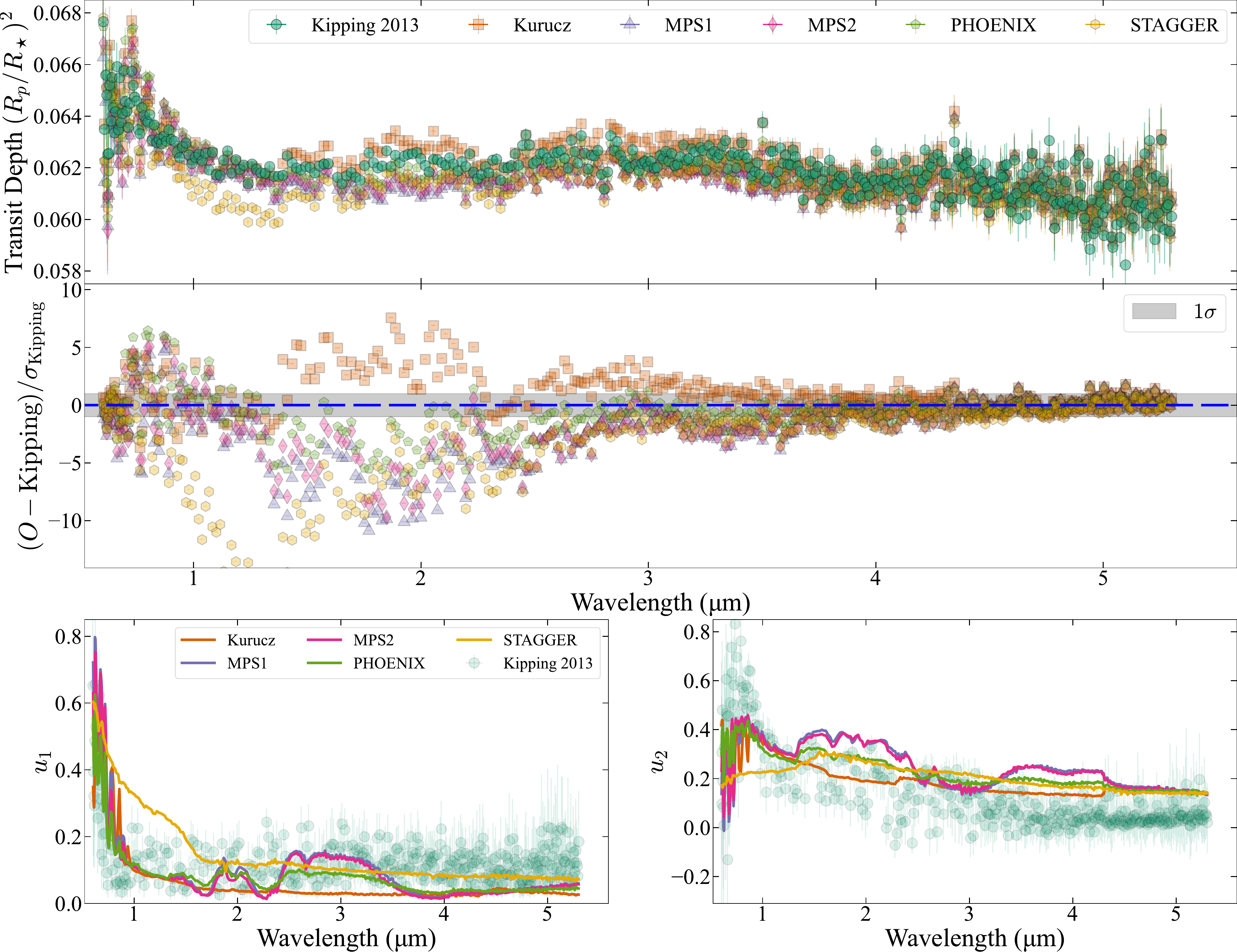}
\figsetgrpnote{Similar to \hyperref[figset:ld]{Fig. Set 17.1} but showing data for visit 2 (Obs. 17).}
\figsetgrpend

\figsetgrpstart
\figsetgrpnum{17.3}
\figsetgrptitle{Visit 3 (Obs. 18)}
\figsetplot{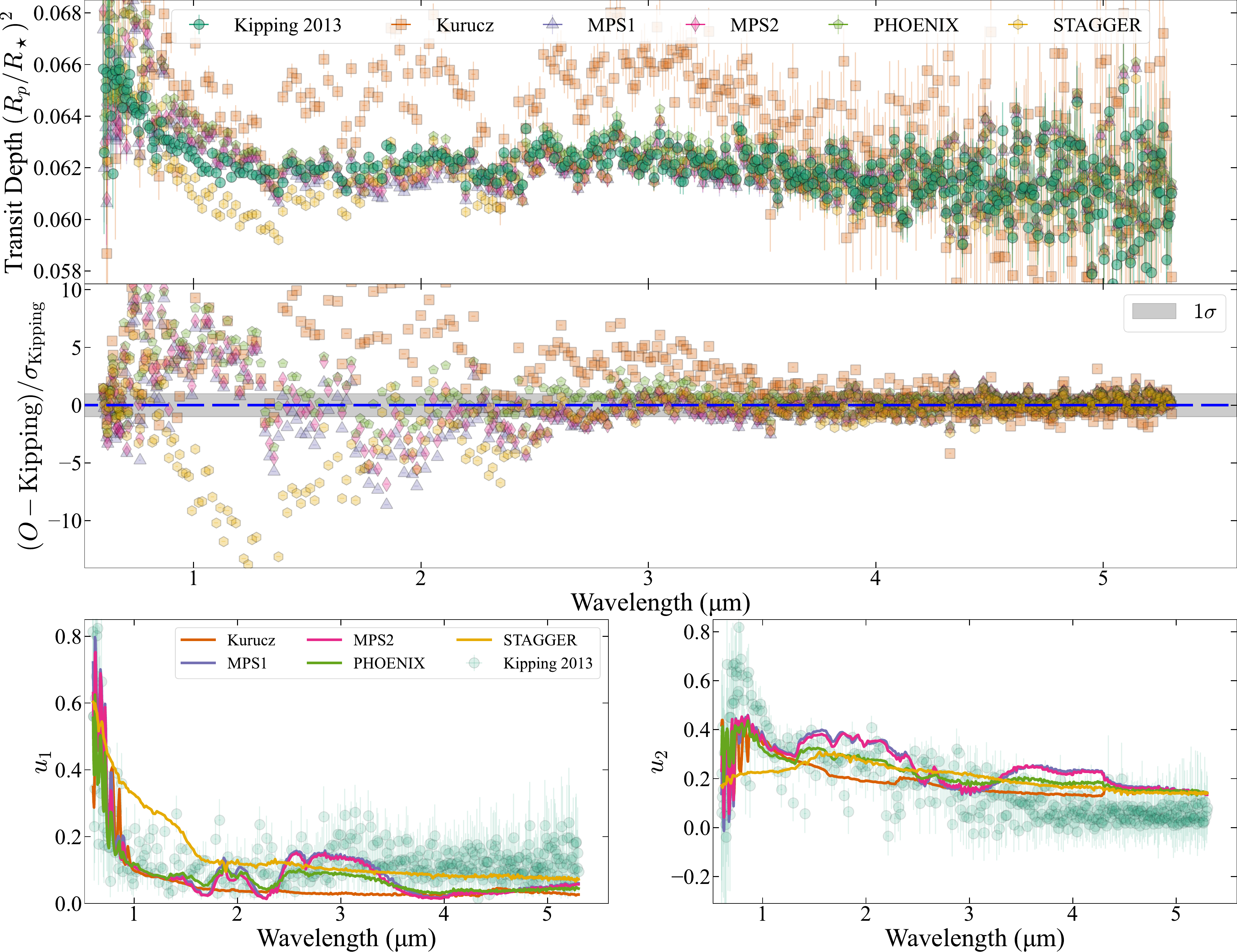}
\figsetgrpnote{Similar to \hyperref[figset:ld]{Fig. Set 17.1} but showing data for visit 3 (Obs. 18).}
\figsetgrpend
\figsetend

\section{Bayesian Equilibrium Chemistry Retrievals}\label{app:eqchem}
\subsection{\texttt{PLATON}}\label{app:platonchemeq}
We performed retrievals on the co-added spectrum, assuming a clear atmosphere (no clouds or hazes) and equilibrium chemistry, using the PLanetary Atmospheric Tool for Observer Noobs modeling package \citep[\texttt{PLATON}\footnote{\url{https://github.com/ideasrule/platon}} v6.3.1;][]{Zhang2019,Zhang2020,platon6,platon631}. \texttt{PLATON} allows for modeling of the TLS effect assuming one stellar heterogeneity (e.g., spots or faculae). The dilution due to unocculted starspots was modeled following the methodology in \cite{Rackham2018} and, similar to our \texttt{PICASO} implementation (see \S\ref{sec:picaso}), was parameterized with a spot coverage fraction and spot temperature (see priors in \autoref{tab:5205chemretrieval}). \texttt{PLATON} computes chemical equilibrium abundances using \texttt{FastChem} with a grid spanning $-2\le \log \mathrm{[M/H]} \le 3$ and $0.001 \le \mathrm{C/O} \le 2$. We adopted an isothermal pressure-temperature profile and included an error multiplication factor in the model. We performed the retrieval using the default \texttt{PLATON} opacity database resolution of $R=20,000$ and sampled the parameter space using the dynamic nested sampler \texttt{dynesty} with $N=2000$ live points and a convergence criterion of $\Delta\ln Z=1.0$. 

\setcounter{figure}{17}
\begin{figure*}[!ht]
\epsscale{1.17}
\plotone{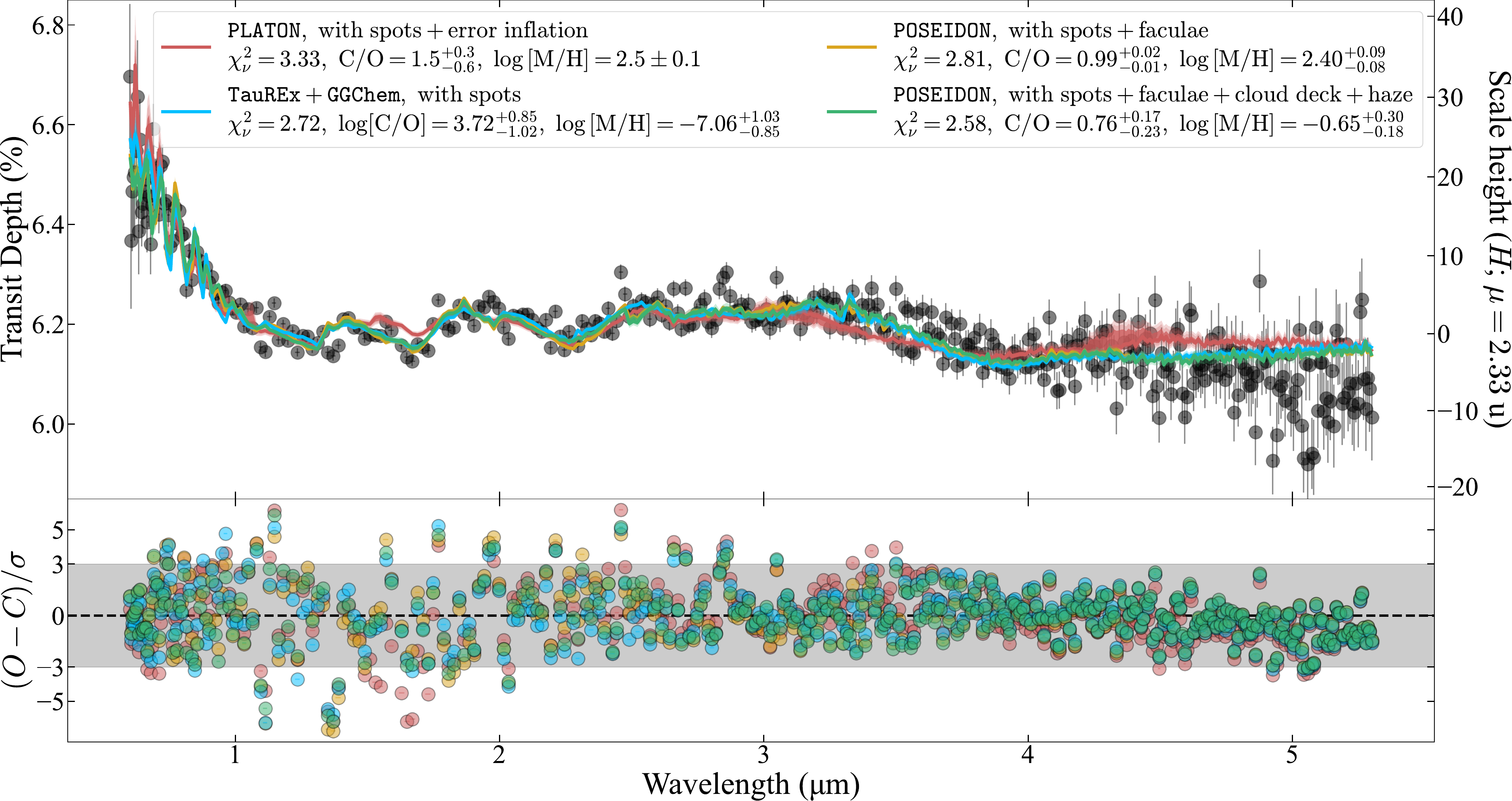}
\caption{\textbf{Top.} Atmospheric retrievals, assuming equilibrium chemistry, on the co-added \texttt{ExoTiC-JEDI} spectrum (black circles). The best-fitting models for \texttt{POSEIDON} (yellow and green lines), \texttt{PLATON} (red line), and \texttt{TauREx} (blue line) are plotted along with the $1-3$ model posteriors (shaded blue or red regions). \textbf{Bottom.} The residuals to the best-fitting solution for \texttt{POSEIDON} (yellow and green circles), \texttt{PLATON} (red circles), and \texttt{TauREx} (blue circles). The $\pm3\sigma$ region is shaded for reference.}
\label{fig:chemeqret}
\end{figure*}

\begin{deluxetable*}{lc|cc|cc}
\tablecaption{Model parameters for retrievals with the \texttt{ExoTiC-JEDI} co-added spectrum assuming equilibrium chemistry and a clear atmosphere with \texttt{PLATON} and \texttt{TauREx}. Empty parameters are not used for that specific model. \label{tab:5205chemretrieval}}
\tablehead{
 &
 &
\multicolumn{2}{c|}{\texttt{PLATON}} &
\multicolumn{2}{c}{\texttt{TauREx + GGChem}}
\\
 &
 &
\multicolumn{2}{c|}{$\chi^2_\nu=3.33$, $\ln Z=2465.36\pm0.09$} &
\multicolumn{2}{c}{$\chi^2_\nu=2.72$, $\ln Z = 2359.88 \pm 0.18$} 
\\
\colhead{Parameter} &
\multicolumn{1}{c|}{Units} &
\colhead{Prior} &
\multicolumn{1}{c|}{Value} &
\colhead{Prior} &
\colhead{Value}
}
\startdata
~~Stellar radius ($R_\star$) & $\mathrm{R_\odot}$ & $\mathcal{N}(0.392,0.015)$ & $0.395 \pm 0.007$ & \nodata & \nodata \\
~~Planetary mass ($M_p$) & $\mathrm{M_J}$ & $\mathcal{N}(1.08,0.06)$ & $1.07 \pm 0.06$ & $\mathcal{N}(1.08,0.06)$ & $1.04 \pm 0.04$\\
~~Planetary radius ($R_p$) & $\mathrm{R_J}$ & $\mathcal{U}(0.84,1.03)$ & $0.94 \pm 0.02$ & $\mathcal{U}(0.79,1.07)$ & $0.929 \pm 0.001$ \\
~~Equilibrium temperature ($T_{\mathrm{eq}}$) & K & $\mathcal{U}(200,1500)$ & $1460_{-40}^{+20}$ & $\mathcal{U}(200,1500)$ & $834^{+15}_{-27}$\\
~~Atmospheric metallicity ($\mathrm{[M/H]}$) & dex & $\mathcal{U}(-2,3)$ & $2.5 \pm 0.1$ & $\mathcal{U}(-10,5)$ & $-7.06^{+1.03}_{-0.85}$ \\
~~Atmospheric carbon-to-oxygen ratio (C/O)$^\ddag$ & \nodata & $\mathcal{U}(0.001,2)$ & $1.5_{-0.6}^{+0.3}$ & $\mathcal{LU}(-10,5)$  & $5215_{-4712}^{+31615}$\\
~~Spot coverage fraction ($f_\mathrm{spot}$) & \nodata &  $\mathcal{U}(0,1)$ & $0.23_{-0.05}^{+0.10}$ & $\mathcal{U}(0,1)$ & $0.097\pm0.002$ \\
~~Spot temperature ($T_\mathrm{spot}$) & K & $\mathcal{U}(2930,3930)$ & $3410_{-50}^{+40}$ & $\mathcal{U}(2300,6000)$ & $3187\pm15$\\
~~Spot surface gravity ($\log g_\mathrm{spot}$) & dex & Fixed & 4.84 & $\mathcal{U}(3.5,6)$ & $3.54^{+0.05}_{-0.03}$\\
~~Photosphere temperature ($T_\mathrm{eff}$) & K & $\mathcal{N}(3430,54)$  & $3560_{-40}^{+30}$ & $\mathcal{N}(3430,54)$ & $3601^{+21}_{-26}$\\
~~Photosphere surface gravity ($\log g_\mathrm{fac}$) & dex & Fixed & 4.84 & $\mathcal{U}(4.34,5.34)$ & $4.84 \pm 0.03$\\
~~Error multiplicative factor$^\dagger$ & \nodata & $\mathcal{U}(0.5,5)$ & $1.82 \pm 0.06$ & \nodata & \nodata \\
\enddata
\tablenotetext{\dagger}{The error inflation term is not used to calculate the reported $\chi^2_\nu$ for \texttt{PLATON}. If error inflation were included in the calculation, then $\chi^2_\nu=1.01$.}
\tablenotetext{\ddag}{For \texttt{TauREx}, C/O was sampled in dex using a uniform logarithmic prior (base 10); this prior is denoted as $\mathcal{LU}(X,Y)$ and represents a lower limit of $10^X$ and upper limit of $10^Y$.}
\end{deluxetable*}

\subsection{\texttt{TauREx}}
We performed a set of Bayesian retrievals assuming equilibrium chemistry with the retrieval package Tau Retrieval for Exoplanets \citep[\texttt{TauREx}\footnote{\url{https://github.com/ucl-exoplanets/taurex3}} v3.2.4;][]{Refaie2021,Refaie2022}. We used the \texttt{GGChem}\footnote{\url{https://github.com/ucl-exoplanets/GGchem}} \citep[Gleich-Gewichts-Chemie;][]{Woitke2018} plugin \citep{Refaie2022} to calculate molecular abundances, assuming a solar chemical abundance, an isothermal profile spanning pressure levels of $10^{-7}-10^3$ bar, and a clear atmosphere. We did not include the effects of ions or condensation equilibrium in the \texttt{GGChem} calculations. The model included all species provided by \texttt{TauREx} \citep{Changeat2025} or ExoMol \citep{Chubb2021} that contained H, He, C, N, O, P, S. We used the mixin\footnote{\url{https://taurex3.readthedocs.io/en/stable/user/taurex/mixins.html}} capability of \texttt{TauREx} to implement the TLS model following the methodology in \cite{Rackham2018} in which we parameterized the TLS model using a spot coverage fraction, spot temperature, and spot surface gravity (see priors in \autoref{tab:5205chemretrieval}). We adopted the NewEra\footnote{\url{https://www.fdr.uni-hamburg.de/record/18108}} synthetic spectra \citep{Hauschildt2025} to model spots and faculae and sampled the opacities at a spectral resolution of R $\gtrsim15,000$ for the radiative transfer calculations. We sampled the posteriors using \texttt{MultiNest} with $N=1000$ live points and a convergence criterion of $\Delta\ln Z=1$.

\subsection{\texttt{POSEIDON}}\label{sec:poseidonchemeq}
We also performed retrievals on the co-added spectrum, assuming chemical equilibrium, using the \texttt{POSEIDON} code \citep{MacDonald2017, MacDonald2023}. \texttt{POSEIDON} uses \texttt{FastChem} to compute equilibrium chemical abundances on a grid that spans a range of $-1\le \log \mathrm{[M/H]} \le 4$ and $0.2 \le \mathrm{C/O} \le 2$. Similarly to the free retrievals with \poseidon (\S\ref{sec:freechemret}), we adopted an isothermal pressure-temperature profile, performed the radiative transfer calculations on an intermediate resolution spectral grid set to $R=20,000$ from $0.5 - 5.2$ \textmu{}m, modeled the TLS effect due to unocculted spots and faculae, and used \texttt{MultiNest} to sample the posteriors with the same convergence criterion. The TLS model was analogous to model M1.4 and included a stellar photosphere with two heterogeneities that have a unique filling factor and where each feature was modeled by PHOENIX spectra assuming free temperature and surface gravity. Our retrievals assumed either a clear atmosphere, an atmosphere with clouds, or an atmosphere with clouds and hazes (comparable to M3.1--3.3 described in \S\ref{sec:retrievals:models}, albeit assuming equilibrium chemistry). The priors and results are listed in \autoref{tab:5205aerosolretrieval}.

\subsection{Limitations with equilibrium chemistry}

The results for all retrievals with the \texttt{ExoTiC-JEDI} co-added spectrum are shown in \autoref{fig:chemeqret} and a summary of the priors and retrieved parameters are listed in \autoref{tab:5205chemretrieval} and \autoref{tab:5205aerosolretrieval}. Under the assumption of a clear atmosphere, the \texttt{POSEIDON} code provides a better fit to the data compared to \texttt{PLATON} because of (i) the additional flexibility in the model, particularly with the TLS effect, which considers two heterogeneities with different temperatures and surface gravities from the host star and (ii) the inclusion of error inflation in \texttt{PLATON} which results in an increased error bar to account for inconsistencies between the model and the data. Both \texttt{POSEIDON} and \texttt{PLATON} recover a consistent atmospheric carbon-to-oxygen ratio ($\mathrm{C/O}\sim0.9-1.5$) and metallicity ($\log \mathrm{[M/H]\sim2.4-2.5}$ dex). These retrievals provided a better fit than the grid results (a lower $\chi^2_\nu$ compared to \S\ref{sec:forwardmodelling}). 

\begin{figure}[!ht]
\epsscale{1.15}
\plotone{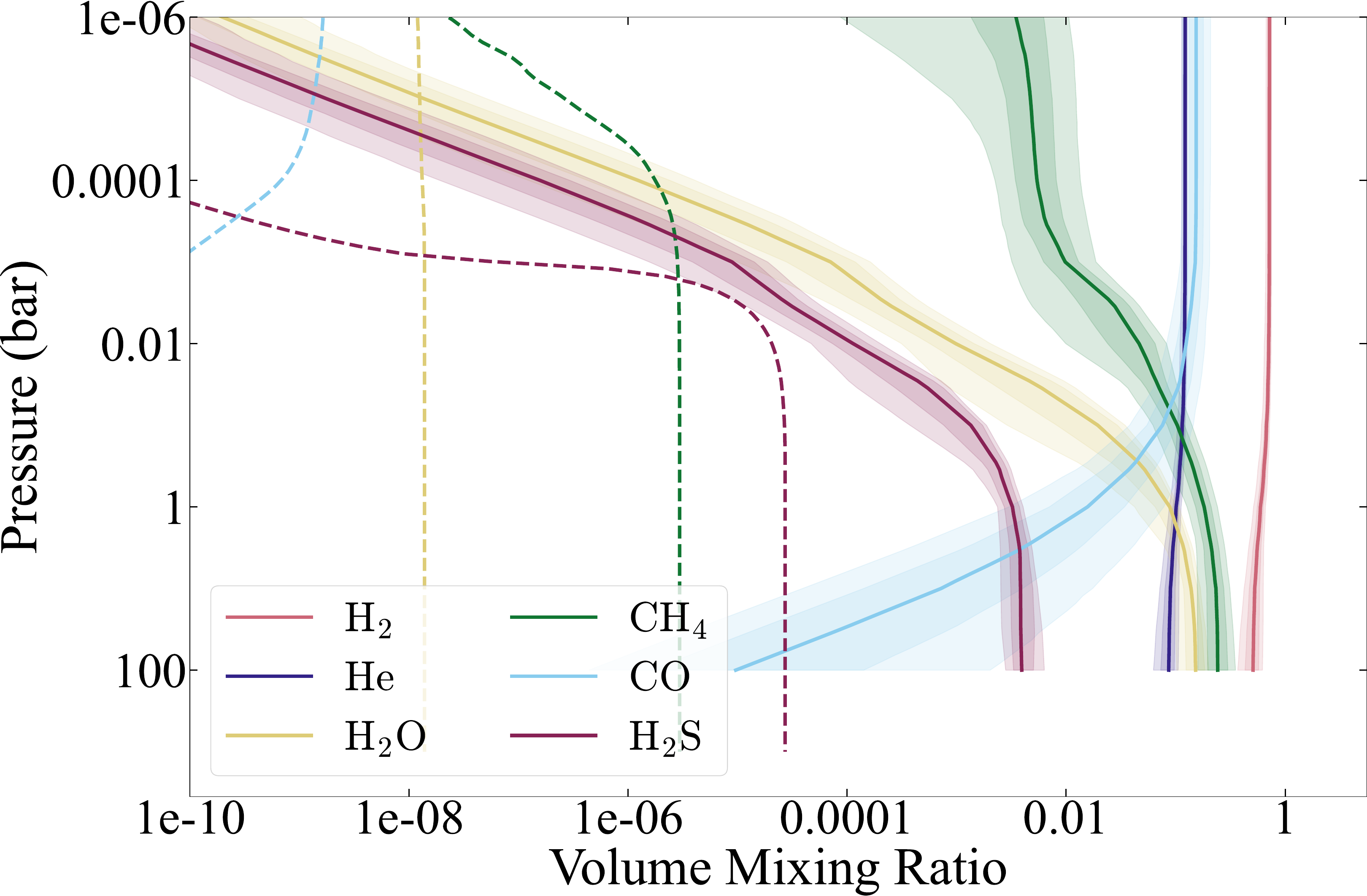}
\caption{Volume mixing ratios for select molecules, assuming chemical equilibrium and a clear atmosphere, obtained from \texttt{POSEIDON} (solid lines) when fitting the co-added spectrum. The shaded regions indicate the $\pm1-2\sigma$ model posteriors. The dashed lines are the abundances for \ce{H2O}, \ce{CH4}, \ce{CO}, and \ce{H2S} from the best-fitting \texttt{VULCAN} model in \autoref{tab:picasovulcan}. From the forward model, these molecules are expected with VMRs $<10^{-4}$ in stark contrast to the \texttt{POSEIDON} values where some trace molecules are found with VMRs $>0.01$.}
\label{fig:poseidon_abundances}
\end{figure}

The solution from \texttt{POSEIDON} and \texttt{PLATON} contrasted sharply with the metal-poor atmosphere that was preferred by forward models (\S\ref{sec:forwardmodelling}) and \texttt{TauREx}. Of the three Bayesian retrievals, the \texttt{TauREx} retrieval was not limited to a grid and appeared to favor an atmosphere that was metal-poor and carbon-rich (qualitatively similar to our preferred model in \S\ref{sec:freechemret}). \autoref{fig:chemeqret} shows that the predicted model from \texttt{TauREx} and \texttt{POSEIDON} appeared to be very similar despite the different atmospheric properties. We examined the abundances of a few molecules (\ce{H2O}, \ce{CO}, \ce{CH4}) in the \texttt{POSEIDON} retrieval (see \autoref{fig:poseidon_abundances}), which revealed that the abundances of carbon-bearing species were orders of magnitude larger than those of the best-fitting forward models ($\mathrm{VMR}\lesssim10^{-4}$). Furthermore, carbon-bearing species (\ce{CO} and \ce{CH4}) were more abundant than helium at almost all pressure levels in the atmosphere for the \texttt{POSEIDON} retrieval and reflected a non-physical solution when compared with measurements of Solar System gas giants or hot Jupiters orbiting FGK dwarfs. The large abundance of carbon-bearing species suggested that the data may be modeled with a carbon-rich atmosphere, but the assumption of chemical equilibrium required high metallicity when limited to the parameter space of the pre-computed grids. When the arbitrary restriction on metallicity is relaxed, as with \texttt{TauREx} retrievals, the atmosphere appeared to qualitatively favor the metal-poor solution from our grid-based analysis. 

We note that the lack of aerosol opacity (either in the form of clouds or hazes) may also bias the results from forward models, \texttt{PLATON}, and \texttt{TauREx} retrievals, which all assumed a clear atmosphere for TOI-5205b. We explored the impact of clouds and hazes on retrievals assuming equilibrium chemistry using \texttt{POSEIDON} (see \autoref{tab:5205aerosolretrieval}). As seen in \S\ref{sec:freechemret}, the addition of clouds or a haze and cloud deck yielded a better solution with a larger Bayesian evidence that favored configurations with an additional component significantly (Bayes factors compared to the clear atmosphere model were $\ln B>10$). Qualitatively, the solution with only clouds (analogous to M3.2 in \autoref{tab:retrieval_models}) was similar to the clear atmosphere results in that the preferred solution favored a metal- and carbon-rich atmosphere. This produced similar non-physical abundances for \ce{CO2} and \ce{CH4} (each with VMRs between $\sim1-6\%$) and the posterior distribution for the C/O ratio was truncated by the upper limit of the \texttt{POSEIDON} grid (C/O$\to 2$). The solution with a cloud deck and haze (analogous to M3.3 in \autoref{tab:retrieval_models}) was the best fitting equilibrium chemistry solution from all Bayesian retrievals, and qualitatively matched the metal-poor and carbon-rich solutions derived from the forward models, \texttt{TauREx}, and free chemistry retrievals. Unfortunately, the metallicity posterior was truncated by the lower limit of the \texttt{POSEIDON} grid ([M/H]$\to-1$). 

\begin{deluxetable*}{lcc|c|c|c}
\tabletypesize{\scriptsize}
\tablecaption{Model parameters for \texttt{POSEIDON} retrievals for the \texttt{ExoTiC-JEDI} co-added spectrum, assuming equilibrium chemistry, and varying aerosol configurations (comparable to M3.1-M3.3 in \autoref{tab:retrieval_models}). Empty parameters are not used for that specific model. \label{tab:5205aerosolretrieval}}
\tablehead{
 &
 &
 &
\multicolumn{1}{c|}{Clear} &
\multicolumn{1}{c|}{Clouds} &
\multicolumn{1}{c}{Clouds + hazes}
\\
 &
 &
 &
\multicolumn{1}{c|}{$\chi^2_\nu=2.81$} &
\multicolumn{1}{c|}{$\chi^2_\nu=2.71$} &
\multicolumn{1}{c}{$\chi^2_\nu=2.58$} 
\\
 &
 &
 &
\multicolumn{1}{c|}{$\ln Z=2371.59 \pm 0.15$} &
\multicolumn{1}{c|}{$\ln Z=2389.69 \pm 0.15$} &
\multicolumn{1}{c}{$\ln Z=2402.19 \pm 0.15$} 
\\
\colhead{Parameter} &
\colhead{Units} &
\multicolumn{1}{c|}{Prior} &
\multicolumn{1}{c|}{Value} &
\multicolumn{1}{c|}{Value} &
\colhead{Value}
}
\startdata
~~Planetary radius ($R_p$) & $\mathrm{R_J}$ & $\mathcal{U}(0.80,1.08)$ & $0.932\pm0.001$ & $0.927\pm0.001$  & $0.931\pm0.001$\\
~~Equilibrium temperature ($T_{\mathrm{eq}}$) & K & $\mathcal{U}(300,1500)$ & $815^{+36}_{-16}$ & $730^{+46}_{-66}$ & $716^{+51}_{-52}$\\
~~Atmospheric metallicity ($\mathrm{[M/H]}$) & dex & $\mathcal{U}(-0.9,3.9)$ & $2.40^{+0.09}_{-0.08}$ & $1.62^{+0.19}_{-0.31}$ & $-0.65^{+0.30}_{-0.18}$ \\
~~Atmospheric carbon-to-oxygen ratio (C/O) & dex & $\mathcal{U}(0.3,1.9)$ & $0.99^{+0.02}_{-0.01}$ & $1.46^{+0.28}_{-0.26}$ & $0.76^{+0.17}_{-0.23}$\\
~~Spot coverage fraction ($f_\mathrm{spot}$) & \nodata &  $\mathcal{U}(0,1)$ & $0.35\pm0.02$ & $0.35\pm0.02$ & $0.29^{+0.03}_{-0.02}$\\
~~Spot temperature ($T_\mathrm{spot}$) & K & $\mathcal{U}(2300,3430)$ & $3420^{+6}_{-7}$ & $3421^{+6}_{-7}$ & $3420\pm6$\\
~~Spot surface gravity ($\log g_\mathrm{spot}$) & dex & $\mathcal{U}(4.34,5.34)$ & $4.37^{+0.06}_{-0.02}$ & $4.36\pm0.02$ & $4.42^{+0.04}_{-0.06}$\\
~~Faculae coverage fraction ($f_\mathrm{fac}$) & \nodata & $\mathcal{U}(0,0.5)$ & $0.07\pm0.01$ & $0.07\pm0.01$ & $0.04\pm0.01$\\
~~Faculae temperature ($T_\mathrm{fac}$) & K & $\mathcal{U}(3430,4802)$ & $4095^{+42}_{-25}$ & $4100^{+23}_{-20}$ & $4274^{+182}_{-64}$\\
~~Faculae surface gravity ($\log g_\mathrm{fac}$) & dex & $\mathcal{U}(4.34,5.34)$  & $4.37\pm0.02$ & $4.37\pm0.02$ & $4.52^{+0.28}_{-0.11}$\\
~~Photosphere temperature ($T_\mathrm{eff}$) & K & $\mathcal{N}(3430,54)$  & $3664^{+15}_{-19}$ & $3657^{+18}_{-19}$ & $3654^{+21}_{-24}$\\
~~Photosphere surface gravity ($\log g_\mathrm{fac}$) & dex & $\mathcal{U}(4.34,5.34)$ & $4.56^{+0.08}_{-0.04}$ & $4.54\pm0.03$ & $4.65^{+0.06}_{-0.05}$\\
~~Cloud top pressure in bars ($\log P_\mathrm{cloud}$) & dex & $\mathcal{U}(-6,2)$ & \nodata & $-3.82^{+0.25}_{-0.21}$ & $-2.06^{+0.24}_{-0.27}$ \\
~~Rayleigh enhancement factor ($\log~a$) & dex & $\mathcal{U}(-4,8)$ & \nodata & \nodata & $6.99^{+0.61}_{-0.79}$ \\
~~Scattering slope ($\gamma$) & \nodata & $\mathcal{U}(-20,2)$ & \nodata & \nodata & $-13.26^{+1.62}_{-1.45}$ \\
\enddata
\end{deluxetable*}

A comparison of equilibrium chemical codes demonstrated the potential for biases in which these retrieval codes converged with confident (tight uncertainties, such as those for C/O) but incorrect solutions \citep[e.g.,][]{Refaie2022}. From our analysis, a significant limitation was the bounds of the atmospheric parameters as this could (i) result in solutions that were qualitatively similar to the results derived assuming free-chemistry (e.g., \texttt{TauREx}) or non-physical abundances for trace molecules such as with retrievals using \texttt{PLATON} and clear or cloudy retrievals with \texttt{POSEIDON} or (ii) impose arbitrary limits on the parameter space as seen with \texttt{POSEIDON} retrievals including clouds and hazes. \textbf{Given the discrepancies between forward modeling and Bayesian retrievals assuming equilibrium chemistry and the parameter boundaries and non-physical nature for a subset of the Bayesian retrievals, we did not enforce the requirement of equilibrium chemistry. We instead make inferences on the atmospheric composition of TOI-5205b using results from  Bayesian retrievals assuming free chemistry that are described in \S\ref{sec:freechemret}.}

\section{Free chemistry retrieval supporting materials}\label{app:retrievalsupplement}

\subsection{Hazy Retrievals}\label{app:hazyret}

\begin{figure*}
    \centering
    \includegraphics[width=0.99\linewidth]{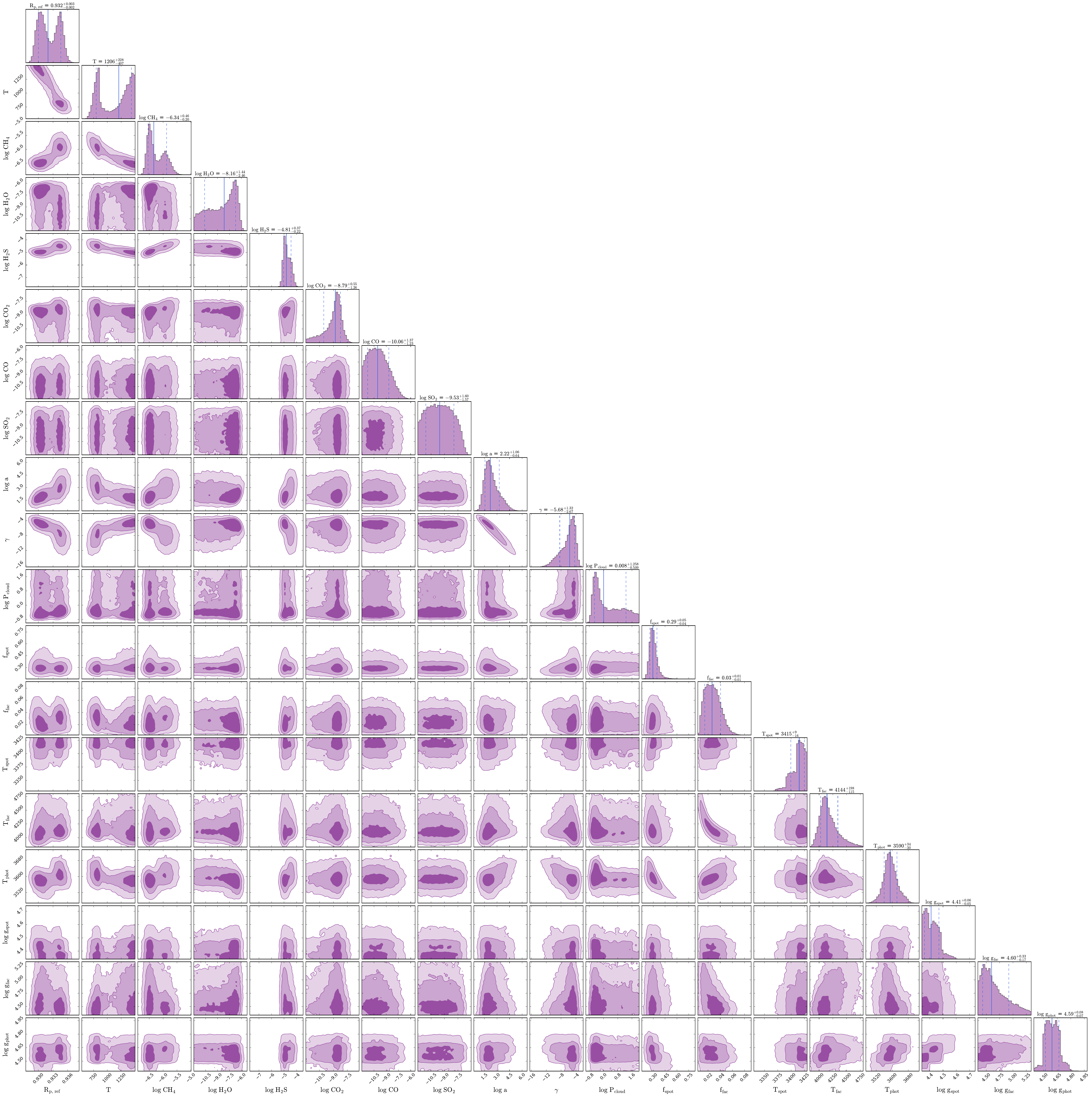}
    \caption{Retrieval results for fits to the coadded spectrum with TLS contamination and an atmosphere with clouds and hazes (M3.3; see \autoref{tab:retrieval_models}).}
    \label{fig:retrieval_corner_hazy}
\end{figure*}

\autoref{fig:retrieval_corner_hazy} shows the results from \poseidon retrievals, assuming free chemistry, on the coadded spectrum using the model with stellar contamination and a planetary atmosphere with a gray cloud deck and a power law haze (M3.3). The addition of the power law haze parameters ($\log a$ and $\gamma$) created a bimodal solution most notably in the atmospheric temperature and planetary radius. Within the higher temperature mode, increases in temperature correlated with larger values of the haze scattering slope, $\gamma$, and smaller starspot covering fractions. This indicated that haze scattering in the planetary atmosphere is trading off with the stellar contamination caused by cool starspots.    

\begin{figure*}
    \centering
    \includegraphics[width=0.99\linewidth]{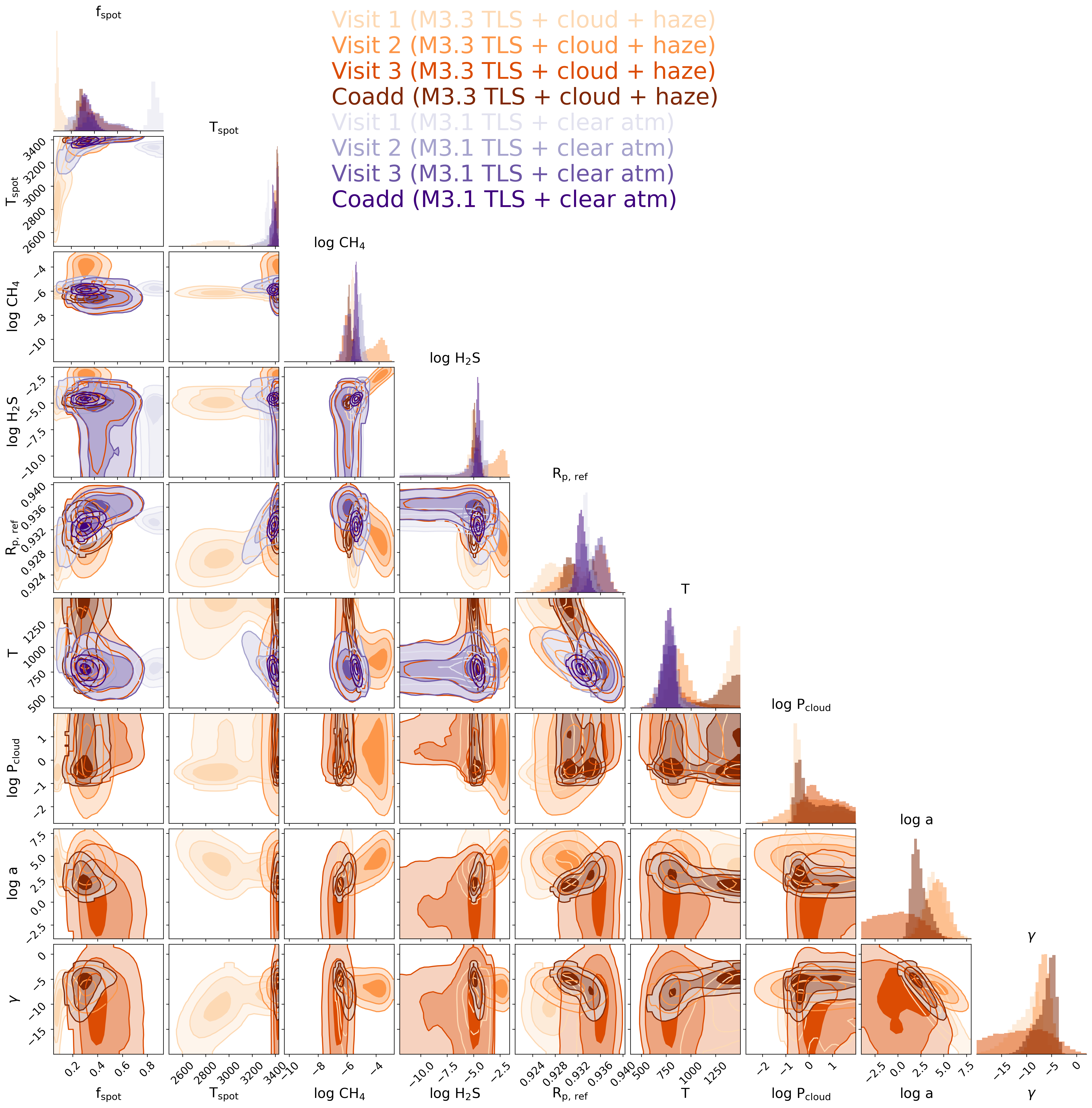}
    \caption{Similar to \hyperref[fig:retrieval_corner_visits]{Fig. Set 4.1} except comparing M3.3 with clouds and hazes to M3.1 with no clouds. Significant correlations are seen across the parameter space in the visit-to-visit variability, particularly for the hazy retrievals.}
    \label{fig:retrieval_corner_hazy_visits}
\end{figure*}

\autoref{fig:retrieval_corner_hazy_visits} is similar to \hyperref[fig:retrieval_corner_visits]{Fig. Set 4.1} and compares the M3.3 retrieval model that has TLS, clouds and hazes, to the M3.1 model that has no atmospheric aerosols. Retrievals with hazes tend to show lower starspot temperatures and fractions, higher atmospheric temperatures, higher retrieved gas abundances, and smaller planet reference radii compared to the M3.1 aerosol-free retrievals. The retrieved starspot temperature also exhibits an anti-correlation with the haze slope amplitude ($\mathrm{\log a}$) for M3.3.  These hazy retrievals also show a large degree of variance in the abundances of retrieved gases (\ce{CH4} and \ce{H2S}) between the separate visits, while the clear atmosphere retrievals are more consistent between visits. Taken together, these correlations and visit-to-visit variances demonstrate that the retrieval is erroneously using the atmospheric haze model to assist in fitting for the stellar contamination signal imposed on the transmission spectrum from unocculted starspots.    

\subsection{Retrieval fit results and residuals}

\autoref{tab:retrievals_clouds} and \autoref{tab:retrievals_clear} provide quantitative results for the cloudy and clear retrievals on the individual visit spectra and the co-added spectrum. The abundances ($\log X$) are the volume mixing ratios for each species.

\begin{deluxetable*}{r|l|l||l|l|l|l}
\tablewidth{0.98\textwidth}
\tabletypesize{\scriptsize}
\tablecaption{Atmospheric retrieval priors and posteriors for the M3.2 model with clouds, analogous to \autoref{tab:retrievals_clear}. \label{tab:retrievals_clouds} }
\tablehead{
\colhead{Parameters} & \colhead{Units} & \colhead{Priors} & \colhead{Visit 1} & \colhead{Visit 2} & \colhead{Visit 3} & \colhead{Visits Co-added}}
\startdata
$\mathrm{R}_{\mathrm{p, \, ref}}$ & $\mathrm{R_J}$ & $\mathcal{U}(0.88, 1.19)$ & $0.9315^{+0.0011}_{-0.0011}$ & $0.9304^{+0.0021}_{-0.0030}$ & $0.9355^{+0.0013}_{-0.0018}$ & $0.93235^{+0.00086}_{-0.00081}$ \\
$\mathrm{T}$ & K & $\mathcal{U}(200.00, 1500.00)$ & $789^{+109}_{-89}$ & $832^{+92}_{-73}$ & $813^{+179}_{-104}$ & $804^{+47}_{-55}$ \\
$\log \, \mathrm{CH_4}$ & dex & $\mathcal{U}(-12.00, -1.00)$ & $-5.81^{+0.30}_{-0.31}$ & $-4.84^{+1.10}_{-0.54}$ & $-6.59^{+0.31}_{-0.31}$ & $-5.87^{+0.17}_{-0.17}$ \\
$\log \, \mathrm{H_2 O}$ & dex & $\mathcal{U}(-12.00, -1.00)$ & $-9.6^{+1.7}_{-1.5}$ & $-9.3^{+1.8}_{-1.7}$ & $-9.7^{+1.6}_{-1.5}$ & $-10.0^{+1.4}_{-1.3}$ \\
$\log \, \mathrm{H_2 S}$ & dex & $\mathcal{U}(-12.00, -1.00)$ & $-4.83^{+0.38}_{-0.72}$ & $-3.57^{+1.14}_{-0.60}$ & $-5.1^{+0.5}_{-2.7}$ & $-4.54^{+0.18}_{-0.19}$ \\
$\log \, \mathrm{CO_2}$ & dex & $\mathcal{U}(-12.00, -1.00)$ & $-9.6^{+1.4}_{-1.5}$ & $-6.4^{+1.2}_{-1.0}$ & $-10.3^{+1.2}_{-1.1}$ & $-8.41^{+0.63}_{-1.27}$ \\
$\log \, \mathrm{CO}$ & dex & $\mathcal{U}(-12.00, -1.00)$ & $-9.4^{+1.8}_{-1.7}$ & $-9.2^{+2.0}_{-1.8}$ & $-9.5^{+1.8}_{-1.6}$ & $-9.9^{+1.5}_{-1.4}$ \\
$\log \, \mathrm{SO_2}$ & dex & $\mathcal{U}(-12.00, -1.00)$ & $-9.4^{+1.8}_{-1.7}$ & $-8.7^{+2.2}_{-2.1}$ & $-9.4^{+1.7}_{-1.7}$ & $-9.3^{+1.8}_{-1.8}$ \\
$\log \, \mathrm{P}_{\mathrm{cloud}}$ & dex (bar) & $\mathcal{U}(-6.00, 2.00)$ & $0.12^{+1.06}_{-0.35}$ & $-1.02^{+0.40}_{-1.04}$ & $0.38^{+0.99}_{-0.55}$ & $-0.17^{+0.17}_{-0.16}$ \\
$\mathrm{f}_{\mathrm{spot}}$ & \nodata & $\mathcal{U}(0.00, 1.00)$ & $0.487^{+0.066}_{-0.088}$ & $0.361^{+0.069}_{-0.062}$ & $0.43^{+0.15}_{-0.09}$ & $0.337^{+0.044}_{-0.042}$ \\
$\mathrm{f}_{\mathrm{fac}}$ & \nodata & $\mathcal{U}(0.00, 0.50)$ & $0.035^{+0.012}_{-0.010}$ & $0.051^{+0.021}_{-0.017}$ & $0.0125^{+0.0062}_{-0.0041}$ & $0.057^{+0.027}_{-0.016}$ \\
$\mathrm{T}_{\mathrm{spot}}$ & K & $\mathcal{U}(2300.00, 3430.00)$ & $3386^{+19}_{-16}$ & $3406^{+16}_{-26}$ & $3412^{+11}_{-19}$ & $3401^{+18}_{-21}$ \\
$\mathrm{T}_{\mathrm{fac}}$ & K & $\mathcal{U}(3430.00, 4802.00)$ & $4088^{+94}_{-87}$ & $4010^{+107}_{-82}$ & $4428^{+190}_{-191}$ & $3983^{+84}_{-76}$ \\
$\mathrm{T}_{\mathrm{phot}}$ & K & $\mathcal{N}(3430.00, 54.00)$ & $3527^{+27}_{-28}$ & $3599^{+26}_{-27}$ & $3532^{+32}_{-32}$ & $3611^{+24}_{-24}$ \\
$\log \, \mathrm{g}_{\mathrm{spot}}$ & dex & $\mathcal{U}(4.34, 5.34)$ & $4.374^{+0.054}_{-0.022}$ & $4.68^{+0.11}_{-0.12}$ & $4.429^{+0.078}_{-0.063}$ & $4.448^{+0.089}_{-0.068}$ \\
$\log \, \mathrm{g}_{\mathrm{fac}}$ & dex & $\mathcal{U}(4.34, 5.34)$ & $4.438^{+0.118}_{-0.071}$ & $4.417^{+0.092}_{-0.055}$ & $4.72^{+0.33}_{-0.25}$ & $4.374^{+0.054}_{-0.022}$ \\
$\log \, \mathrm{g}_{\mathrm{phot}}$ & dex & $\mathcal{U}(4.34, 5.34)$ & $4.574^{+0.085}_{-0.054}$ & $4.80^{+0.11}_{-0.10}$ & $4.72^{+0.13}_{-0.11}$ & $4.654^{+0.086}_{-0.056}$ \\
$\log~\mathrm{C/O}$ & dex & \nodata & $1.8_{-0.8}^{+1.0},~2\sigma>0.31$ & $1.1_{-0.5}^{+0.6},~2\sigma>0.27$ & $1.2_{-0.9}^{+1.0},~2\sigma>-0.32$ & $1.7_{-0.6}^{+0.7},~2\sigma>0.49$ \\
$\log~[\mathrm{M/H}]$ & dex & \nodata & $-2.1_{-0.6}^{+0.4}$ & $-0.8_{-0.6}^{+1.1}$ & $-2.3_{-1.2}^{+0.5}$ & $-1.8\pm0.2$ \\
$\log~\mathrm{C/H}$ & dex & \nodata & $-2.5\pm0.3$ & $-1.5_{-0.5}^{+1.1}$ & $-3.2\pm0.3$ & $-2.6\pm0.2$\\
$\log~\mathrm{O/H}$ & dex & \nodata & $-4.5_{-1.0}^{+0.9},~2\sigma<-2.95$ & $-2.8_{-0.9}^{+1.1},~2\sigma<-0.74$ & $-4.8_{-1.0}^{+0.9},~2\sigma<-3.04$ & $-4.5\pm0.7,~2\sigma<-3.28$\\
$\log~\mathrm{S/H}$ & dex & \nodata & $-0.2_{-0.7}^{+0.4}$ & $1.1_{-0.6}^{+1.1}$ & $-0.4_{-2.2}^{+0.5}$ & $0.1\pm0.2$\\
\hline
$\chi^2_{\nu}$ & \nodata & \nodata & 1.71 & 1.44 & 1.37 & 2.38 \\
$\ln Z$ & \nodata & \nodata & 2378.74 & 2412.23 & 2326.50 & 2422.58 \\
\enddata 
\end{deluxetable*}

\hyperref[app:fig:retrieval_visits]{Fig. Set 22} shows the cloud-free M3.1 retrieval model fit to the co-added TOI-5205b transmission spectrum with 1 and 2$\sigma$ credibility envelopes (upper panel) and residuals between the median fit and the data (lower panel). There are an excessive number of poorly fit data points between $1-3$ \textmu{}m that are present in all three visits. This spectral region drives our $\chi^2_{\nu} > 1$ and may be due to underestimated uncertainties or insufficient stellar spectral models that cannot completely reproduce the contaminated transmission spectrum of TOI-5205b. Our model fits also systematically overestimate the transit depths at the reddest wavelengths $\lambda > 4.8$ \textmu{}m. This could be due to spot crossing events or TLS contamination, which still have an impact beyond 4 \textmu{}m, or potentially due to nightside emission from the planet during transit, which can decrease transit depths relative to models without contamination.   

\figsetstart
\figsetnum{22}
\figsettitle{Retrieved Spectra for Individual Visits}

\figsetgrpstart
\figsetgrpnum{22.1}
\figsetgrptitle{Overview}
\figsetplot{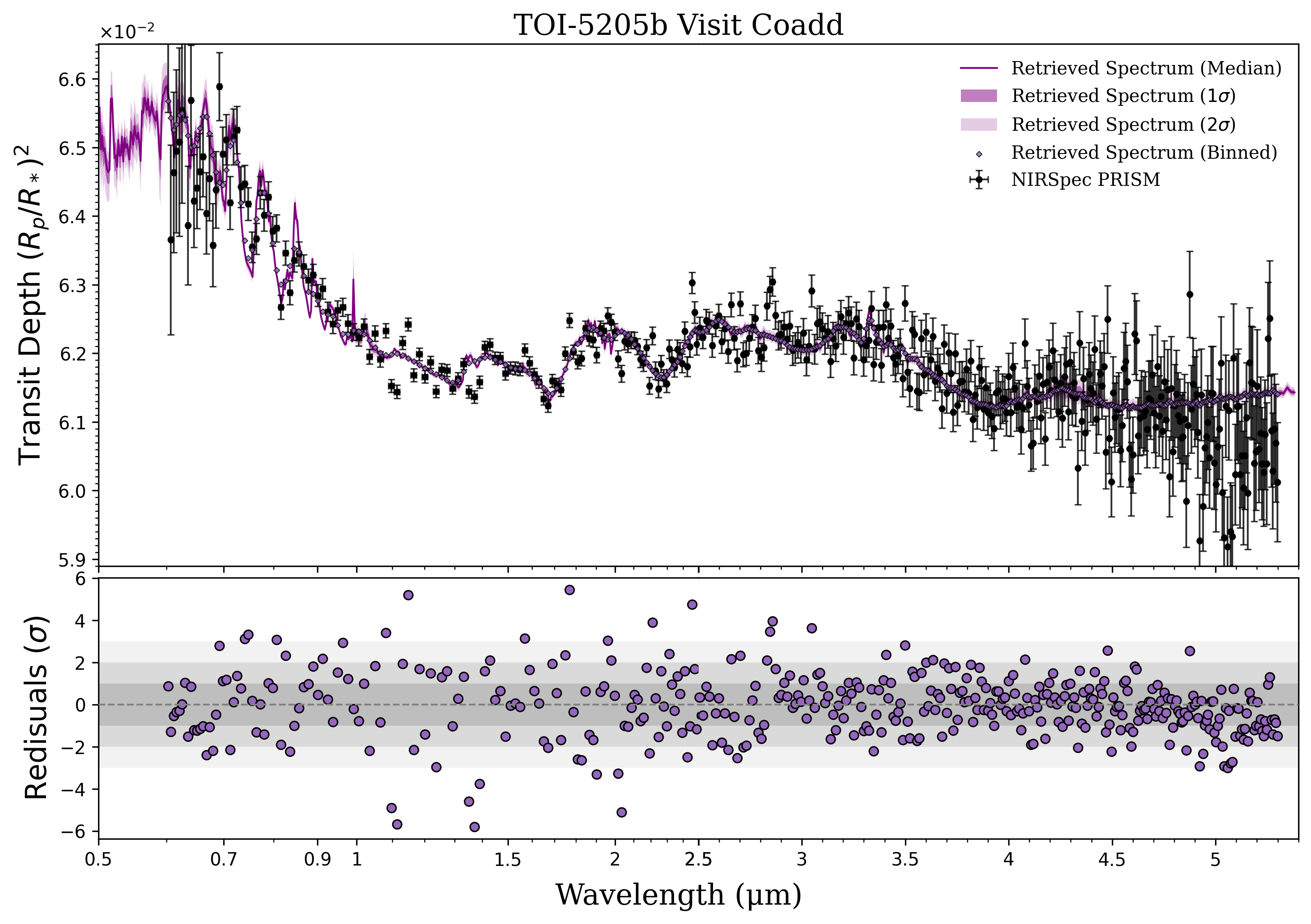}
\figsetgrpnote{The three-visit co-added transmission spectrum of TOI-5205b fitted with the cloud free M3.1 retrieval model (top panel) and the residuals on the fit (bottom panel). A logarithmic x-axis is used from 0.5-2.5 $\mu$m and a linear x-axis $>2.5$ $\mu$m. The cloudy retrieval fits using model M3.2 are indistinguishable from those shown here. The complete figure set (4 images, for three visits plus the combined spectrum) is available in the online journal.}\label{app:fig:retrieval_visits}
\figsetgrpend

\figsetgrpstart
\figsetgrpnum{22.2}
\figsetgrptitle{Visit 1}
\figsetplot{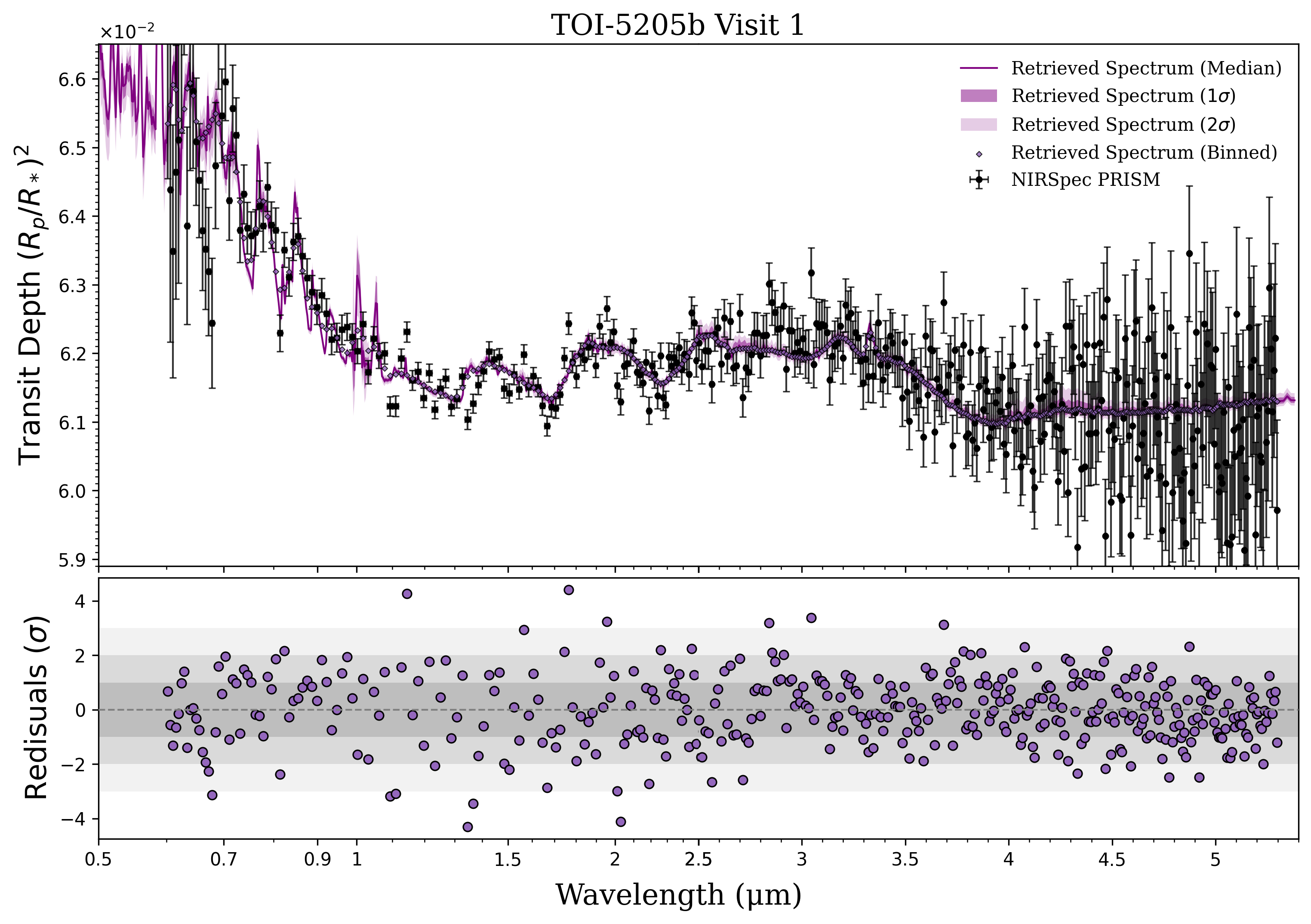}
\figsetgrpnote{Fitted spectrum and residuals to the Visit 1 spectrum using model M3.1 (as shown in \hyperref[app:fig:retrieval_visits]{Fig. Set 22.1}.}
\figsetgrpend

\figsetgrpstart
\figsetgrpnum{22.3}
\figsetgrptitle{Visit 2}
\figsetplot{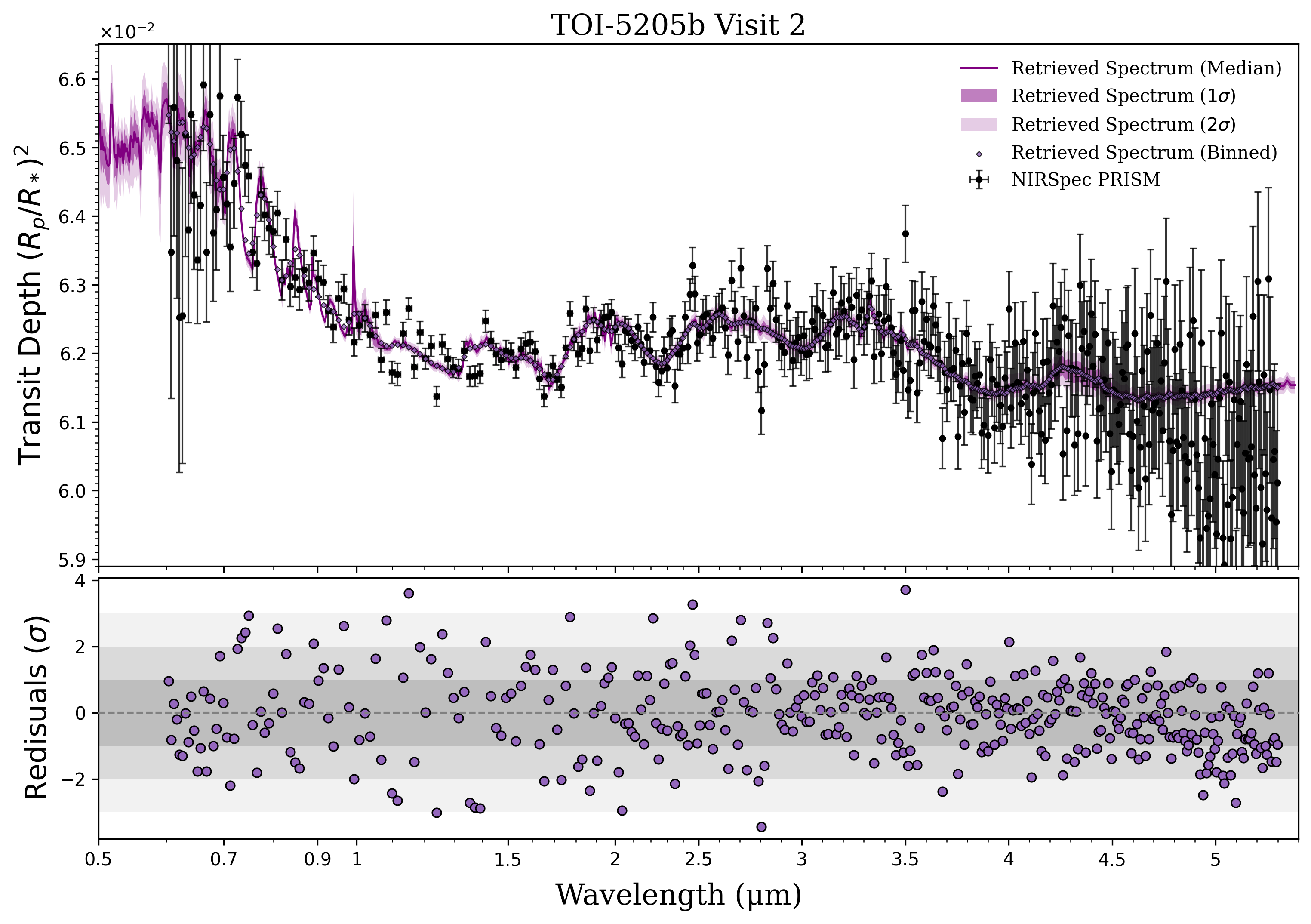}
\figsetgrpnote{Fitted spectrum and residuals to the Visit 2 spectrum using model M3.1 (as shown in \hyperref[app:fig:retrieval_visits]{Fig. Set 22.1}.}
\figsetgrpend

\figsetgrpstart
\figsetgrpnum{22.4}
\figsetgrptitle{Visit 3}
\figsetplot{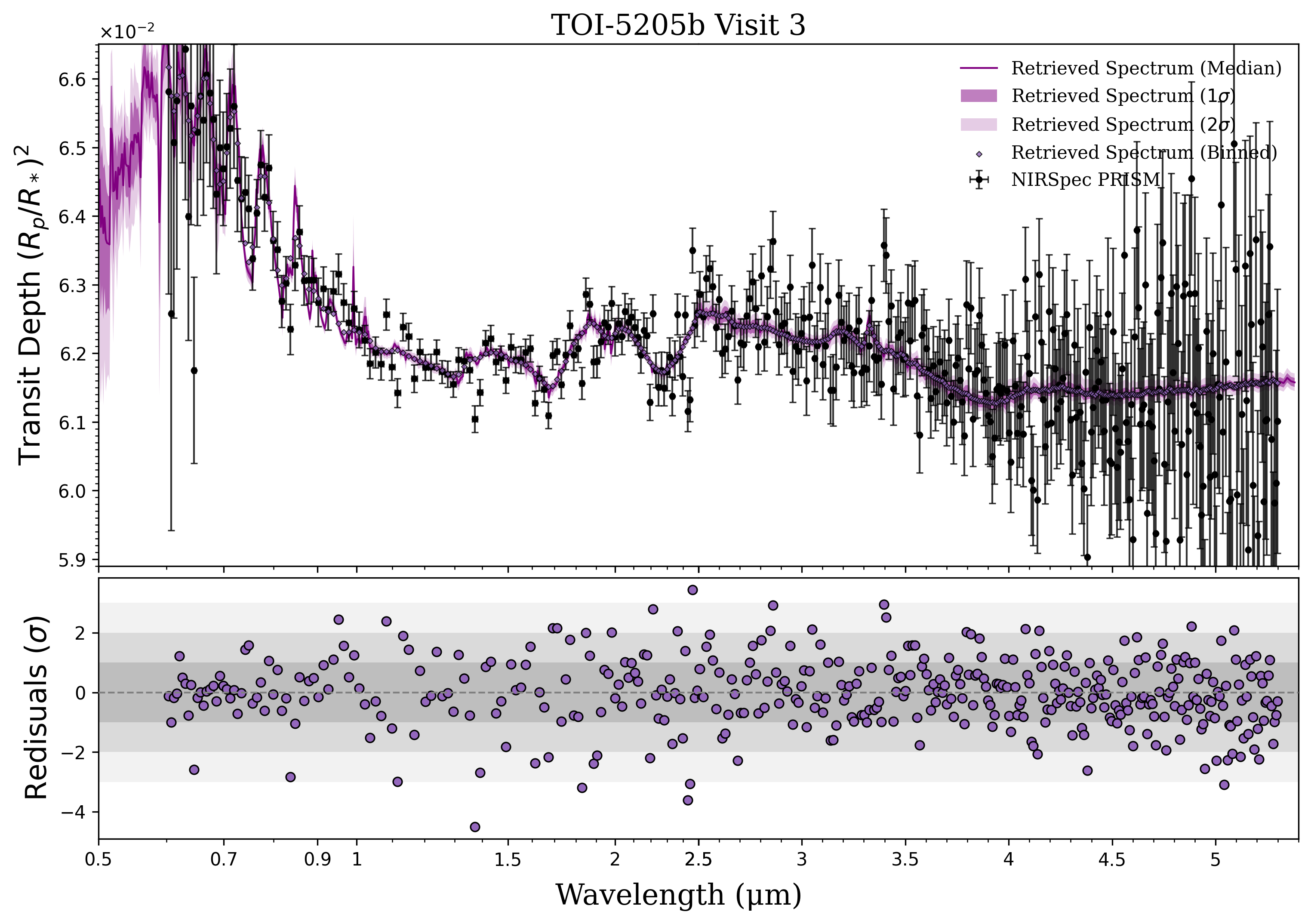}
\figsetgrpnote{Fitted spectrum and residuals to the Visit 3 spectrum using model M3.1 (as shown in \hyperref[app:fig:retrieval_visits]{Fig. Set 22.1}.}
\figsetgrpend

\figsetend

\subsection{Retrievals on Partial Spectra} \label{sec:retrievals:partial}

To understand how different portions of the transmission spectrum drive the retrieval towards constraints on the planet atmosphere and the contaminating star, we made several wavelength cuts to the co-added transmission spectrum and repeated our previous retrieval simulations using the cloud-free M3.1 model from \S\ref{sec:retrievals}. \autoref{fig:retrieval_corner_wavelength} presents the spectrum with the various models (upper right inset) and 1-D and 2-D marginalized posterior distributions for five different retrievals covering wavelengths of $\lambda>$ 3.5, 2.5, 1.5, and 0.5 \textmu{}m. The choice to focus on the spectra at wavelengths longer than these cuts is based on the stronger impact of stellar contamination at bluer wavelengths \citep[see \autoref{fig:retrieval_spectrum_breakdown}; e.g.,][]{Seager2024}. 

Markedly different results are obtained using data from different wavelength ranges, including many statistically incompatible constraints for the atmospheric temperature, gas abundances, and reference radius. Results for $\lambda > 3.5$ \textmu{}m show that a high temperature atmosphere with abundant \ce{H2S} (a volume mixing ratio of $-2.03^{+0.43}_{-0.60}$) and no \ce{CH4} constraint fits the downward sloped spectrum. For $\lambda > 2.5$ \textmu{}m a dichotomous result is obtained with a low temperature atmosphere ($561.0^{+96.1}_{-85.6}$ K) with abundant \ce{CH4} and no constraint on \ce{H2S}. These two longwave only datasets prove to be poor predictors of the shortwave data as they immediately diverge beyond the cutoff. Results for $\lambda > 1.5$ \textmu{}m are similar to the $\lambda > 3.5$ \textmu{}m case with high $T$, but have moderate abundances of \ce{CH4}, \ce{H2S}, and \ce{H2O}, and it offers a sensible prediction of the shorter wavelength data, albeit offset from the true slope in below 1 \textmu{}m. 
The $\lambda > 0.5$ \textmu{}m full wavelength constraints show modest temperature atmospheres that are within $1\sigma$ of $T_{\rm eq}$. 
We note that all cases show non-zero starspot covering fractions, indicating the need for TLS modeling. However, the shortwave data offer significantly refined constraints on the stellar parameters. 

\setcounter{figure}{22}
\begin{figure*}
    \centering
    \includegraphics[width=0.99\linewidth]{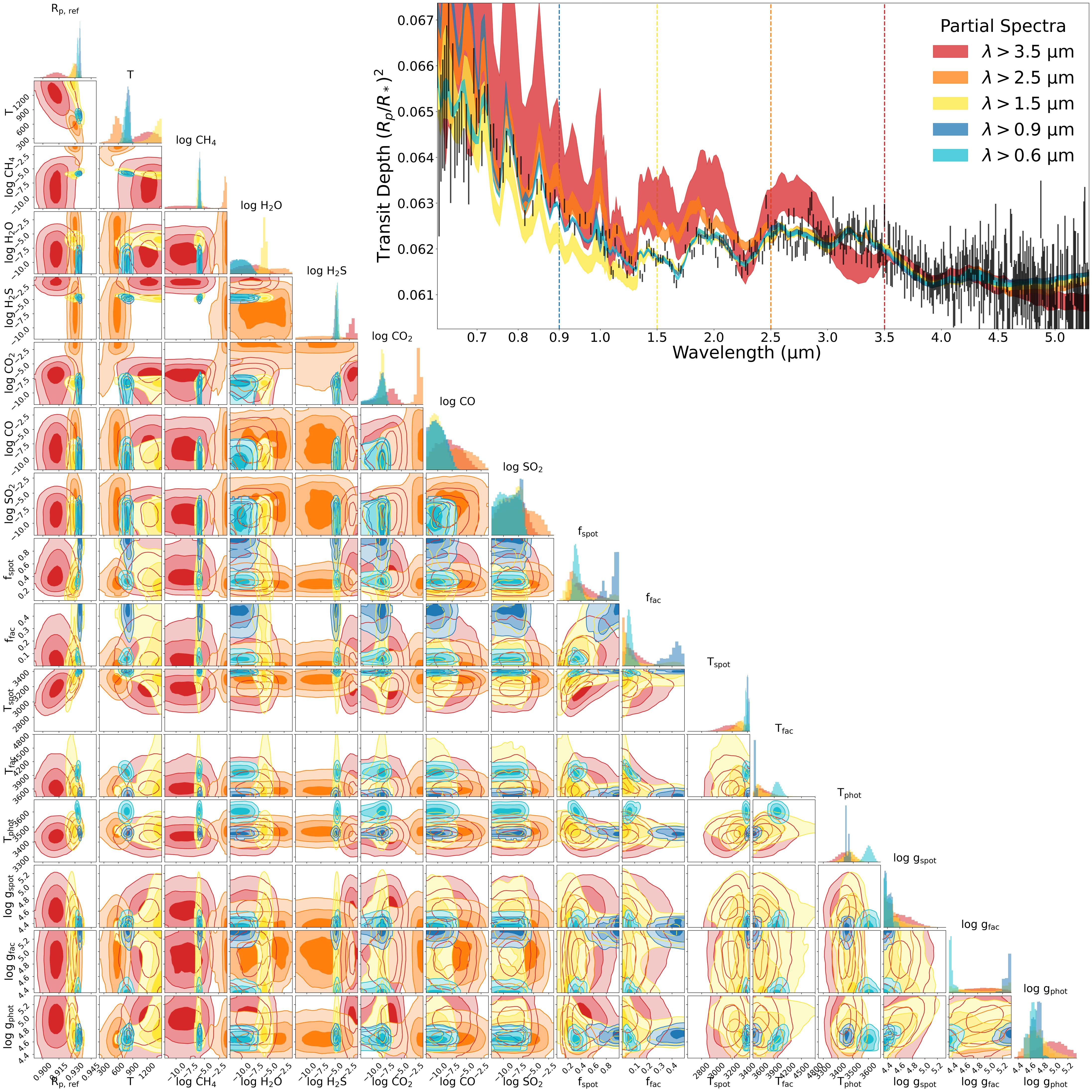}
    \caption{Posterior probability distributions for stellar and planetary parameter (lower left corner of subplots) and 1$\sigma$ range of fits to the spectrum (upper right panel) from retrievals covering only selected wavelength ranges of the transmission spectra. Colors indicate the short wavelength limit for the different cases. Transmission spectrum envelopes extend shortward of the dashed line corresponding to the respective case wavelength limit to visualize the posterior predictive accuracy of the retrieval constraints on the omitted portion of the data. These predictions illustrate the many possible (mis)interpretations that could emerge from access to limited wavelength data.}
    \label{fig:retrieval_corner_wavelength}
\end{figure*}

\subsection{Limited sensitivity to $\mathit{CS_2}$ and $\mathit{NH_3}$ in the atmosphere} \label{app:exploreotherspecies}
A carbon-rich atmosphere and a sulfur reservoir present an opportunity to explore more complex sulfur chemistry. Recent tentative detections of carbon disulfide (\ce{CS2}) in the atmosphere of TOI-270d ($350~\mathrm{K}\le T_{\mathrm{eq}}\le380~\mathrm{K}$) using JWST \citep{Benneke2024,Holmberg2024} motivated an investigation in other sulfur reservoirs beyond \ce{H2S} and \ce{SO2}. In atmospheres that are oxygen-poor, \cite{Mukherjee2024} demonstrate that carbon sulfides may be significant reservoirs of the S-inventory in the upper atmosphere. The importance of carbon sulfides is expected in planetary atmospheres cooler than TOI-5205b ($T_{eq}\le600$ K), however, because we do not detect \ce{SO2}, we investigated the expected presence of \ce{CS2} using the best-fitting \texttt{VULCAN} grid (see \S\ref{sec:vulcan}). For TOI-5205b, the sulfur-enhanced atmospheres ($\mathrm{S/H}$=100) provide the maximum VMRs of $\log\mathrm{[CS_2]}=-8.56$ ($\log K_{zz}=9$) or $\log\mathrm{[CS_2]}=-8.22$ ($\log K_{zz}=6$). \ce{H2S} remains the most abundance sulfur-bearing molecule for TOI-5205b at a VMR consistent with the free chemistry retrieval. Even in the most sulfur-enhanced model allowing for disequilibrium chemistry, \ce{CS2} is not expected to be a significant component in the atmosphere of TOI-5205b. 

We were also motivated to investigate the presence of \ce{NH3} by recent JWST observations of WASP-107b, a warm Neptune orbiting a Sun-like star with an equilibrium temperature ($T_{eq}\sim750$ K) comparable to TOI-5205b, which has a detection of ammonia in its atmosphere \citep{Welbanks2024,Sing2024}. \ce{NH3}, like \ce{CH4}, is expected to become readily observable in the atmospheres of warm gas giants and a significant reservoir for nitrogen \citep{Fortney2020,Ohno2023nitro1,Ohno2023nitro2}. From the \poseidon free chemistry retrieval for the modified M3.1 model that included ammonia (see \S\ref{sec:retrievals:combined}), we did not recover any significant detection ($\ln B=0.4$) of $\log \mathrm{[NH_3]}=-7.06^{+0.44}_{-1.98}$. For reference, the best-fitting \texttt{VULCAN} models predict a value of $\log \mathrm{[NH_3]}=-5.86$. At the equilibrium temperature of TOI-5205b, the dominant spectral feature of \ce{NH3} is expected between $2.9-3.1$ \textmu{}m. \cite{Welbanks2024} demonstrate the difficulty in detecting \ce{NH3} with multiple instruments (HST/WFC3, JWST/NIRCam, and JWST/MIRI) and used a cross-validation technique to identify this wavelength range as the most important for detecting \ce{NH3} in WASP-107b. In TOI-5205b, this is a region where stellar contamination, \ce{H2S}, and \ce{CH4} all contribute to the spectrum (see \autoref{fig:retrieval_spectrum_breakdown}), such that the models do not differ significantly with the inclusion of \ce{NH3}. We conclude that while TOI-5205b is very similar in temperature to WASP-107b, the modeling of stellar contamination and the transmission spectrum does not provide evidence for \ce{NH3} in its atmosphere.

\section{Retrieval results with different reductions}\label{app:eurposeidon}
To ensure consistency across data reductions and binning schemes, we also performed Bayesian retrievals for the pixel-level \texttt{Eureka!} spectra and 120 channel \texttt{ExoTiC-JEDI} spectra (40 nm bins). These retrievals were limited to
\begin{enumerate*}[label=(\roman*)]
\item the co-added \texttt{Eureka!} spectrum to investigate the preference for model series M3.X in \autoref{tab:retrieval_models},
\item the visit and co-added \texttt{Eureka!} spectra to investigate the differences for the preferred model series, model M3.1, and
\item the binned visit and co-added \texttt{ExoTiC-JEDI} spectra to investigate the differences for the preferred model series, model M3.1.
\end{enumerate*}
For all reductions presented below, we use the same configuration to generate the transmission spectrum as described in \hyperref[app:specfit]{Appendix \ref*{app:specfit}} and fit the derived spectrum as detailed in \S\ref{sec:freechemret} (e.g., same priors, live points, and model components), such that the only difference is the data reduction. For brevity, we do not show the plots in this appendix but only report the posterior values.

\subsection{M3.X on the co-added \texttt{Eureka!} spectrum}
With the co-added \texttt{Eureka!} spectrum, the configurations with clouds (M3.2) or hazes (M3.3) were significantly favored because the Bayes factors were $\ln B<10$ relative to the clear atmosphere (with $\ln Z_{\mathrm{M3.1}}=2377.29$, $\ln Z_{\mathrm{M3.2}}=2386.35$, and $\ln Z_{\mathrm{M3.3}}=2387.21$). Qualitatively, the retrievals on the co-added \texttt{Eureka!} spectrum agree with a metal-poor and carbon-rich atmosphere where the dominant species were \ce{CH4} and \ce{H2S} (see \S\ref{sec:freechemret}). The biggest difference was that TOI-5205b was much hotter in M3.1 than the equilibrium temperature ($T=1078\pm82$ K compared to $T_{eq}=742$ K) and the values retrieved from the \texttt{ExoTiC-JEDI} reductions. For models M3.2 and M3.3, the temperature approached the upper bound ($T\to1500$ K). 

Furthermore, there was a nominal detection of \ce{CO2} compared to \texttt{ExoTiC-JEDI}: $\log \mathrm{CO_{2,~M3.1}} = -7.83\pm0.31$, $\log \mathrm{CO_{2,~M3.2}} = -8.17^{+0.21}_{-0.23}$, and $\log \mathrm{CO_{2,~M3.3}} = -8.19^{+0.16}_{-0.17}$, respectively. We tested the significance of the \ce{CO2} detection with \texttt{Eureka!} by performing the same retrievals without \ce{CO2} for each model. Models without \ce{CO2} resulted in a lower Bayesian evidence ($\ln B_{\mathrm{M3.1}}=6.5$, $\ln B_{\mathrm{M3.2}}=6.75$, and $\ln B_{\mathrm{M3.3}}=12.75$ for models M3.X relative to the models without \ce{CO2}). 

The overall quality of the fits to the \texttt{Eureka!} reduction ($\chi^2_\nu$ and $\ln Z$) was lower ($\chi^2_{\nu,~\mathrm{M3.1}}=2.76$, $\chi^2_{\nu,~\mathrm{M3.2}}=2.70$, and $\chi^2_{\nu,~\mathrm{M3.1}}=2.65$) compared to \texttt{ExoTiC-JEDI} ($\chi^2_\nu\le2.4$ and $\ln~Z>2420$, see \autoref{tab:retrieval_models}), suggesting that none of these models provided a comparably good fit for the \texttt{Eureka!} reduction. When coupled with the persistent high retrieved temperature, we caution that, although the spectra agree at the $1-2\sigma$ level (see \hyperref[fig:spectracomp]{Fig. Set 2}), there may be some additional minor systematics in the $\texttt{Eureka!}$ dataset which result in a hotter temperature and may contribute to the \ce{CO2} detection. Despite these differences, the M3.X model fits with the \texttt{Eureka!} reduction resulted in comparable detections of \ce{CH4} ($\ln B=77$ between M3.1 and M3.1 without \ce{CH4}) and \ce{H2S} ($\ln B=16$ between M3.1 and M3.1 without \ce{H2S}). The consistent retrieval of these species corroborated the inferences from the \texttt{ExoTiC-JEDI}, namely that TOI-5205b appears to have a highly sub-solar atmospheric metallicity and super-solar carbon-to-oxygen ratio regardless of the data reduction algorithm.

\subsection{M3.1 for \texttt{Eureka!} visit spectra and co-added spectrum}
The results for retrievals, adopting model M3.1, to individual visits and the co-added spectrum derived with \texttt{Eureka!} are detailed in \autoref{tab:eureka_retrievals}. This table is a direct comparison to \autoref{tab:retrievals_clear}. The hot temperature and varied \ce{CO2} abundance are evident. The VMRs for \ce{CH4} and \ce{H2S} derived with the co-added spectrum are comparable (within $1\sigma$ uncertainties) to those derived with the \texttt{ExoTiC-JEDI} reduction (see \S\ref{sec:freechemret}). 

\begin{deluxetable*}{r|l|l||l|l|l||l}
\tablewidth{0.98\textwidth}
\tabletypesize{\scriptsize}
\tablecaption{Atmospheric retrieval priors and posteriors for the M3.1 model fit to the pixel-level \texttt{Eureka!} reductions, which assumed a clear atmosphere and a TLS effect component. This is analogous to \autoref{tab:retrievals_clear} for \texttt{ExoTiC-JEDI}. \label{tab:eureka_retrievals}}
\tablehead{
\colhead{Parameters} & 
\colhead{Units} &
\multicolumn{1}{c||}{Priors} & 
\colhead{Visit 1} & 
\colhead{Visit 2} & 
\multicolumn{1}{c||}{Visit 3} & 
\colhead{Visits co-added}}
\startdata
$\mathrm{R}_{\mathrm{p, \, ref}}$ & $\mathrm{R_J}$ & $\mathcal{U}(0.80, 1.08)$ & $0.924\pm0.001$ & $0.932\pm0.001$ & $0.932\pm0.001$ & $0.930\pm0.001$  \\
$\mathrm{T}$ & K & $\mathcal{U}(200.00, 1500.00)$ & $1391^{+69}_{-83}$ & $920^{+92}_{-79}$ & $1187^{+124}_{-120}$ & $1078\pm82$ \\
$\log \, \mathrm{CH_4}$ & dex & $\mathcal{U}(-12.00, -1.00)$ & $-5.99^{+0.14}_{-0.13}$ & $-5.58\pm0.24$ & $-6.66^{+0.22}_{-0.21}$ & $-6.12^{+0.18}_{-0.16}$  \\
$\log \, \mathrm{H_2 O}$ & dex & $\mathcal{U}(-12.00, -1.00)$ & $-9.04^{+1.57}_{-1.88}$ & $-9.60^{+1.65}_{-1.51}$ & $-9.64^{+1.57}_{-1.52}$ & $-9.79^{+1.51}_{-1.44}$  \\
$\log \, \mathrm{H_2 S}$ & dex & $\mathcal{U}(-12.00, -1.00)$ & $-6.03^{+0.70}_{-3.83}$ & $-4.61^{+0.28}_{-0.31}$ & $-4.83^{+0.26}_{-0.27}$ & $-4.73\pm0.17$  \\
$\log \, \mathrm{CO_2}$ & dex & $\mathcal{U}(-12.00, -1.00)$ & $-8.15^{+0.32}_{-0.38}$ & $-7.32^{+0.50}_{-0.52}$ & $-9.96^{+1.26}_{-1.34}$ & $-7.83\pm0.31$ \\
$\log \, \mathrm{CO}$ & dex & $\mathcal{U}(-12.00, -1.00)$ & $-9.89^{+1.60}_{-1.36}$ & $-9.64^{+1.69}_{-1.50}$ & $-9.63^{+1.71}_{-1.53}$ & $-9.85^{+1.62}_{-1.42} $  \\
$\log \, \mathrm{SO_2}$ & dex & $\mathcal{U}(-12.00, -1.00)$ & $-7.06^{+0.62}_{-2.95}$ & $-8.69^{+1.71}_{-2.06}$ & $-9.49^{+1.70}_{-1.60}$ & $-9.35^{+1.74}_{-1.75}$  \\
$\mathrm{f}_{\mathrm{spot}}$ & \nodata & $\mathcal{U}(0.00, 1.00)$ & $0.14^{+0.04}_{-0.02}$ & $0.43^{+0.11}_{-0.08}$ & $0.46^{+0.16}_{-0.09}$ & $0.42\pm0.06$  \\
$\mathrm{f}_{\mathrm{fac}}$ & \nodata & $\mathcal{U}(0.00, 0.50)$ & $0.10^{+0.12}_{-0.04}$ & $0.05^{+0.03}_{-0.02}$ & $0.02\pm0.01$ & $0.06^{+0.03}_{-0.02}$  \\
$\mathrm{T}_{\mathrm{spot}}$ & K & $\mathcal{U}(2300, 3430)$ & $2879^{+64}_{-51}$ & $3394^{+23}_{-25}$ & $3412^{+12}_{-20}$ & $3379^{+12}_{-16}$  \\
$\mathrm{T}_{\mathrm{fac}}$ & K & $\mathcal{U}(3430, 4802)$ & $3445^{+19}_{-10}$ & $4032^{+123}_{-101}$ & $4251^{+178}_{-163}$ & $3967^{+94}_{-78}$  \\
$\mathrm{T}_{\mathrm{phot}}$ & K & $\mathcal{N}(3430, 54)$ & $3354^{+36}_{-43}$ & $3562\pm26$ & $3534\pm32$ & $3557\pm24$  \\
$\log \, \mathrm{g}_{\mathrm{spot}}$ & dex & $\mathcal{U}(4.34, 5.34)$ & $5.21^{+0.10}_{-0.14}$ & $4.51^{+0.12}_{-0.10}$ & $4.44^{+0.10}_{-0.07}$ & $4.37^{+0.05}_{-0.02}$ \\
$\log \, \mathrm{g}_{\mathrm{fac}}$ & dex & $\mathcal{U}(4.34, 5.34)$ & $4.69^{+0.33}_{-0.25}$ & $4.43^{+0.12}_{-0.07}$ & $4.56^{+0.27}_{-0.15}$ & $4.37^{+0.06}_{-0.02}$ \\
$\log \, \mathrm{g}_{\mathrm{phot}}$ & dex & $\mathcal{U}(4.34, 5.34)$ & $5.18^{+0.09}_{-0.16}$ & $4.70^{+0.12}_{-0.09}$ & $4.73^{+0.14}_{-0.11}$ & $4.57^{+0.08}_{-0.05}$ \\
$\log~\mathrm{C/O}$ & dex & \nodata & $0.7_{ - 0.5}^{ + 1.0},~2\sigma>-0.12$ & $1.2_{ - 0.5}^{ + 0.4},~2\sigma>0.28$ & $1.2_{ - 0.8}^{ + 0.9},2\sigma>-0.19$ & $1.2_{-0.4}^{+0.3},~2\sigma>0.34$ \\
$\log~[\mathrm{M/H}]$ & dex & \nodata & $-2.8\pm0.3$ & $-1.9\pm0.3$ & $-2.1\pm0.3$ & $-2.0\pm0.2$ \\
$\log~\mathrm{C/H}$ & dex & \nodata & $-2.7\pm0.1$ & $-2.3\pm0.2$ & $-3.3\pm0.2$ & $-2.8\pm0.2$ \\
$\log~\mathrm{O/H}$ & dex & \nodata & $-3.6_{ - 1.0}^{ + 0.5},~2\sigma<-2.73$ & $-3.7_{ - 0.5}^{ + 0.5},~2\sigma<-2.70$ & $-4.8\pm0.9,~2\sigma<-3.23$ & $-4.3_{ - 0.3}^{+ 0.4},~2\sigma<-3.33$\\
$\log~\mathrm{S/H}$ & dex & \nodata & $-1.3_{ - 1.0}^{ + 0.6}$ & $0.0\pm0.3$ & $-0.2\pm0.3$ & $-0.1\pm0.5$\\
\hline
$\chi^2_{\nu}$ & \nodata & \nodata  & 2.42 & 1.67 & 1.55 & 2.76 \\
$\ln Z$ & \nodata & \nodata & 2269.41 & 2401.75 & 2322.32 & 2377.29 \\
\enddata 
\end{deluxetable*}

\subsection{M3.1 for binned \texttt{ExoTiC-JEDI} visit spectra and co-added spectrum}
We also investigated the impact of binning on the reported results. For this test, at Stage 4 (see \hyperref[app:exotic]{Appendix \ref*{app:exotic}}) we binned the light curves using 40 nm bins (4-column bins), resulting in 120 channels spanning $0.6-5.3$ \textmu{}m. All other data reduction choices remained identical to the pixel-level \texttt{ExoTiC-JEDI} reductions. We then performed identical fits using model M3.1 described in \S\ref{sec:freechemret} and present the results in \autoref{tab:binned_exotic_retrievals}. This table is intended to be a direct comparison to the results with pixel-level spectra presented in \autoref{tab:retrievals_clear}. We note that, despite some differences in the planetary temperature or TLS model, the reported VMRs from the visit spectra and co-added spectrum are consistent (within $1\sigma$) with the results from the pixel-level reduction and do not change the conclusion that TOI-5205b appears to have a metal-poor and carbon-rich atmosphere as described in \S\ref{sec:poseidonresult}. 

\begin{deluxetable*}{r|l|l||l|l|l||l}
\tablewidth{0.98\textwidth}
\tabletypesize{\scriptsize}
\tablecaption{Atmospheric retrieval priors and posteriors for the M3.1 model fit to the 40 nm binned \texttt{ExoTiC-JEDI} reductions, which assumed a clear atmosphere and a TLS effect component. This is analogous to \autoref{tab:retrievals_clear} for the pixel-level \texttt{ExoTiC-JEDI} reductions. \label{tab:binned_exotic_retrievals}}
\tablehead{
\colhead{Parameters} & 
\colhead{Units} & 
\multicolumn{1}{c||}{Priors} & 
\colhead{Visit 1} & 
\colhead{Visit 2} & 
\multicolumn{1}{c||}{Visit 3} & 
\colhead{Visits co-added}}
\startdata
$\mathrm{R}_{\mathrm{p, \, ref}}$ & $\mathrm{R_J}$ & $\mathcal{U}(0.80, 1.08)$ & $0.933\pm0.001$ & $0.936\pm0.001$ & $0.935^{+0.001}_{-0.002}$ & $0.934\pm0.001$  \\
$\mathrm{T}$ & K & $\mathcal{U}(200.00, 1500.00)$ & $822^{+111}_{-102}$ & $798^{+73}_{-74}$ & $923.2^{+147}_{-120}$ & $792^{+56}_{-58}$ \\
$\log \, \mathrm{CH_4}$ & dex & $\mathcal{U}(-12.00, -1.00)$ & $-6.09^{+0.29}_{-0.31}$ & $-5.69^{+0.24}_{-0.23}$ & $-6.70^{+0.29}_{-0.30}$ & $-6.05\pm0.16$ \\
$\log \, \mathrm{H_2 O}$ & dex & $\mathcal{U}(-12.00, -1.00)$ & $-9.58^{+1.59}_{-1.53}$ & $-9.76^{+1.54}_{-1.41}$ & $-9.70^{+1.58}_{-1.48}$ & $-9.93^{+1.40}_{-1.34}$ \\
$\log \, \mathrm{H_2 S}$ & dex & $\mathcal{U}(-12.00, -1.00)$ & $-5.36^{+0.62}_{-3.60}$ & $-4.46^{+0.31}_{-0.36}$ & $-4.62^{+0.32}_{-0.39}$ & $-4.61^{+0.18}_{-0.19}$ \\
$\log \, \mathrm{CO_2}$ & dex & $\mathcal{U}(-12.00, -1.00)$ & $-9.63^{+1.28}_{-1.47}$ & $-7.39^{+0.62}_{-0.81}$ & $-10.09^{+1.27}_{-1.22}$ & $-8.36^{+0.61}_{-1.22}$ \\
$\log \, \mathrm{CO}$ & dex & $\mathcal{U}(-12.00, -1.00)$ & $-9.15^{+1.97}_{-1.79}$ & $-9.31^{+1.86}_{-1.69}$ & $-8.85^{+2.10}_{-2.00}$ & $-9.43^{+1.84}_{-1.66}$  \\
$\log \, \mathrm{SO_2}$ & dex & $\mathcal{U}(-12.00, -1.00)$ & $-9.38^{+1.69}_{-1.66}$ & $-9.10^{+1.89}_{-1.84}$ & $-9.49^{+1.70}_{-1.61}$ & $-9.46^{+1.71}_{-1.64}$  \\
$\mathrm{f}_{\mathrm{spot}}$ & \nodata & $\mathcal{U}(0.00, 1.00)$ & $0.55^{+0.07}_{-0.13}$ & $0.34^{+0.12}_{-0.10}$ & $0.49^{+0.16}_{-0.11}$ & $0.43^{+0.07}_{-0.09}$  \\
$\mathrm{f}_{\mathrm{fac}}$ & \nodata & $\mathcal{U}(0.00, 0.50)$ & $0.02\pm0.01$ & $0.04^{+0.03}_{-0.01}$ & $0.02\pm0.01$ & $0.04\pm0.01$  \\
$\mathrm{T}_{\mathrm{spot}}$ & K & $\mathcal{U}(2300, 3430)$ & $3379\pm20$ & $3369^{+34}_{-52}$ & $3406^{+16}_{-23}$ & $3396^{+19}_{-22}$  \\
$\mathrm{T}_{\mathrm{fac}}$ & K & $\mathcal{U}(3430, 4802)$ & $4169^{+127}_{-96}$ & $4057^{+155}_{-134}$ & $4275^{+181}_{-176}$ & $4088^{+73}_{-77}$  \\
$\mathrm{T}_{\mathrm{phot}}$ & K & $\mathcal{N}(3430, 54)$ & $3493\pm30$ & $3550^{+33}_{-35}$ & $3515^{+30}_{-25}$ & $3546\pm23$  \\
$\log \, \mathrm{g}_{\mathrm{spot}}$ & dex & $\mathcal{U}(4.34, 5.34)$ & $4.37^{+0.06}_{-0.02}$ & $4.53^{+0.15}_{-0.12}$ & $4.46^{+0.11}_{-0.08}$ & $4.43^{+0.10}_{-0.07}$ \\
$\log \, \mathrm{g}_{\mathrm{fac}}$ & dex & $\mathcal{U}(4.34, 5.34)$ & $4.47^{+0.15}_{-0.09}$ & $4.47^{+0.16}_{-0.09}$ & $4.57^{+0.28}_{-0.16}$ & $4.38^{+0.07}_{-0.03}$ \\
$\log \, \mathrm{g}_{\mathrm{phot}}$ & dex & $\mathcal{U}(4.34, 5.34)$ & $4.59^{+0.09}_{-0.06}$ & $4.83^{+0.18}_{-0.14}$ & $4.75^{+0.14}_{-0.12}$ & $4.66^{+0.09}_{-0.07}$ \\
$\log~\mathrm{C/O}$ & dex & \nodata & $1.5_{ - 0.8}^{ + 0.9},~2\sigma>0.11$ & $1.1_{ - 0.5}^{ + 0.6},~2\sigma>0.17$ & $0.9_{-0.8}^{+1.1},~2\sigma>-0.28$ & $1.5_{-0.6}^{+0.7},~2\sigma>0.34$ \\
$\log~[\mathrm{M/H}]$ & dex & \nodata & $-2.5_{ - 0.9}^{ + 0.6}$ & $-1.7\pm0.3$ & $-1.9_{-0.4}^{+0.3}$ & $-1.9\pm0.2$ \\
$\log~\mathrm{C/H}$ & dex & \nodata & $-2.8\pm0.3$ & $-2.4_{-0.2}^{+0.3}$ & $-3.3_{-0.3}^{+0.4}$ & $-2.7\pm0.2$ \\
$\log~\mathrm{O/H}$ & dex & \nodata & $-4.5_{ - 1.0}^{ + 0.9},~2\sigma<-2.77$ & $-3.7_{ - 0.7}^{ + 0.6},~2\sigma<-2.58$ & $-4.5\pm1.1,~2\sigma<-2.27$ & $-4.5_{ - 0.7}^{+ 0.8},~2\sigma<-3.19$\\
$\log~\mathrm{S/H}$ & dex & \nodata & $-0.7_{ - 2.7}^{ + 0.6}$ & $0.2_{ - 0.4}^{ + 0.3}$ & $0.0_{-0.4}^{+0.3}$ & $0.0\pm0.2$\\
\hline
$\chi^2_{\nu}$ & \nodata & \nodata & 2.74 & 1.61 & 1.44 & 3.39 \\
$\ln Z$ & \nodata & \nodata & 712.72 & 761.49 & 745.07 & 725.97 \\
\enddata 
\end{deluxetable*}

\end{document}